\documentclass[twocolumn, 12pt]{aa}
\pdfoutput=1

\usepackage[hidelinks]{hyperref}
\usepackage[flushleft]{threeparttable}
\DeclareUnicodeCharacter{2212}{-}
\usepackage[export]{adjustbox}
\usepackage{xcolor}
\hypersetup{
    colorlinks,
    linkcolor={blue!60!black},
    citecolor={blue!60!black},
    urlcolor={blue!80!black}
}
\usepackage[switch]{lineno}

\newcommand{\lb}{\label}
\newcommand{\adeg}{^\circ\!\!}
\newcommand{\Fermi}{\textit{Fermi}\xspace}
\newcommand{\LAT}{\textsl{LAT}\xspace}

\begin{document}

\title{Hunting for gamma-ray emission from Fast Radio Bursts}
  


  \author{G.~Principe \thanks{\email{giacomo.principe@ts.infn.it}}
          \inst{1,2,3}
          \and
          L.~Di~Venere 
          \inst{4}
          \and
          M.~Negro  
          \inst{5, 6, 7}
          \and
          N.~Di Lalla 
          \inst{8}
          \and
           N.~Omodei 
          \inst{8}
          \and
          R.~Di Tria 
          \inst{9,4}
          \and
          M.N.~Mazziotta
          \inst{4}
          \and
		  F.~Longo 
          \inst{1,2}
          }

   \institute{
            {Dipartimento di Fisica, Universit\'a di Trieste, I-34127 Trieste, Italy}
            \and 
            {Istituto Nazionale di Fisica Nucleare, Sezione di Trieste, I-34127 Trieste, Italy}
            \and
            {INAF - Istituto di Radioastronomia, I-40129 Bologna, Italy}
            \and
             {Istituto Nazionale di Fisica Nucleare, Sezione di Bari, I-70126 Bari, Italy}
            \and
            {University of Maryland, Baltimore County, Baltimore, MD 21250, USA}
            \and
            {NASA Goddard Space Flight Center, Greenbelt, MD 20771, USA}
            \and
            {Center for Research and Exploration in Space Science and Technology, NASA/GSFC, Greenbelt, MD 20771, USA}
            \and
             {W. W. Hansen Experimental Physics Laboratory, Kavli Institute for Particle Astrophysics and Cosmology, Department of Physics and
SLAC National Accelerator Laboratory, Stanford University, Stanford, CA 94305, USA}
            \and
            {Dipartimento Interateneo di Fisica dell'Universit\'a e del Politecnico di Bari, I-70125 Bari, Italy}
            }
            
            
   \date{Received March 24, 2023; accepted May 15, 2023}


  \abstract
   {Fast radio bursts (FRBs) are a recently discovered class of GHz-band, ms-duration, Jy-level-flux astrophysical transients. Although hundreds of models have been proposed so far for FRB progenitors (most popular ones involve magnetars), their physical origin and emission mechanism are still a mystery, making them one of the most compelling problems in astrophysics.
    } 
   {FRBs are caused by astrophysical processes that are not yet understood. Exploring their high-energy counterpart is crucial for constraining their origin and emission mechanism.}
   {Thanks to more than 13 years of gamma-ray data collected by the Large Area Telescope on board the \textit{Fermi} Gamma-ray Space Telescope, and to more than 1000 FRB events (from 561 non-repeating and 22 repeating sources), one of the largest sample created as of today, we perform the largest and deepest search for high-energy emission from FRB sources to date, between 100 MeV and 1 TeV.
   In addition to the analysis involving individual FRB events on different time-scales (from few seconds up to several years), we performed, for the first time, a stacking analysis on the full sample of FRB events as well as a search for triplet photons in coincidence with the radio event.}
   {We do not detect significant emission, reporting the most stringent constraints, on short time scales, for the FRB-like emission from SGR 1935+2154 with E$_{\gamma}<10^{41}$ erg, corresponding to a factor $\eta < 10^{7}$ with respect to the emitted radio energy. 
   Similarly, for the stacked signal of steady emission from all repeaters, the obtained upper limit (UL) on the FRBs luminosity (L$_\gamma < 1.6 \times 10^{43}$ erg s$^{-1}$) is more than two orders of magnitudes lower than those derived from the individual sources.
   Finally, no individual or triplet photons have been significantly associated with FRB events. We derived the LAT ms-sensitivity to be $\sim$0.3 ph cm$^{−2}$ s$^{−1}$ and constrained the gamma-ray energy $E_{\gamma, \delta_T = 1 \textrm{ms}} \lesssim 10^{47} (\textrm{D}_L/150 \textrm{Mpc})^2$ erg, ruling out a gamma-ray-to-radio energy ratio greater than $10^{9}$ on ms timescales.
   }
   {The results reported here represent the most stringent UL reported so far on the high-energy emission from FRBs on short and long time scales, as well as on cumulative emission and individual photon searches. While the origin of FRBs is still unclear, our work provides important constraints for FRB modeling, which might shed light on their emission mechanism.
   }
   {}


\maketitle
   

\section{Introduction}
\lb{sec:intro}

Fast radio bursts (FRBs) are a new transient phenomenon 
consisting of bright and short-duration radio pulses that flash unpredictably in the sky \citep{2013Sci...341...53T,2019A&ARv..27....4P,2019ARA&A..57..417C}.
They usually present large dispersion measures\footnote{Column density of free electrons along a line of sight which is measured with delay of pulse arrival time as a function of frequency.} (DMs) exceeding  Galactic values, suggesting a possible extra-galactic origin \citep{2020Natur.581..391M}.
Despite being discovered more than 15 years ago \citep{2007Sci...318..777L}, their origin and emission mechanism are not clearly understood.


After their discovery in 2007, more than a thousand FRB events (from about 500 published FRB sources) have so far been reported \citep[see e.g. FRB catalogs][]{2016PASA...33...45P,2021ApJS..257...59C}. The first repeating FRB (FRB\,20121102) was discovered by \cite{2016Natur.531..202S}. Since then, many FRBs have been found to exhibit repeating bursts randomly in time.
Despite several attempts \citep[e.g.,][]{2016MNRAS.457.2530B,2018NatAs...2..839C}, no clear indications for physically different populations distinguishing repeating and non-repeating sources have been identified. 
Very recently, \citet{2021ApJ...923....1P} showed that bursts from repeating sources, on average, have larger widths and are narrower in bandwidth.
A couple of repeaters present a periodic pattern in their activity cycle. This is the case for FRB\,20180916 \citep{2020Natur.582..351C}, and for the first repeater FRB\,20121102 \citep{2020MNRAS.495.3551R}, for which only a $\sim$3$\sigma$ significance periodicity has been reported.
More recently, considering also the multi-components of individual bursts, a significant (6.5$\sigma$) periodicity of 216.8 ms has been observed in the pulse profile of the FRB\,20191221A. Such short periodicity provides
strong evidence for a neutron-star origin of the event \citep{2022Natur.607..256C}.

While a wide variety of theoretical models have been proposed for FRBs \footnote{see frbtheorycat.org} \citep[][]{2019PhR...821....1P}, their origin and emission mechanism are still an open question \citep[see  e.g.][]{2020Natur.587...45Z,2021SCPMA..6449501X,2022arXiv221203972Z}. In April 2020 an FRB-like emission was first associated with the Galactic magnetar SGR\,1935+2154 \citep{2020Natur.587...54C}. An X-ray burst, recorded by the INTEGRAL, AGILE, Konus-Wind and Insight-HMXT telescopes \citep[][respectively]{2020ApJ...898L..29M,2021NatAs...5..401T,2021NatAs...5..372R,2021NatAs...5..378L} was seen in coincidence with a very bright FRB-like emission. This event supports the magnetar scenario for the origin of FRBs and opens the possibility of detecting their high-energy emission counterpart \citep{2020ApJ...899L..27M}.
Recently (October 2022), a couple of more coincident radio and X-ray events
were detected from the same source \citep{2022ATel15681....1D,2022ATel15686....1F,2022ATel15698....1L,2022ATel15682....1W,2022ATel15794....1Y}. Although this does not prove that FRB events are due to magnetar activity, magnetars seem to be the current best candidate for FRB origin.

Besides the FRB-like emission from a Galactic magnetar, only two other FRB sources (FRB\,121102 and FRB\,20190520B) have been unambiguously associated with compact persistent radio sources with faint emission $\sim$200$\mu$Jy and no clear counterpart \citep{2017Natur.541...58C,2022Natur.606..873N}. 
In addition, about 20 FRBs have been accurately localised (at arcsec or milli-arcsec scale) to a host galaxy \citep{2020ApJ...903..152H}.
While the localization of some FRBs in a star-forming region within their host galaxies \citep[e.g. FRB\,20121102, FRB\,20180916][]{2017Natur.541...58C,2017ApJ...834L...8M,2017ApJ...834L...7T,2020Natur.577..190M} is consistent with active repeating FRBs associated with young extreme objects such as magnetars, a precise localization of the FRB\,20200120E repeater to a Globular Cluster (GC) within the M81 galaxy challenges this scenario \citep{2021ApJ...910L..18B,2022Natur.602..585K}. 
The extragalactic origin of the FRB events is also in agreement with their large values of DM. The luminosity distances of these host galaxies ranges from 3.6 Mpc \citep[for FRB\,20200120E located in the M81 galaxy, ][]{2021ApJ...910L..18B} to 4 Gpc \citep[corresponding to z=0.66, for FRB\,20190523A, ][]{2019Natur.572..352R}, giving a solid basis to their cosmological origin.


Many attempts to search for multi-wavelength and multi-messenger emission both for neutrinos and gravitational waves (GW) from FRBs have been performed in order to constrain the origin of these radio phenomena among the hundreds of different models predicted \citep[see, e.g.,][]{2018ApJ...857..117A,2019PhR...821....1P,2021Univ....7...76N,2023ApJ...946...80A,2022arXiv220312038T}.
The predicted neutrino and GW emissions are usually very faint and an event needs to be close enough for them to be detected. Besides the case of SGR\,1935+2154, the rarity of nearby FRB sources may be the main reason for the absence of further multi-wavelength and multi-messenger counterparts.

The most popular models for the origin of FRBs invoke strongly magnetized neutron stars (i.e., magnetars) and predict gamma-ray emission associated with these radio bursts \citep[see e.g.][]{2013arXiv1307.4924P,2014MNRAS.442L...9L,2014ApJ...780L..21Z,2017ApJ...836L...6M,2021ApJ...919...89Y,2023arXiv230110184W}.
In particular, accordingly to \citet{2014MNRAS.442L...9L,2019MNRAS.485.4091M,2020ApJ...899L..27M} FRBs are expected to arise from synchrotron maser emission at ultra-relativistic magnetized shocks, such as those produced by flare ejecta from young magnetars. In this scenario FRBs are produced by magnetar flares presenting an afterglow emission peaking at gamma-ray energies
E$_{\textrm{peak}} >$ MeV–GeV with luminosities of $L_\gamma \sim 10^{45}-10^{46}$ erg s$^{-1}$ on a time-scale of 0.1$-$10 ms, i.e. comparable to the FRB event itself  \citep{2019MNRAS.485.4091M}. Moreover, for \citet{2020ApJ...899L..27M} the expected gamma-ray energy is expected to be at least a factor $\eta \geq 10^{4}$ larger than the emitted radio energy, while an even larger fluence ratio (10$^{5}$--10$^{6}$) 
has been predicted by \citet{2014MNRAS.442L...9L}.


The recent detection of high-energy emission, at GeV energies, from a magnetar giant flare in the Sculptor galaxy by \citet{2021NatAs.tmp...11F} motivates the search for gamma-ray counterparts to the known FRB events.
In the last years, a few number of searches for FRB counterparts at gamma-ray energies have been performed without finding any significant detection but only analysing a limited sample of FRB events \citep[up to a few dozen; see, e.g.,][]{2019ApJ...879...40C,2020A&A...637A..69G,2020ApJ...893L..42T,2021ApJ...915..102V}, 

Thanks to its large field of view and good sensitivity, the \textit{Fermi} Large Area Telescope \citep[LAT,][]{2009ApJ...697.1071A} is an ideal instrument for monitoring transient events that occur at unpredictable position in the sky.
Taking advantage of over 13 years of data collected by \textit{Fermi}-LAT, and to 1020 published FRBs events (see Section \ref{sec:sample} for more details on the selected sample), we perform the largest and deepest systematic search for high-energy counterparts of the reported repeating and non-repeating FRB sources.

We use different analysis techniques with the intent of unveiling temporal coincidences between FRBs and gamma-ray events at different timescales, ranging from a few seconds to days/weeks and up to several years, with \textit{Fermi}-LAT \citep[for preliminary results on the periodic FRB\,20180916 see][]{2022icrc.confE.624P}.
In addition to the study of each individual FRB source, we perform a stacking analysis to investigate the cumulative emission for the several events from each repeating FRB, as well as from all the events in our sample. Finally we also apply a triplet photon counting method, previously used for magnetar-flares searches \citep{2021NatAs.tmp...11F}, to search for clusters of photons coming from the FRB sources.

Throughout this article, we report the radio flux density at 600 MHz and we assume the following cosmological values: $H_{0} = 70$ km s$^{-1}$ Mpc$^{-1}$ , $\Omega_{M} = 0.3$, and $\Omega_{\Lambda}=0.7$ in a flat Universe.
\section{Sample of FRB events}
\label{sec:sample}
In this work we considered all the published FRB events reported up to September, 2021.
The FRBs included in our sample were selected from the following resources: 
\begin{itemize}
    \item 118 events from the FRBCAT\footnote{\url{https://frbcat.org}} \citep{2016PASA...33...45P};
    \item 535 repeating and non-repeating FRBs reported in the first CHIME/FRB catalog \citep{2021ApJS..257...59C};
    \item 230 bursts\footnote{\url{http://www.chime-frb.ca/repeaters}} from the 20 repeating FRBs reported by the CHIME/FRB collaboration as of September 15, 2021, including 73 bursts from the periodic FRB\,20180916;
    \item 235 bursts from FRB\,121102 collected by \citet{2020MNRAS.495.3551R}.
\end{itemize}

Because several events are reported in more than one of the above-listed samples, we removed the repetitions. Six FRB events were not considered ``a priori'' in our analysis because they occurred before the launch of the \textit{Fermi}-LAT satellite in June, 2008 (FRB\,20010724, FRB\,20010621, FRB\,20010312, FRB\,20010305) or during the period at the end of March 2018 (FRB\,20180324, FRB\,20180321), when the LAT was in safety mode for over 23 days without taking any data due to a problem with a solar panel orientation \citep{2021ApJS..256...12A}.
In addition to the listed samples of FRB events, we considered also the FRB-like emission from the Galactic magnetar SGR\,1935+2154. 
For most of the FRB events in our sample, we have collected information on the name, dedispersed time\footnote{Time of arrival with reference to infinite frequency for the specific burst.} (UTC), position (RA, Dec) with relative uncertainties, DM, width, flux and fluence with relative uncertainties. 
Regarding the FRB name we adopted for the non-repeating FRB events from the first CHIME/FRB catalog the TNS name format: FRB\,YYYYMMDDx, which includes also the information of the number of source reported on a given UTC day in the final letter 'x', while for all the others events we used a general format FRB\,YYYYMMDD.

Among the 20 FRB sources with known host galaxy as of today, 14 have been included in our sample \citep{2017Natur.541...58C,2019Sci...366..231P,2019Natur.572..352R,2020ApJ...903..152H,2020ApJ...895L..37B,2021ApJ...917...75M,2022Natur.609..685X}; the radio properties of these FRBs are presented in Table \ref{tab:frb_host}. A possible galaxy association has been reported for only two of the repeaters included in our sample: FRB\,20180814A and FRB\,20190303A\citep[z$\sim$0.068 and z$\sim$0.064,][respectively]{2022arXiv221211941M}.

For many of the events from repeating FRBs reported in the CHIME/FRB website\footnote{\url{http://www.chime-frb.ca/repeaters}}, only preliminary time and positions are available and further information will be provided in a forthcoming publication by the CHIME/FRB collaboration.
All the information on the FRB events in our sample can be found at following link\footnote{\url{https://www-glast.stanford.edu/pub\_data/1807/}}, accompanied with the results obtained in our analysis of their gamma-ray counterpart emission.

\begin{table*}
\caption{\small FRB sources with known host galaxy.  
\label{tab:frb_host}} 
\centering
\begin{threeparttable}
\small
\centering
\begin{tabular}{ccccccccc}
\hline \hline
Name & RA & Dec & DM & D$_{L}$ & z & S$_r$\tnote{$a$} & L$_r$\tnote{$b$} & E$_r$\tnote{$c$} \\
& deg & deg & pc cm$^{-3}$ & Mpc & & Jy & erg s$^{-1}$ & erg \\
\hline
FRB\,20190711 & 329.42 & -80.36 & 593.1 & 3000 & 0.522 & 4.5 & 4.9$\times 10^{43}$ & 2.09$\times 10^{41}$\\
FRB\,20190611 & 320.75 & -79.40 & 321.4 & 2050 & 0.3778 & 10 & 5.03$\times 10^{43}$ & 1.14$\times 10^{41}$\\
FRB\,20190608 & 334.02 & -7.90 & 338.7 & 550 & 0.1179 & 2.7 & 9.88$\times 10^{41}$ & 7.52$\times 10^{39}$\\
FRB\,20190523 & 207.06 & 72.47 & 760.8 & 4000 & 0.66 & - & - & -\\
FRB\,20190102 & 322.42 & -79.48 & 363.6 & 1500 & 0.291 & 73 & 2.0$\times 10^{44}$ & 3.01$\times 10^{40}$\\
FRB\,20181112 & 327.35 & -52.97 & 589.3 & 2700 & 0.4755 & 310 & 2.69$\times 10^{45}$ & 1.54$\times 10^{41}$\\
FRB\,20180924 & 326.11 & -40.9 & 361.4 & 1700 & 0.3214 & 9.5 & 3.27$\times 10^{43}$ & 4.30$\times 10^{40}$\\
FRB\,20180916 & 29.50 & 65.73 & 349.2 & 150 & 0.0337 & 1.7 & 4.52$\times 10^{40}$ & 7.02$\times 10^{38}$\\
FRB\,20121102$^{r}$ & 82.99 & 33.15 & 560.0 & 950 & 0.19273 & 0.31 & 3.33$\times 10^{41}$ & 8.72$\times 10^{38}$\\
SGR\,1935+2154$^{r}$ & 293.73 & 21.90 & 332.7 & 0.014 & - & 110000 & 7.00$\times 10^{36}$ & 3.00$\times 10^{34}$\\ 
FRB\,20201124A$^{r}$ & 77.0 & 26.05 & 413.5 & 450 & 0.09795 & - & - & -\\
FRB\,20200120E$^{r}$ & 149.25 & 68.82 & 87.8 & 3.6 & 0.00014 & 1.8 & 2.79$\times 10^{37}$ & 2.66$\times 10^{34}$\\
FRB\,20190520A & 273.52 & 26.32 & 431.9 & 1200 & 0.241 & 1.10 & 1.96$\times 10^{42}$ & 2.72$\times 10^{39}$\\
FRB\,20171020 & 333.75 & -19.67 & 114.1 & 38 & 0.0087 & 52 & 8.84$\times 10^{40}$ & 3.15$\times 10^{38}$\\
\hline
\end{tabular}
\begin{tablenotes}
    \item{$^a$} Radio flux density at peak time.
    \item{$^b$} Radio luminosity.
    \item{$^c$} Radio energy.
    \item{$^r$} FRB sources known to repeat, in their case we report the observed maximum value of S$_r$, L$_{r}$ and E$_{r}$.
\end{tablenotes}
\end{threeparttable}
\end{table*}

Our final sample counts 1020 FRB events from 583 different celestial objects: 561 non repeating FRB sources, and 459 events from 22 repeaters.
Among the repeaters, most of the events (232) are related to the first repeater \citep[FRB\,20121102,][]{2017Natur.541...58C}, while 72 events are from the periodic FRB\,20180916 \citep{2020Natur.582..351C}, and 32 events from FRB\,20201124A located in a Milky-way sized, barred-spiral galaxy \citep{2022Natur.609..685X}. 
The remaining repeaters have between 2 and 21 events included in our sample (see also Table \ref{tab:frb_rep}).

\begin{figure}
\centering
\includegraphics[width=\columnwidth]{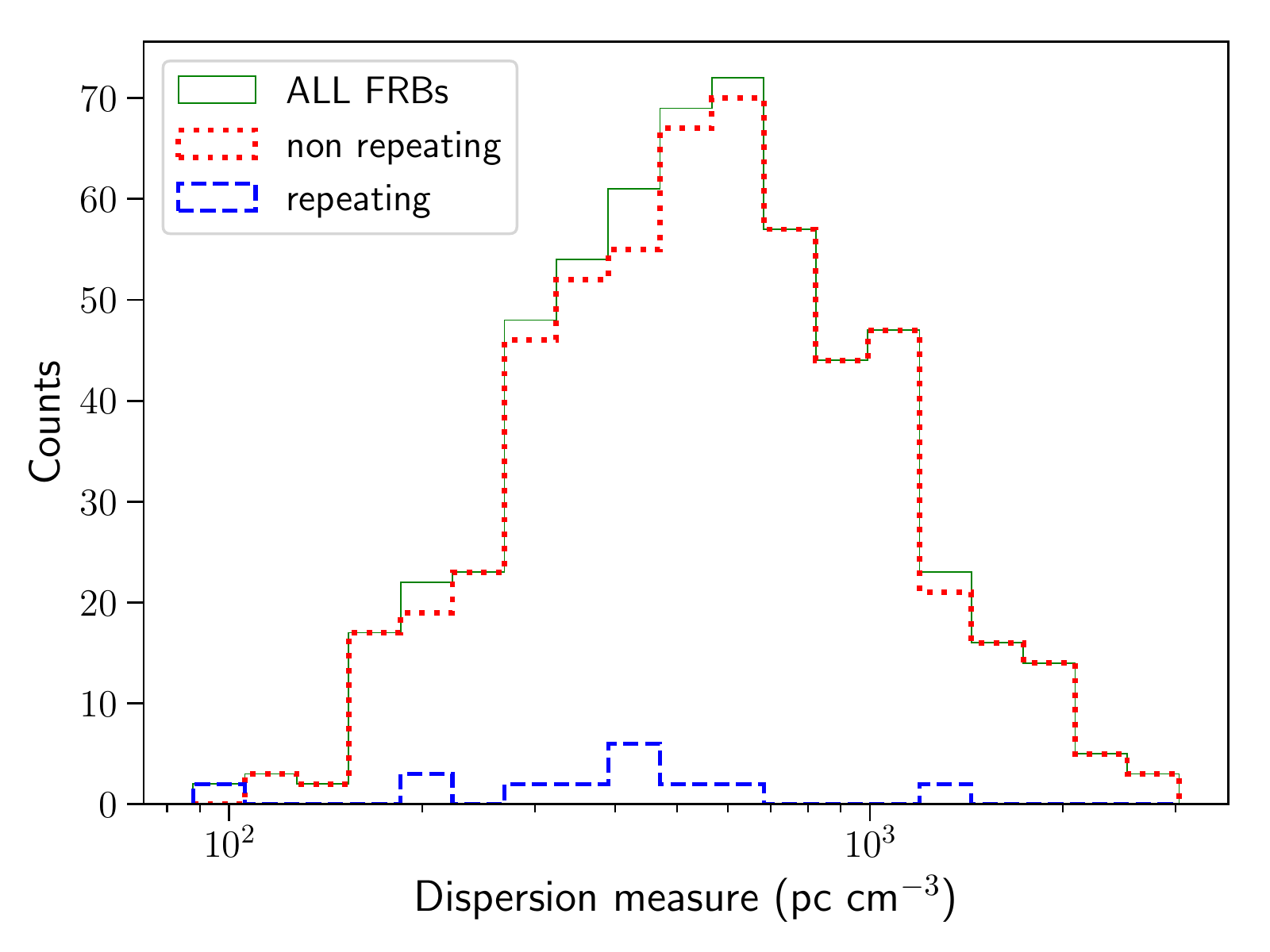}
\caption{\small \label{fig:dm_distrib}
Dispersion measure (DM) distribution of the FRB sources contained in our sample.}
\end{figure}

The distribution of the DM of the FRB sources contained in our sample is shown in Fig. \ref{fig:dm_distrib}. 

\begin{figure}
\centering
\includegraphics[width=1.1\columnwidth]{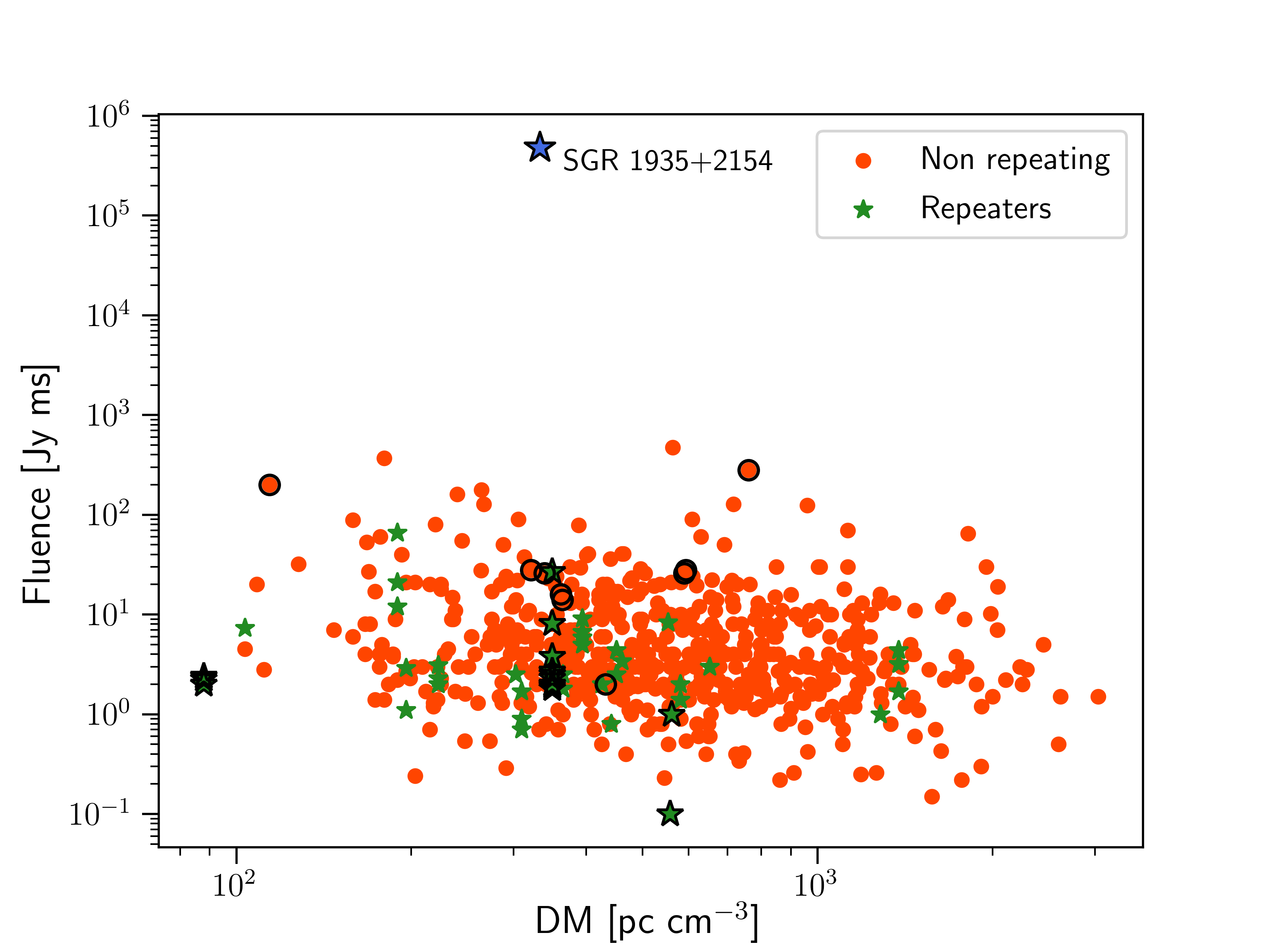}
\caption{\small \label{fig:fluence_dm}
The distribution of radio fluence vs DM for the FRB events in our sample. FRBs with known host-galaxy are outlined in black.}
\end{figure}

\noindent The fluence of the FRB events in our sample presents a median value of a few Jy ms, with a broad range of values starting at 0.1 Jy ms up to over 100000 Jy ms for the FRB-like emission from SGR\,1935+2154 (see Fig. \ref{fig:fluence_dm}). 

\subsection{FRB distance and luminosity derivation}
\label{sec:fruitbat}
We use FRUITBAT\footnote{\url{https://fruitbat.readthedocs.io/}} \citep{2019JOSS....4.1399B}, an open source python package, in order to estimate the redshift of FRB sources with unknown host galaxies  in our sample using their dispersion measures and Galactic coordinates. The latter are used to correct for the Galactic contribution on the DM value, which depends on the source position (particularly relevant for the sources located in the Galactic plane).   
For the redshift derivation we adopted the method derived by \citet{2018ApJ...867L..21Z} and the cosmological parameters obtained by 
\citet{2020A&A...641A...6P}.

Apart from the three FRBs with known hosts, namely SGR\,1935+2154, FRB\,20200120E and FRB\,20171020 which are located in our Galaxy (at 14 kpc), in the nearby M81 (3.6 Mpc) and Sc (37.7 Mpc) galaxies, respectively, all the other FRBs present a luminosity distance larger than 100 Mpc, with a maximum value of about 21 Gpc (z$\sim$ 2.55) for FRB\,20180906B.
For two FRB sources in our sample, namely  
FRB\,20180430 (DM=264.1, l=$221.75^{\circ}$, b=-4.61$^{\circ}$)
and FRB\,20170606 (DM=247.0, l=167.87$^{\circ}$, b=4.79$^{\circ}$), it was not possible to derive a reliable distance estimate, since both of them are located along the Galactic plane (b$<5^{\circ}$) and present low DM values (DM$<$300 pc cm$^{-3}$), which are compatible with that of the Galaxy.

\begin{figure}
\centering
\includegraphics[width=1.1\columnwidth]{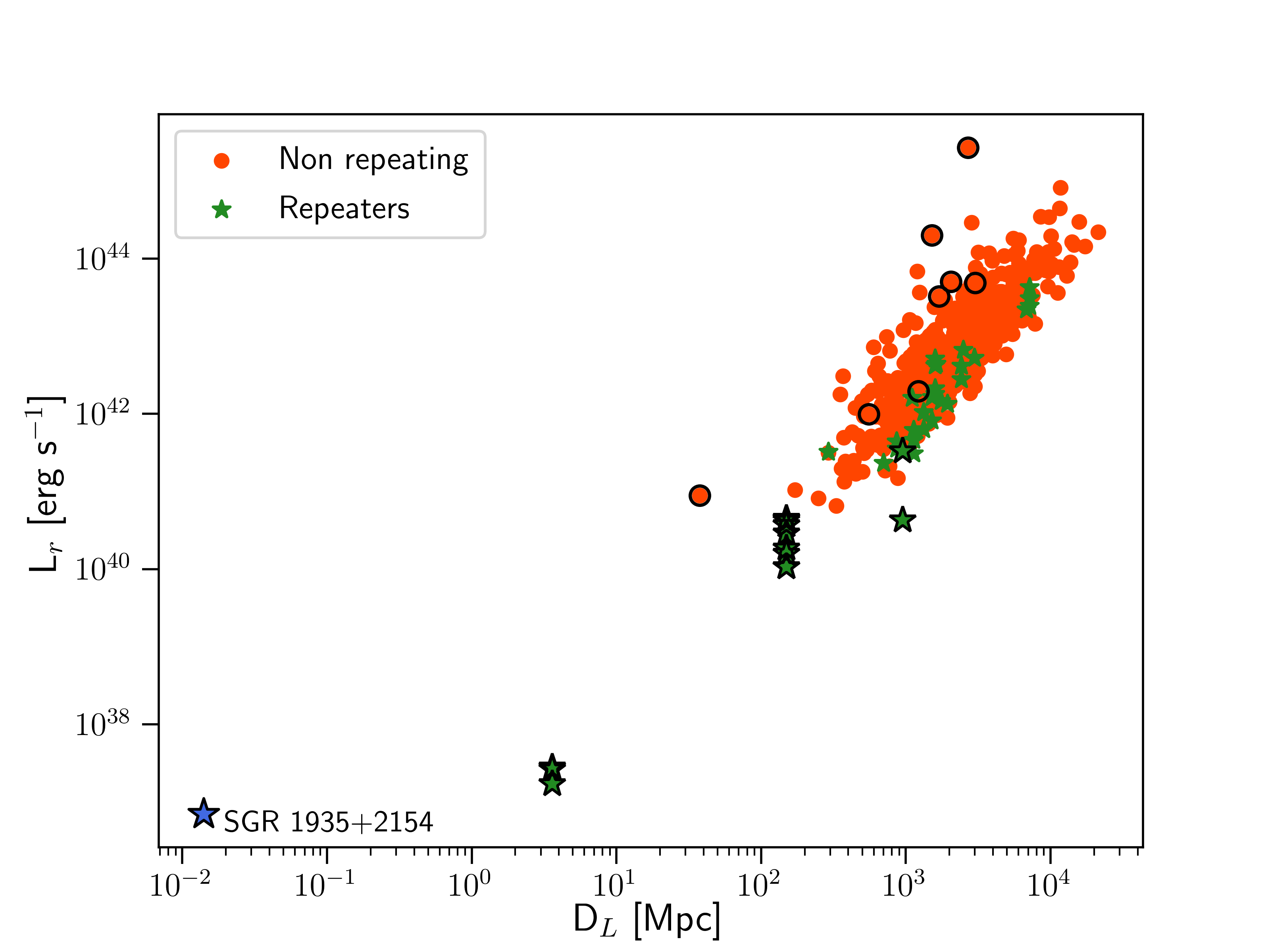}
\includegraphics[width=1.1\columnwidth]{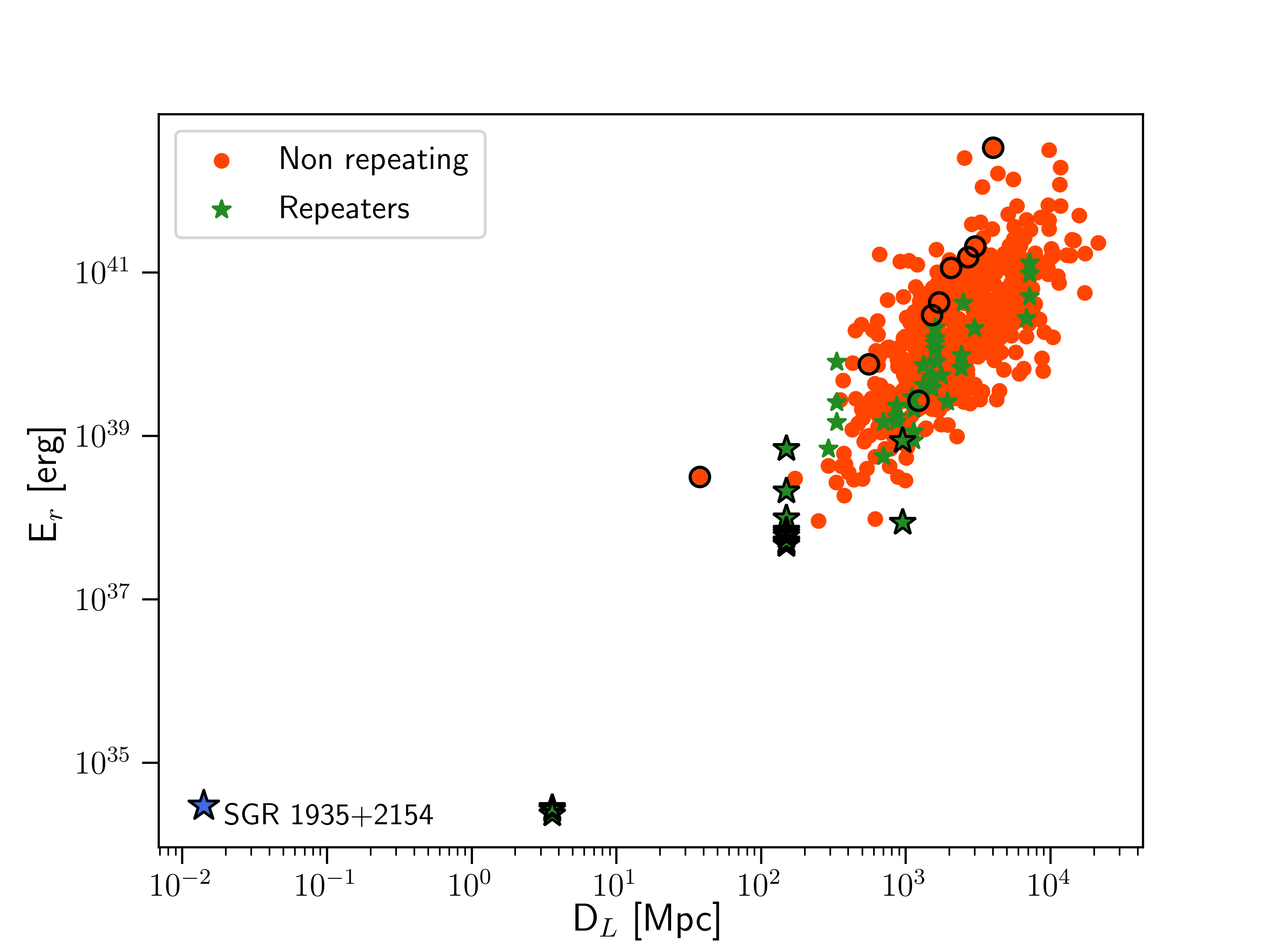}
\caption{\small \label{fig:lum_en_dist}
The radio luminosity (upper plot) and energy (bottom) vs. luminosity distance for all of the FRBs in our sample that have flux measurements. FRBs with known host-galaxy are outlined in black.}
\end{figure}

In addition to the redshift and distance estimation, we used the flux and fluence information to derive the total energy emitted and luminosity of each individual FRB event. Fig. \ref{fig:lum_en_dist} shows the radio luminosity (upper plot) and energy (bottom) as a function of distance for all of the FRB events in our sample that have flux measurements. The FRBs' luminosities (energies) range between $10^{34}$ and $10^{45}$ erg s$^{-1}$ ($10^{32}$ and $10^{43}$ erg). 

\subsection{FRBs-events sample and 4FGL-DR2 comparison}
\begin{figure*}
\centering
\includegraphics[trim=1.5cm 2.5cm 1.5cm 3.3cm,clip,width=14cm]{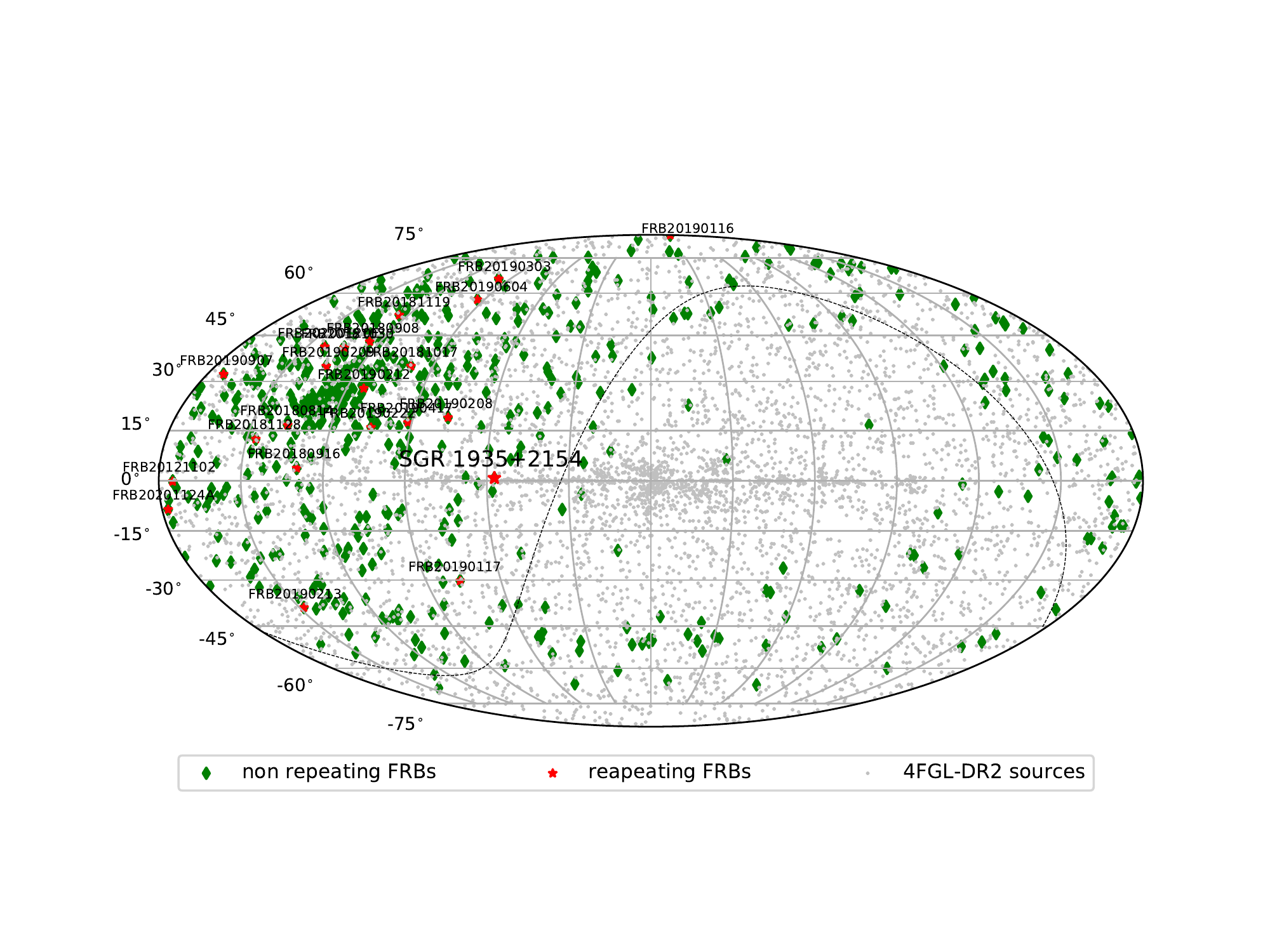}
\caption{\small \label{fig:sky_map}
Sky map, in Galactic coordinates and Mollweide projection, showing the sky coordinates of the FRB sources in our sample: 561 non repeating and 22 repeating FRB sources. Repeating FRBs are labeled in the plot. All the 4FGL-DR2 sources \citep{2020ApJS..247...33A} are also plotted, with gray points, for a comparison.}
\end{figure*}

Fig. \ref{fig:sky_map} shows the sky map of the selected FRB sources, compared with the gamma-ray sources in the second data release of the fourth \textit{Fermi}-LAT source catalog \citep[4FGL-DR2,][]{2020ApJS..247...33A}. 
The spatial distribution is highly asymmetric between the northern and southern hemispheres, due to the large number of bursts detected by the CHIME/FRB telescope \citep{2018ApJ...863...48C}, which is located in Canada. 

\noindent Comparing the location of the FRB sources with the 4FGL-DR2 sources we found that sixteen of them have compatible positions, falling within the large positional error ($0.2 ^{\circ}$--0.4$ ^{\circ}$) of CHIME/FRB. 
Among these, three lie within the positional uncertainty of the 4FGL-DR2 sources ($\sim 0.05^{\circ}$) and are therefore the most promising for an association with a \textit{Fermi}-LAT source: FRB\,20180309, FRB\,20180924A and FRB\,20190220A.
Assuming that all the 4FGL-DR2 extragalactic sources ($\sim$3000) are homogeneously distributed in the sky and considering a tolerance radius of 0.05$^{\circ}$ (similar to the averaged localisation uncertainty in 4FGL-DR2), as well as the number of different celestial objects in our sample (583), we would expect to have three associations by chance with a probability of about 10--15\%

FRB\,20180309 is located in the vicinity of the milli-second pulsar 4FGL\,J2124.7-335 \citep[PSR\,J2124-3358 DM=4.6152,][]{2005AJ....129.1993M}. While neutron stars are one of the most popular candidates for the progenitors of FRB  \citep{2014A&A...562A.137F,2019PhR...821....1P}, this FRB event, which is among the brightest detected so far (with a fluence of $F > 83.5$ Jy ms), presents a DM of 263.42 pc cm$^{-3}$ favoring an extragalactic origin ($z_{est}=0.200$). Multi-wavelength follow-up observations with Gemini, VLA, and ATCA did not yet find a clear host galaxy for this burst \citep{2021ApJ...913...78A}.
FRB\,20180924A lies within the localisation uncertainty of the LAT source 4FGL\,J0221.8+3730. This gamma-ray source has been classified as a blazar candidate of an unknown type (BCU) and has no redshift information available. FRB\,20180924A presents quite a high dispersion measure (DM=1114.5 pc cm$^{-3}$, indicating a distant origin with an estimated redshift of about $z_{est}\sim 0.85$, and had a fluence of 3.5$\pm$1.2 Jy ms during its burst episode.
Finally, the position of FRB\,20190220A (DM = 216.120 pc cm$^{-3}$, $z_{est}=0.1561$, fluence of 0.68$\pm$0.44 Jy ms) is close to the gamma-ray source 4FGL\,J1549.4+7409, a BCU with no redshift information available.
Both the BCUs present a gamma-ray spectrum described by a power law with no sign of an energy cutoff below 100\,GeV suggesting z$\lesssim$1 \citep{2010ApJ...712..238F}, still compatible with the expected luminosity distance of the FRBs. Further MWL observations will be needed to either confirm or reject the possible association of FRB\,20180924A and FRB\,20190220A with the two BCUs 4FGL\,J0221.8+3730 and 4FGL\,J1549.4+7409, respectively.

\subsection{The periodic FRB\,20180916}
Among the repeaters, FRB\,20180916 (J0158+65) has recently been reported to present a significant periodicity of $16.35\pm 0.15$ day, obtained from the first 38 events recorded by  \citet{2020Natur.582..351C}. In particular all bursts arrive in a 5.4-day phase window, while 50\% of them are constrained to a 0.6-day phase window.
This source has been localised to a star-forming region in a nearby massive spiral galaxy at redshift $z=0.0337\pm0.0002$ \citep{2020Natur.577..190M}.

Our sample contains 72 radio burst episodes from this periodic FRB source. Particular attention will be given to the analysis of this source. Besides the search for gamma-ray emission for each individual event and a cumulative analysis of all the events reported for this FRB source, we also perform a folding analysis on the 0.6-day and 5.4-day phase windows.

\section{Analysis description and results}
\label{sec:analysis_desription}
The LAT detects photons by converting them into electron-positron pairs and has an operational energy range from 20\,MeV to more than 300 GeV \citep{2009ApJ...697.1071A}.
Thanks to its large field of view and continuous sky survey, the LAT represents an optimal instrument for the search for transient events flashing at random locations in the sky, such as FRB events. 
 
In this work we applied several analysis methods in order to search for the gamma-ray emission from FRB events on different time scales, ranging from 10 s (short-time scale emission) to 13 years (steady emission from the repeaters).
For the search for the short-time scale emission we considered four time windows $\delta_T$= 10, 100, 1000, 10000 s in order to extend the searches to time windows large enough to take into account the radio dispersion delays which can amount to many seconds. Finally we investigate possible triplets of photons close in time with the FRB event.

The analysis applied in this work is divided into five parts. First, we analysed the gamma-ray data from each individual FRB event on a short time scale ($\delta_T$ = 10--10000 s, see Sect. \ref{sec:analysis_single_source}). Secondly, we  searched for possible steady emission from the 22 repeaters in our sample (see Sect. \ref{sec:analysis_13years}).
Particular attention was paid to the periodic FRB\,20180916, for which also the 0.6 and 5.4 days active phase windows were investigated using a folding analysis (see Sect. \ref{sec:folding_analysis}). 
Subsequently we performed a stacking analysis on the undetected sources in the individual study in order to search for cumulative emission from the population of FRB events (see Sect. \ref{sec:stacking_analysis}). 
Finally, we used a photon counting method to search for photon triplets possibly associated with the FRB events (see Sect. \ref{sec:triplets_analysis}).
In the following part we present the description of each analysis method accompanied by the relative results.


\subsection{Short-time scale analysis of individual FRB events}

\label{sec:analysis_single_source}
For each FRB event, we performed the analysis on 10, 100, 1000 and 10000 s time windows ($\delta_T$) centred on the FRB-event time.
For our analysis we considered only the FRBs that occurred during the first 13 years of \textit{Fermi}-LAT operation: between August 5, 2008, and August 2, 2021 (MJD 54683 -- 59428). We selected events which have been reprocessed with the P8R3\_SOURCE\_V3 instrument response functions \citep[IRFs,][]{2018arXiv181011394B}, in the energy range between 100\,MeV and 1\,TeV. The low energy bound is justified by the large uncertainties in the arrival directions of photons below 100 MeV, leading to a possible confusion between point-like sources and the Galactic diffuse component \citep[see ][for a different analysis implementation to solve this and other issues at low energies with \textit{Fermi}-LAT]{2018A&A...618A..22P, 2019RLSFN.tmp....7P}, while the upper bound is related to poor effective area and photons collection above 1 TeV \citep{2021ApJS..256...12A}.
To reduce the contamination from the low-energy Earth limb emission \citep{2009PhRvD..80l2004A} we applied a zenith angle cut of $\theta<105^{\circ}$ to the data.

The binned likelihood analysis (which consists of model optimisation, localisation, spectrum and variability analyses) was performed with \textit{Fermipy}\footnote{version 1.0.1, \url{http://fermipy.readthedocs.io/en/latest/}} \citep{2017arXiv170709551W}, a python package that facilitates the analysis of LAT data with the \textit{Fermi} Tools\footnote{https://github.com/fermi-lat}, of which the version 2.0.18 was used. 
In order to determine whether an FRB is detected or not we used a test statistic\footnote{The test statistic (TS) is the logarithmic ratio of the likelihood $\mathcal{L}$ of a model with the source being at a given position in a grid to the likelihood of the model without the source, TS=$2 \log \frac{\mathcal{L}_\mathrm{src}}{\mathcal{L}_\mathrm{null}}$ \citep{1996ApJ...461..396M}.} TS$>25$ as the threshold. The resulting significance is $\sim(TS)^{0.5} \sigma$ \footnote{\url{https://fermi.gsfc.nasa.gov/ssc/data/analysis/documentation/Cicerone/Cicerone_Likelihood/TS_Maps.html}}, and thus TS$>$25 is equivalent to a significance of $>5\sigma$.

For each source in our sample we selected a region of interest (ROI) of 10$^{\circ}$ radius centred on the source position, and each ROI was analysed separately. 
We binned the data with a pixel size of $0.1^{\circ}$ and 8 energy bins per decade.
The model used to describe the sky includes all point-like and extended sources located at a distance $<15^{\circ}$ from the source position listed in the 4FGL-DR2 \citep{2020ApJS..247...33A}, as well as the Galactic diffuse and isotropic emission. For the two latter contributions, we use the same templates\footnote{\url{https://fermi.gsfc.nasa.gov/ssc/data/access/lat/BackgroundModels.html}} adopted to derive the 4FGL catalog.
A point-like source was also always added at the position of the FRB event.

For the likelihood analysis, we first optimised the model of the ROI, then searched for possible additional faint sources in each ROI, not included in 4FGL-DR2, by looking for bright excesses in the TS maps (significance maps). Subsequently, we re-localised the sources of our sample corresponding to the FRB event with TS $>$ 10 ($\sim 3\sigma$).
We performed the spectral analysis leaving free to vary the diffuse background template normalisation and the spectral parameters of the sources within 2$^{\circ}$ of our targets.
For the sources within a radius between 2$^{\circ}$--4$^{\circ}$ and all variable sources only the normalisation was fit, while we fixed at the parameters of all the remaining sources within the ROI.

For the refined spectral energy distribution (SED) plot of the detected sources, we repeated the spectral analysis dividing the photons into seven energy bands: six logarithmically equally-spaced bands between 100 MeV and 100 GeV and one band between 100 GeV and 1 TeV.
We modeled the spectrum of each source with a power-law (PL) function
\begin{equation} \label{eq:power_law}
  \dfrac{dN}{dE} = N_{0} \times \left(\frac{E}{E_{b}}\right)^{- \Gamma} ;  
\end{equation}
\noindent using $E_b=1$ GeV.
Upper limits at 95\% confidence level are reported for the sources with no significant gamma-ray emission (TS < 25).

\subsubsection*{Results on short time scale on individual FRB events}
\label{sec:results_short_time_scale}
Depending on the FRB position and date as well as on the duration of the time windows considered in the analysis, some of the events may happen outside the LAT field of view or may have a low exposure. 
For the short-time analyses the following numbers of events have non-negligible exposure, with at least one photon in the LAT field of view (FoV), allowing us to perform a likelihood analysis:
\begin{itemize}
    \item 258 events ($\delta_T$=10 s),
    \item 276 events ($\delta_T$=100 s),
    \item 415 events ($\delta_T$=1000 s),
    \item 958 events ($\delta_T$=10000 s).
\end{itemize}



\begin{figure}
\centering
\includegraphics[width=1.1\columnwidth]{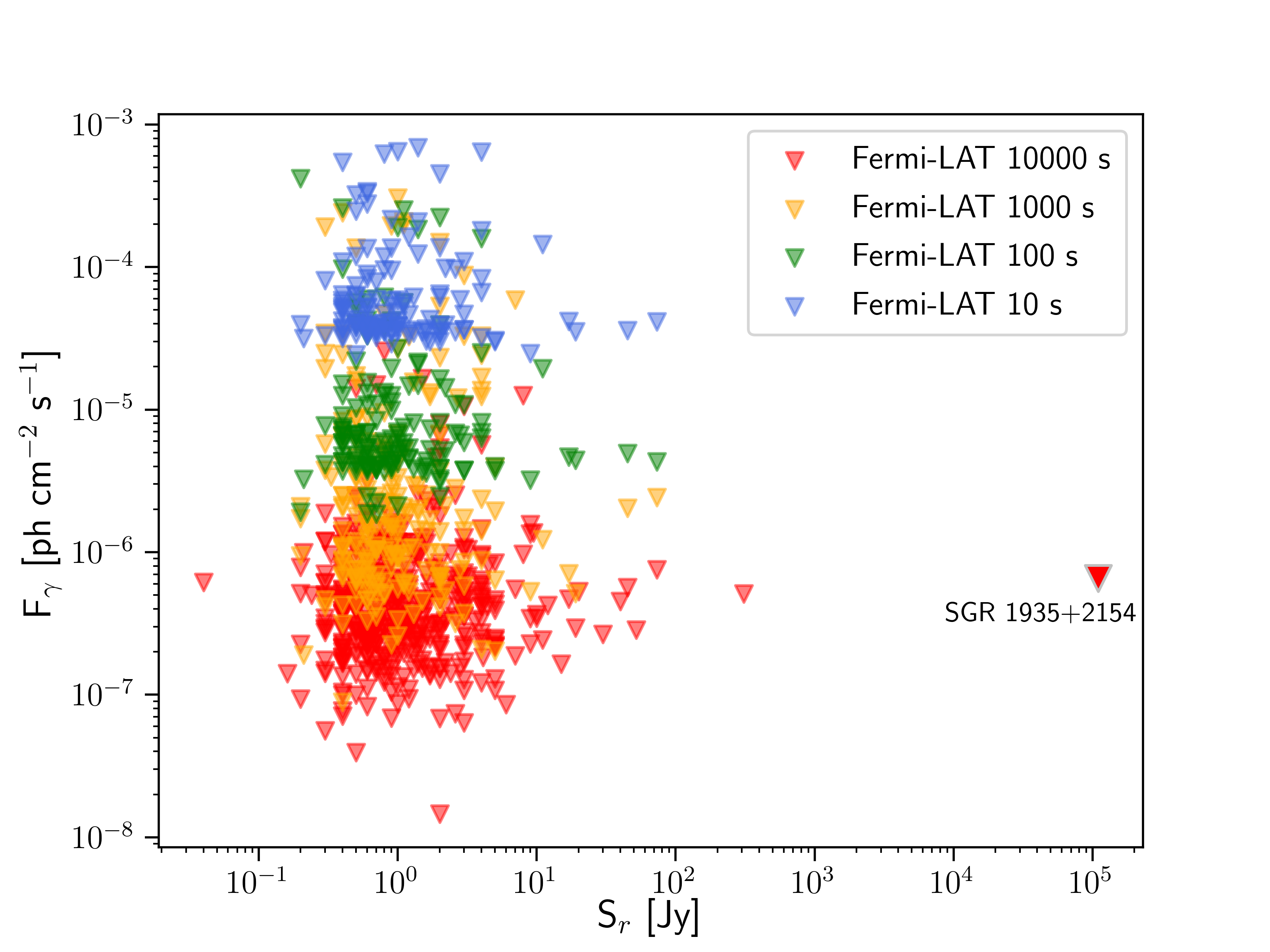}
\caption{\small \label{fig:flux_gamma_radio}
Diagram of the 95\% confidence level UL on the gamma-ray flux as a function of radio  fluxes. The UL from the LAT analysis on $\delta_T$ 10--10000 s are plotted in blue, green, orange and red, respectively.}
\end{figure}

\noindent No significant detection was found from the analysis of the individual bursts on the various short-time scales (see Fig. \ref{fig:flux_gamma_radio}), therefore only upper limits (UL) corresponding to the 95\% confidence level were provided. 

\noindent The most stringent ULs on the flux values
are 2.4$\times10^{-5}$, 1.9$\times10^{-6}$, 8.9$\times10^{-8}$ and 1.5$\times10^{-8}$ ph cm$^{-2}$ s$^{-1}$ for the analysis over $\delta_T=$10, 100, 1000 and 10000 s; respectively. 

\begin{figure}
\centering
\includegraphics[width=1.1\columnwidth]{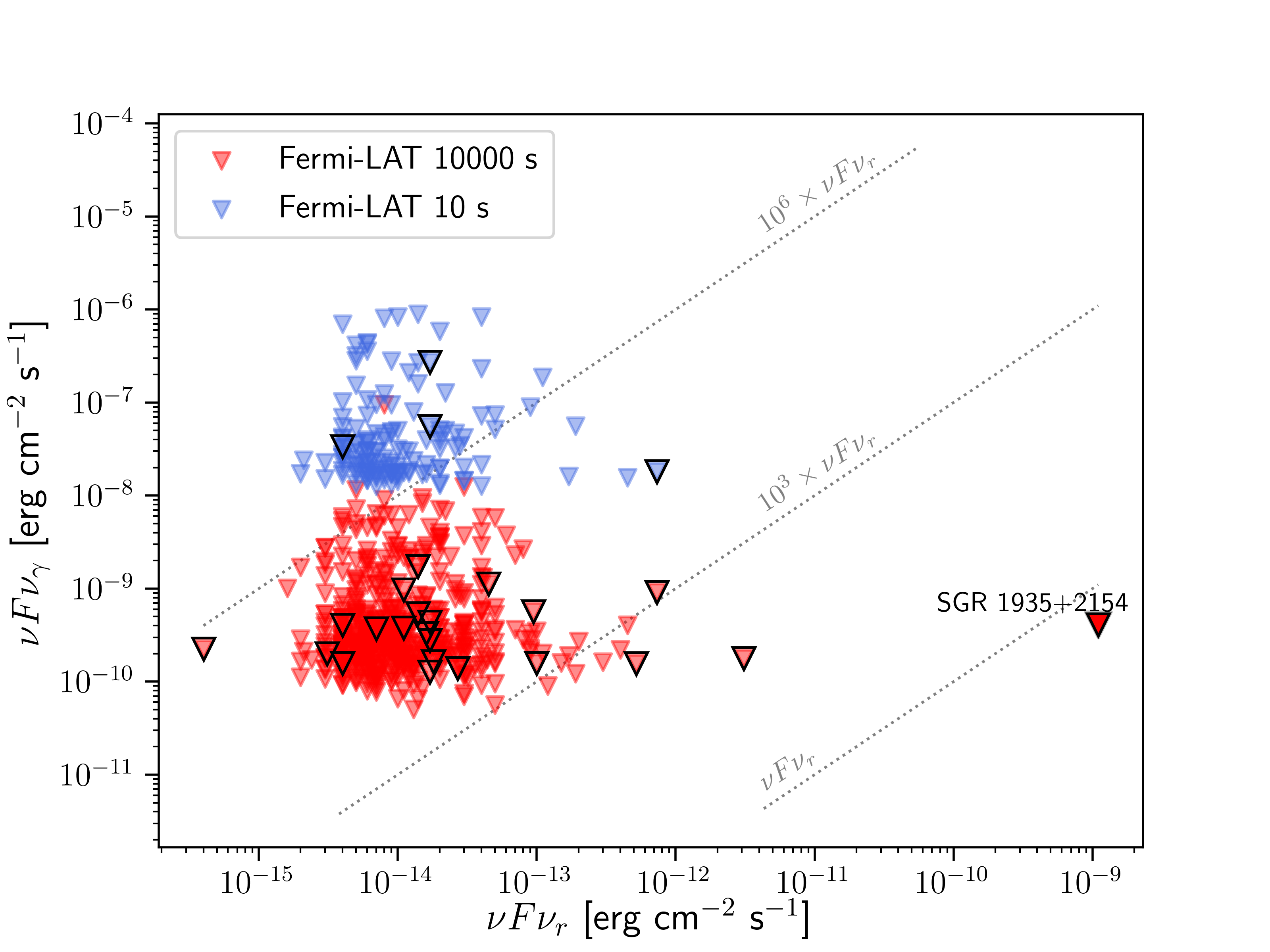}
\caption{\small \label{fig:energy_flux_gamma_radio}
Diagram of the upper limits on the gamma-ray energy flux as a function of radio energy fluxes ($\nu F \nu$). The FRB events reported are those with available information on their radio flux density and with a correctly estimated gamma-ray flux UL (i.e. sufficient gamma-ray exposure). Gamma-ray upper limits from the analysis on 10 s (10000 s) are plotted in blue (red). To avoid further confusion we plot only the $\delta_T=$10 and 10000 s results, the results on $\delta_T=$100 and 1000 s provide UL in between the two extreme time windows (10 and 10000 s). FRB sources with known host-galaxy are highlighted with a black outline.}
\end{figure}

In order to perform a direct comparison between the radio emission and the UL derived for the gamma rays we estimated the energy flux ($\nu F_{\nu}$).
Fig. \ref{fig:energy_flux_gamma_radio} shows a comparison of the UL on the energy flux obtained in the analysis over the $\delta_T=$10 s and $\delta_T=$10000 s time windows as a function of the measured energy flux observed in the radio band.   

\noindent The limit on the energy flux for 10 s ranges between 1.3$\times 10^{-8}$ -- 8.9$\times 10^{-7}$ erg cm$^{-2}$ s$^{-1}$, while for 10000 s it ranges between 5.0$\times 10^{-11}$ -- 9.5$\times 10^{-7}$ erg cm$^{-2}$ s$^{-1}$.
The lowest limits obtained here are one order of magnitude more stringent than all the values previously reported over the same time scale, 10 s \citep[AGILE-GRID UL=$1.8 \times 10^{-7}$ erg cm$^{-2}$ s$^{-1}$,][]{2021ApJ...915..102V}.

Assuming a direct proportionality between the radio and gamma-ray energy fluxes, the most stringent limit is obtained for SGR\,1935+2154 for $\delta_T=$10000 s, ruling out a gamma-ray energy flux larger than 0.4 times the radio one. 
For this source we do not have constraints over shorter time scales as it entered the LAT field of view only about 2000 s after the burst.
For the analysis over $\delta_T=$10 s the most stringent limit is for FRB\,20190102, ruling out a gamma-ray energy flux larger than 2.5$\times 10^{4}$ times the radio one.

\subsection{Long term analysis of repeating FRBs}
\label{sec:analysis_13years}

For the search for steady emission from the repeaters, we considered about 13 years of \textit{Fermi}-LAT data between August 5, 2008, and August 2, 2021 (MJD 54683 -- 59428).
In order to optimise the analysis, we applied different cuts for the data selections at low energies and selected event types with the best point spread function\footnote{A measure of the quality of the direction reconstruction is used to assign events to four quartiles. Gamma rays in Pass 8 data can be separated into 4 PSF event types: 0, 1, 2, 3, where PSF0 has the largest point spread function and PSF3 has the best one.} (PSF). In particular, for energies below 300 MeV we excluded events with zenith angle larger than 85$^{\circ}$, as well as photons from the PSF0 and PSF1 event types. Between 300 MeV and 1 GeV we excluded events with zenith angle larger than 95$^{\circ}$, as well as photons from the PSF0 event type. Above 1 GeV we used all events with zenith angles less than 105$^{\circ}$. 
This time, for the spectral analysis we left free to vary the sources within 5$^{\circ}$ of our targets, we fit only the normalisation of sources in a radius between 5$^{\circ}$ and 10$^{\circ}$ and we fixed the parameters of all the sources within the ROI at larger angular distances from our targets.

In addition to the likelihood and spectral analysis (see Sect. \ref{sec:analysis_single_source}), we also extracted a light curve for each repeating FRB source using temporal bins of 3 months in order to search for gamma-ray emission on this time scale. 
The fluxes in each time window were obtained by leaving only the normalisation free to vary and freezing the other spectral parameters to the best fit values obtained from the full range analysis. 

\subsubsection*{Results of long term analysis on repeating FRBs}
\label{sec:results_long_time_scale}
We searched for steady gamma-ray emission from the 22 repeaters. We reduced the analysed period considered for FRB\,20190117 (J2207+17) because it lies in proximity ($<0.85^{\circ}$) to a bright and strongly variable 4FGL-DR2 source \citep[4FGL J2203.4+1725, sign.$>60 \sigma$, $\Gamma_{var}=940$,][]{2020ApJS..247...33A}, which had strong flaring activity between 2008 and 2015. Therefore we limited the analysis of this repeater between Jan. 2016 and Aug. 2021. For more information on the analysis of this source see Appendix \ref{appendix:190117_analysis}.

We did not detect significant emission from any of the repeaters. We therefore provide ULs at 95\% confidence level on their steady emission. Table \ref{tab:frb_rep} contains the results of the gamma-ray analysis with the ULs on their gamma-ray flux and luminosity.

\begin{table*}
\centering
\begin{threeparttable}
\caption{\small Results of the search for steady gamma-ray emission from repeating FRBs.  \label{tab:frb_rep}}
\small
\centering
\begin{tabular}{cccccc|ccccc}
\hline \hline
Name & DM & D$_{L}$ & N.ev.\tnote{$a$} & S$_{r}$\tnote{$b$} & L$_{r}$\tnote{$c$} & TS\tnote{$d$} &  F$_{\gamma}$\tnote{$e$} & $\nu F\nu _{\gamma}$\tnote{$f$} & L$_{\gamma}$\tnote{$g$} \\
& pc cm$^{-3}$ & Mpc & & Jy & erg s$^{-1}$ & & & & \\
\hline
FRB\,20190907 & 309.6 & 1150 & 4 & 0.4 & 6.15$\times 10^{41}$ & 0.2 & 1.06 & 3.82 & 5.73$\times 10^{43}$\\
FRB\,20181030 & 103.5 & 290 & 9 & 3.2 & 3.25$\times 10^{41}$ & 0.0 & 2.63 & 1.95 & 2.79$\times 10^{42}$\\
FRB\,20180916 & 349.2 & 150 & 72 & 1.7 & 4.52$\times 10^{40}$ & 3.0 & 29.3 & 7.06 & 4.53$\times 10^{42}$\\
FRB\,20180908 & 195.7 & 700 & 4 & 0.6 & 3.50$\times 10^{41}$ & 0.0 & 5.62 & 1.38 & 1.90$\times 10^{43}$\\
FRB\,20181017  & 1281.6 & 6800 & 2 & 0.4 & 2.22$\times 10^{43}$ & 0.0 & 7.65 & 1.41 & 2.81$\times 10^{45}$\\
FRB\,20190604 & 552.6 & 2500 & 2 & 0.9 & 6.69$\times 10^{42}$ & 3.2 & 2.63 & 4.37 & 3.57$\times 10^{44}$\\
FRB\,20181128 & 450.5 & 1350 & 4 & 0.5 & 1.06$\times 10^{42}$ & 0.2 & 16.1 & 3.13 & 1.86$\times 10^{44}$\\
FRB\,20190117\tnote{*}  & 393.6 & 1600 & 6 & 1.7 & 5.14$\times 10^{42}$ & 3.5 & 55.1 & 3.42 & 7.08$\times 10^{44}$\\
FRB\,20200120E**  & 87.8 & 3.6 & 7 & 1.8 & 2.79$\times 10^{37}$ & 0.0 & 2.13 & 0.72 & 6.34$\times 10^{36}$\\
FRB\,20190417 & 1378.2 & 7150 & 12 & 0.7 & 4.30$\times 10^{43}$ & 0.2 & 2.38 & 4.45 & 1.63$\times 10^{45}$\\
FRB\,20181119 & 364.0 & 1500 & 8 & 0.6 & 1.65$\times 10^{42}$ & 0.7 & 1.21 & 3.59 & 9.76$\times 10^{43}$\\
FRB\,20190116 & 441.0 & 1950 & 2 & 0.4 & 1.79$\times 10^{42}$ & 1.4 & 24.0 & 1.10 & 4.50$\times 10^{44}$\\
FRB\,20190213 & 651.4 & 2950 & 2 & 0.5 & 5.26$\times 10^{42}$ & 0.0 & 10.7 & 1.35 & 5.98$\times 10^{44}$\\
FRB\,20190212 & 302.0 & 1100 & 10 & 1.1 & 1.61$\times 10^{42}$ & 7.4 & 0.57 & 8.38 & 1.04$\times 10^{44}$\\
FRB\,20180814 & 189.0 & 330 & 21 & - & - & 0.0 & 9.80 & 0.76 & 4.94$\times 10^{42}$\\
FRB\,20190303 & 222.4 & 860 & 19 & 0.5 & 4.43$\times 10^{41}$ & 0.0 & 1.68 & 1.36 & 1.65$\times 10^{43}$\\
FRB\,20190208 & 580.1 & 2400 & 7 & 0.6 & 4.19$\times 10^{42}$ & 0.1 & 1.80 & 3.66 & 2.67$\times 10^{44}$\\
FRB\,20190209 & 425.0 & 1750 & 2 & 0.6 & 2.21$\times 10^{42}$ & 0.0 & 7.24 & 0.57 & 2.93$\times 10^{43}$\\
FRB\,20201124A** & 413.5 & 450 & 32 & - & - & 0.0 & 2.64 & 0.77 & 4.21$\times 10^{42}$\\
FRB\,20190222 & 460.6 & 1600 & 2 & 1.9 & 5.98$\times 10^{42}$ & 4.5 & 5.67 & 6.94 & 2.64$\times 10^{44}$\\
FRB\,20121102** & 557.0 & 950 & 232 & 0.31 & 3.33$\times 10^{41}$ & 3.2 & 4.81 & 7.71 & 2.72$\times 10^{44}$\\
SGR\,1935+2154** & 332.7 & 0.014 & 1 & 110000 & 7.0$\times 10^{36}$ & 0.6 & 5.59 & 13.4 & 3.70$\times 10^{34}$\\

\hline
\hline
\end{tabular}
\begin{tablenotes}
\item{$^a$} Number of events listed in our sample for each individual repeater.
\item{$^b$} Maximum value of the radio flux density at peak time for a given FRB.
\item{$^c$} Maximum value of the radio luminosity for a given FRB.
\item{$^d$} Test Statistic. 
\item{$^e$} ULs at 95\% confidence level on the gamma-ray flux in units of $10^{-10}$ ph cm$^{-2}$ s$^{-1}$. 
\item{$^f$} ULs at 95\% confidence level on the energy flux in units of $10^{-13}$ erg cm$^{-2}$ s$^{-1}$.
\item{$^g$} ULs at 95\% confidence level on the luminosity in units of erg s$^{-1}$.
\item{$^*$} The analysis of this repeater
was limited between Jan. 2016 and Aug. 2021 (see Sect. \ref{sec:analysis_13years}). 
\item{$^{**}$} FRBs with known host galaxy and therefore measured luminosity distance (D$_{L}$). 
\end{tablenotes}
\end{threeparttable}
\end{table*}

\noindent Our flux and energy flux ULs on the steady gamma-ray emission from the repeaters are of the order of $10^{-10}$ ph cm$^{-2}$ s$^{-1}$ and $10^{-13}$ erg cm$^{-2}$ s$^{-1}$, respectively .
These results represent the most stringent ULs so far on the search for steady gamma-ray emission from repeating FRB sources.

\subsection{Folding analysis on the periodic FRB\,20180916}
\label{sec:folding_analysis}
In addition to studying each individual event, we also performed a cumulative folding analysis of all the events reported for a given repeating FRB. 
In this procedure all the individual time windows (e.g. 1000 s duration centred on the FRB event) related to FRB\,20180916 source were stacked together as a single window, having a length equal to the sum of each single period. 
We then used the previously described analysis (see Sect. \ref{sec:analysis_single_source}) to look for a gamma-ray signal.
This method was also applied to analyse the active phase (0.6 days / 5.4 days) of the periodic FRB\,20180916. 

\subsubsection*{Results on the periodic FRB\,20180916}
We searched for gamma-ray emission from the periodic FRB\,20180916 with the \textit{Fermi}-LAT, but we did not find any significant emission both on the individual events (on 10, 100, 1000 and 10000s) as well on the cumulative phase-folded analysis on all the events.
Based on the measured radio phase of the periodic activity, we performed a folded analysis of LAT data on the 5.4-day phase.
Table \ref{table_ul_frb180916} lists the 95\% integrated ULs on the FRB energy flux, obtained using a power-law spectral model with fixed $\Gamma= 2$.

\begin{table}
\centering
\begin{threeparttable}
\caption{\small \label{table_ul_frb180916} Results on the periodic FRB\,20180916.}
\small
\centering
\begin{tabular}{c|c}
\hline \hline
Analysis &  $\nu F\nu _{\gamma}$\tnote{$a$}\\
 & erg cm$^{-2}$ s$^{-1}$\\
\hline
$\delta_T =10$ s\tnote{$b$} & $<\,7.8 \times 10^{-8}$\\
$\delta_T =1000$ s\tnote{$b$} & $<\,1.4 \times 10^{-9}$\\
folding 1000 s of all bursts\tnote{$c$} & $<\,1.7 \times 10^{-10}$\\
5.4-days active ph. (13 years)\tnote{$d$} & $<\,2.1 \times 10^{-12}$\\
\hline
\end{tabular}
\begin{tablenotes}
    \item{$^a$} The 95\% ULs on the energy flux.
    \item{$^b$} Time window centred on the first observed radio burst (MJD=58377.42972096).
    \item{$^c$} Folding analysis of the cumulative gamma-ray emission on 1000 s time windows on the 72 detected bursts.
    \item{$^d$} Analysis on the 5.4-day active phase windows for 13 years of LAT data. 
\end{tablenotes}
\end{threeparttable}
\end{table}

\begin{figure}[ht]
\centering
\includegraphics[width=\columnwidth]{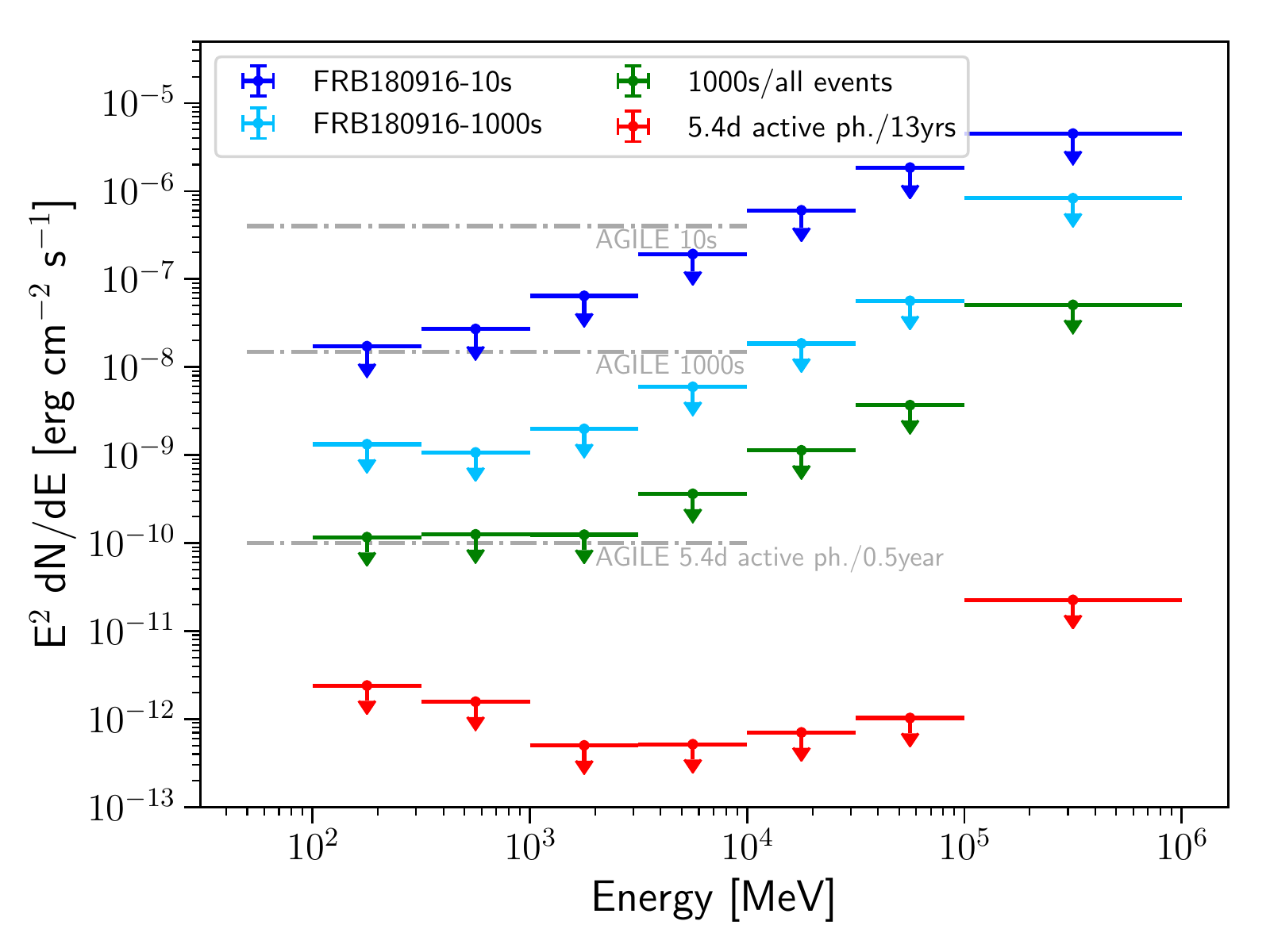}
\caption{\small \label{fig:sed_frb180916} \textit{Fermi}-LAT spectrum of the periodic FRB\,20180916. We plot the 95\% upper-limits for the 10 s and 1000 s time windows centred on the first FRB event, as well as for the folding analysis on the 72 detected bursts using 1000 s time windows, and for the 5.4-day active phase windows obtained for the 13 years of LAT data. For a comparison, in gray are plotted the $2\sigma$ ULs obtained with AGILE-GRID in the 0.5--10\,GeV energy band \citep{2020ApJ...893L..42T}.
}
\end{figure}

Fig. \ref{fig:sed_frb180916} shows the spectral ULs at 95\% confidence level obtained on the periodic FRB\,20180916 for the different time scales used in our analysis.

\noindent Although no significant detection was observed for FRB\,20180916, our results represent the most stringent ULs on its gamma-ray emission so far \citep[for previous results see][]{2020ApJ...893L..42T,2021Univ....7...76N}.

\subsection{Stacking analysis}
\label{sec:stacking_analysis}

As discussed in Sect. \ref{sec:analysis_single_source} and Sect. \ref{sec:analysis_13years}, no FRB event was significantly detected. 
We therefore look for collective emission from all 1020 FRB events and for this purpose we perform a stacking analysis of the sources using the likelihood results of each object as described in Sect. \ref{sec:analysis_single_source}. 
For each source $i$ and energy bin $k$, we derived a log-likelihood profile $\log \mathcal{L}_{i,k}$, i.e. the log-likelihood value as a function of the photon flux.
Assuming a common spectral shape for all sources $dN/dE$, we calculated the corresponding log-likelihood value for each source at a given energy. The total log-likelihood was obtained by summing over all the energy bins and sources:

\begin{equation} \label{eq:stack_logl}
    \log \mathcal{L} = \sum_i \sum_k \log \mathcal{L}_{i,k}|_{dN/dE(E_k)}
\end{equation}

\noindent We assumed a simple power-law spectrum (see Eq. \ref{eq:power_law}) for all the sources in our sample. By varying the normalisation $N_0$ and photon index $\Gamma$ we created a 2-dimensional likelihood profile in order to search for the parameter values which maximize the log-likelihood. In order to avoid convergence problems in the fitting procedure, we fitted each individual burst with a power-law spectrum with spectral index fixed to -2.5 in all the four time windows analysed from 10 s to 10000 s. The procedure was repeated for different values of spectral index (-2.0, -2.5, -3.0), obtaining compatible results. 

The significance of the potential detection is derived by comparing the maximum log-likelihood value with that of the null hypothesis ($\log \mathcal{L}_{null}$), i.e., the hypothesis in which the flux of the gamma-ray emitter is zero. We estimated the significance using the TS value, defined as TS = $2(\log \mathcal{L} - \log \mathcal{L}_{null}$).

In the case of no detection, we estimated the UL at 95\% confidence level on the photon flux by deriving the 2-dimensional contour corresponding to a $\Delta \log \mathcal{L}=4.61/2$, with two additional free parameters ($N_0$ and $\Gamma$ in Eq. \ref{eq:power_law}) in our model compared to the null hypothesis. In \citet{2021MNRAS.507.4564P}, this method was tested by performing MC simulations, and it was verified that a $\Delta \log \mathcal{L}>4.61$ occurs by chance only 5\% of the time, corresponding to a 5\% false positive detection rate.

Finally, we performed a stacking analysis using the luminosity as a free parameter, assuming that all FRB events have the same intrinsic luminosity. Flux values used in Eq. \eqref{eq:stack_logl} were converted into luminosities using the luminosity-distance values derived in Sect. \ref{sec:fruitbat} and the stacking procedure was carried out assuming that the gamma-ray emission is still described by a simple power-law spectrum (as in Eq. \eqref{eq:power_law}) expressed in luminosity units. With this procedure, the results are indeed dominated by the closest FRB sources, for which the expected flux is higher if the luminosities at the source are assumed to be the same.

\subsubsection{Results of the stacking analysis for all FRB events}

\begin{figure}
\centering
\includegraphics[width=1.1\columnwidth]{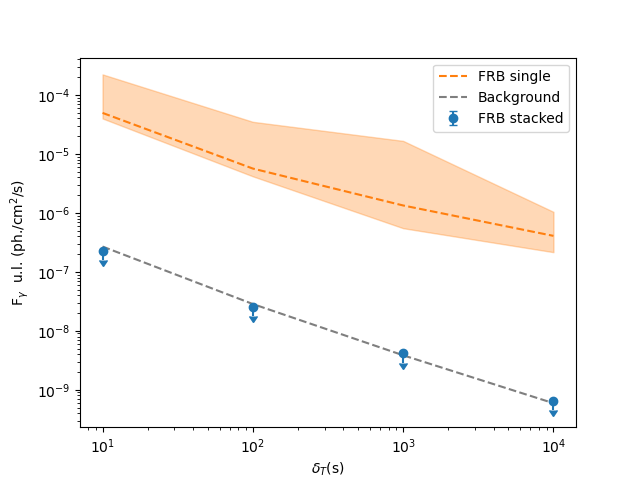}
\caption{\small \label{fig:stacking_ul_flux}
Upper limits on the gamma-ray flux at 95\% confidence level for the time windows considered in our analysis (10, 100, 1000 and 10000 s). The orange dashed line represents the median value of ULs on individual FRB events, while the orange band is obtained from the 10\% and 90\% quantiles of these values., while the blue points are the ULs obtained with the stacking analysis. The gray dashed line shows the upper limits derived from the stacking analysis performed on a sample of background sources.}
\end{figure}

We performed the stacking analysis on all the FRB events in our sample. We did not find any significant emission and we calculated ULs at 95\% confidence level on the gamma-ray flux.
Fig. \ref{fig:stacking_ul_flux} shows the results from the stacking of all FRB events in our sample, compared with the average ULs obtained on individual bursts. As expected, the ULs obtained with the stacking analysis are about two orders of magnitude more stringent than individual sources.

\noindent It can also be noted that ULs are more stringent for longer time windows ($\delta_T$). This is compatible with the increased photon statistics when increasing the width of the time window analysed.

\subsubsection{Comparison with stacking of background sources}
We repeated the stacking analysis on a control sample of bursts, defined so that only signal from background is included, to ensure that this analysis gave results compatible with the non detection of the stacked FRB events. We analysed data from random locations in the sky with the same time windows used for the analysis of the FRB-event sample. The positions were selected to be at least 1 degree from all known gamma-ray sources in the 4FGL-DR2 catalog. The number of selected positions was chosen to match the number of events in our sample. 
We performed the same likelihood analysis described in Sect. \ref{sec:analysis_single_source} on all the background sources, followed by the stacking analysis on this sample. Results are shown as a function of the time window in Fig. \ref{fig:stacking_ul_flux} (gray line). Results are compatible with those obtained from our FRB-event sample (blue points), showing that the upper limits are fully consistent with background fluctuations.

\subsubsection{Limits on energy and luminosity with the stacking analysis}
\label{sec:stacking_energy_luminosity}
\begin{figure}
\centering
\includegraphics[width=1.1\columnwidth]{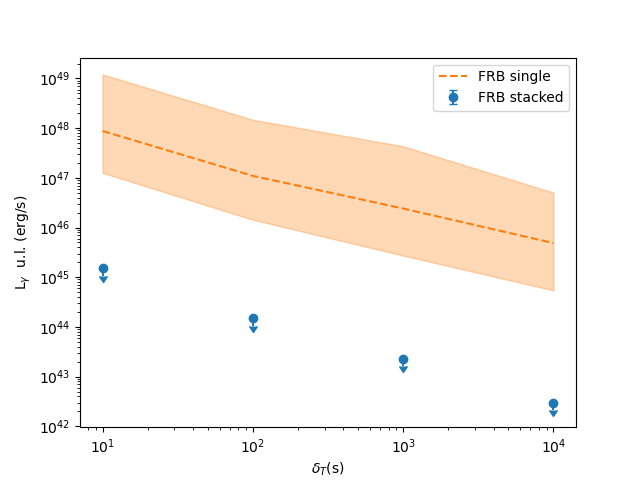}
\includegraphics[width=1.1\columnwidth]{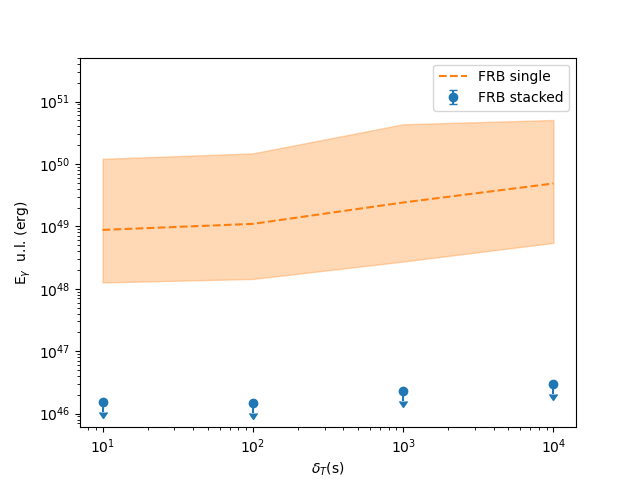}
\caption{\small \label{fig:lum_stacking}
Top: 95\% CL upper limits on the gamma-ray luminosity as a function of the observed time window. Bottom: 95\% CL upper limits on the gamma-ray energy as a function of the observed time window. The ULs have been obtained for the energy range 0.1--1000 GeV. In both plots, the orange line represents the median value of upper limits obtained from individual FRB events, while the orange band is obtained from the 10\% and 90\% quantiles of these values. Blue arrows represent the upper limits from the stacking analysis on all bursts.}
\end{figure}

As described in Sect. \ref{sec:stacking_analysis}, we repeated the stacking procedure using the luminosity of sources as a free parameter. The results are reported in the top panel of Fig. \ref{fig:lum_stacking}. The gamma-ray emitted energy was also calculated by multiplying the luminosity by the time window analysed.

\subsubsection{Stacking of the repeating FRBs}

Since no significant detection resulted from the search for steady emission from the repeaters, we searched for a cumulative signal by stacking data from all repeating FRB sources analysed with 13 years of LAT data. We applied the method described in Sect. \ref{sec:stacking_analysis} in the full energy range and repeated the analysis in 8 energy bins (2 bins per decade). 
The cumulative UL on the photon flux was found to be $8.1 \times 10^{-11}$ ph cm$^{-2}$ s$^{-1}$. 
This value is about one order of magnitude lower than the ULs derived from the individual repeating FRB analysis reported in Table \ref{tab:frb_rep}.

\begin{figure}[ht]
\centering
\includegraphics[width=\columnwidth]{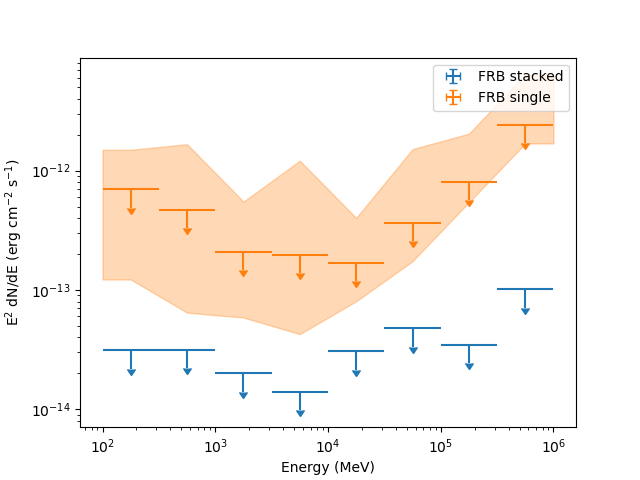}
\caption{\small \label{fig:sed_stack_repeating} \textit{Fermi}-LAT spectrum of the repeating FRBs. Orange: average UL at 95\% C.L. obtained from the single FRB-source analysis. The colored band represents the envelope for all ULs from the individual repeating FRB sources. Blue: UL at 95\% C.L. from stacking analysis.}
\end{figure}

Fig. \ref{fig:sed_stack_repeating} shows the SED obtained from the stacking analysis of the repeating FRBs, compared to the average UL obtained from the single FRB-source analysis.
The stacking procedure was also performed using the luminosity as a free parameter, similarly to what was done for the individual bursts. The UL on the cumulative luminosity is $1.6 \times 10^{43}$ erg s$^{-1}$.

\subsection{Photon triplets search analysis}
\label{sec:triplets_analysis}
Finally, we applied the analysis method for searching for photon triplets described in detail in \citet{2021NatAs.tmp...11F}. For each FRB source, we selected all the SOURCE class photons between 100 MeV and 300 GeV detected by the LAT in 13 years in a ROI of radius of 1$^{\circ}$ centred on the FRB position. Then we computed the time interval $\Delta t_i$ for each triplet of photons $i$ formed by three consecutive events:
\begin{equation}
    \Delta t_i = t_{i+2} - t_i,
\end{equation}
and corrected this quantity for the effect of bad time intervals by subtracting from each $\Delta t_i$ the period of time in which the ROI was not observable by the LAT.

With this dataset, we initially examined the first triplet following each FRB event trigger-time $T_\mathrm{FRB}$ and studied its potential association with the FRB emission. We call this analysis the \textit{next triplet} study.
To compute the probability that three photons cluster by chance, due to statistical fluctuations of the background, in the time range $\Delta t_\mathrm{FRB} = t_\mathrm{FRB,3} - T_\mathrm{FRB}$, where $t_\mathrm{FRB,3}$ is the time of the third photon of the triplet subsequent to the FRB event, we applied the Likelihood Ratio method defined in \citet{LiMa}. The maximum likelihood ratio, testing the presence of a new source, is defined as:
\begin{equation}
\rm \lambda = \left[ \dfrac{\alpha}{1 + \alpha} \left(1 + \dfrac{N_{B}}{N_{FRB}}\right)\right]^{N_{FRB}} \left[ \dfrac{1}{1 + \alpha} \left(1 + \dfrac{N_{FRB}}{N_{B}} \right) \right]^{N_{B}},
\end{equation}
where $N_{FRB} = 3$, $N_{B}$ is the number of LAT photons observed outside the time range $\Delta t_\mathrm{FRB}$, and $\alpha$ is the ratio between $\Delta t_\mathrm{FRB}$ and the total livetime for each ROI over 13 years of observation. The significance $S$ of the triplet in consideration, measured in units of $\sigma$, can hence be calculated as:
\begin{equation}
    \rm S \approx \sqrt{-2 \ln \lambda}.
\end{equation}
Assuming the significance of the Li\&Ma test follows Gaussian statistics with one degree of freedom, the sigma of the Li\&Ma test can be expressed in terms of probability, $p_{{\rm value}}^{{\rm pre}}$, where ``pre'' denotes the \textit{pre-trial} probability. For repeaters, the \textit{post-trial} probability, accounting for the number of triggers, can be estimated as
\begin{equation}
    p_{{\rm value}}^{{\rm post}} = 1-e^{-p_{{\rm value}}^{{\rm pre}}N_{{\rm trials}}}.
\end{equation}



\subsubsection{Triplet delay analysis for non-repeating FRBs}
For the given arrival time of each FRB event, one could study the delay of the arrival time of the most significant subsequent triplet of photons detected at the same location as the FRB. We call this analysis the \textit{next-best triplet} study.

Using the triplet dataset produced in Sect.~\ref{sec:triplets_analysis}, we focused on the analysis of the sample of non-repeating FRB sources, for which there is no confusion in associating a triplet with an FRB event. For each FRB source, we selected the triplet with the shortest $\Delta t$ (highest $p$-value), detected after the FRB trigger-time $T_\mathrm{FRB}$, and calculated the triplet delay $D_\mathrm{i}$ as:
\begin{equation}
    D_i = t_\mathrm{i,1} - T_\mathrm{FRB},
\end{equation}
with $t_\mathrm{i,1}$ indicating the time of arrival of the first photon within the triplet. We then built the distribution of $D_i$, to be compared to the null-hypothesis distribution, which is obtained by averaging 100 random realizations shuffling, each time, the trigger times of the FRB sample.
In this way, we can determine whether the two distributions are statistically different or consistent with each other. A statistically significant deviation from the null-distribution would indicate that FRBs are followed by gamma-ray emission after a preferential amount of time. 


\subsubsection{Results for triplet photon search}
\label{sec:results_triplets}
\begin{figure}
    \centering
    \includegraphics[width=\columnwidth]{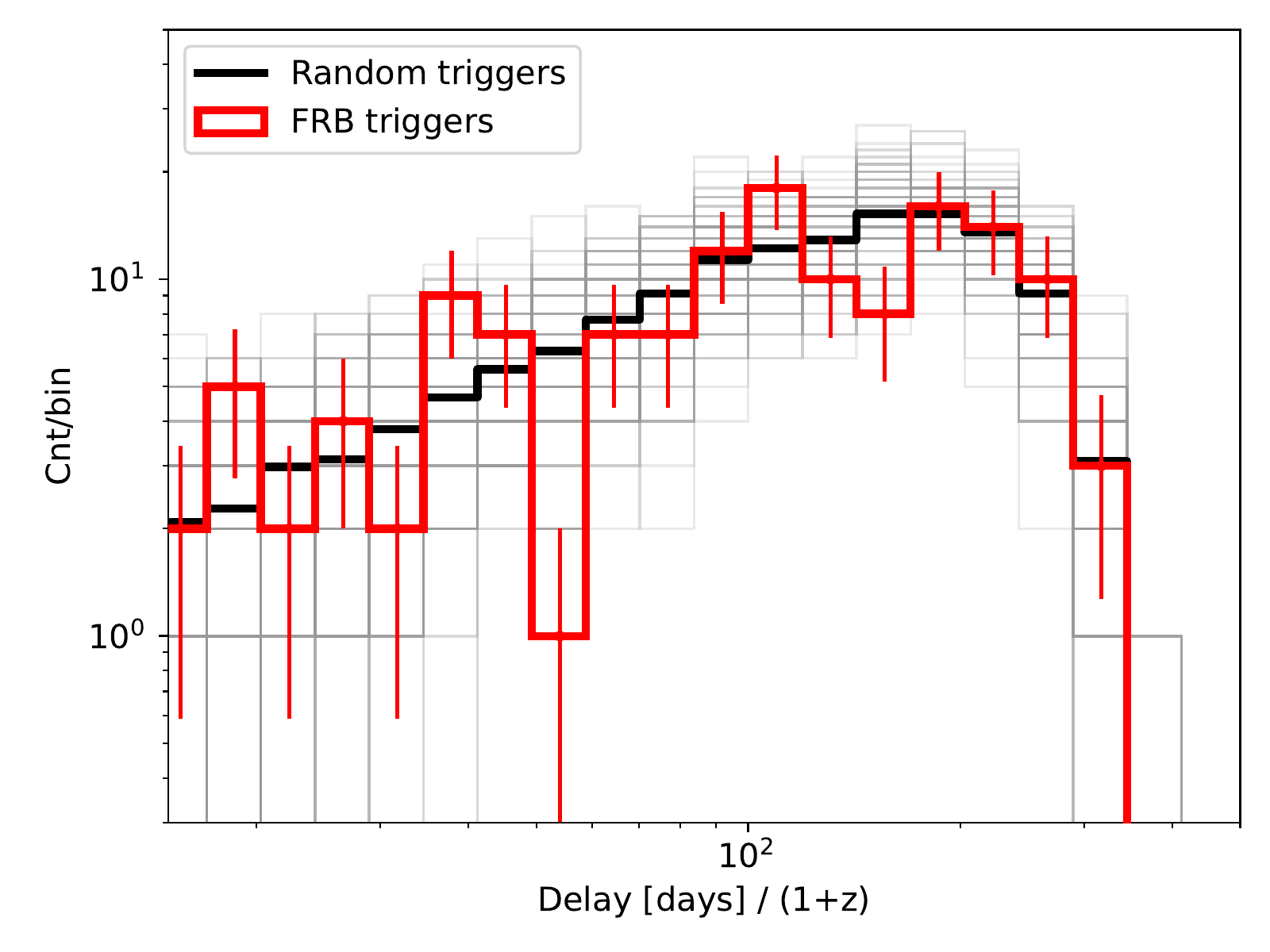}
    \caption{Distribution of the delay of the most significant triplet with respect to the trigger time of the FRB event. This distribution only includes non-repeating FRBs and is compared to 100 background distributions (gray lines) obtained by shuffling the trigger times for each FRB event. The black distribution is the mean background distribution obtained by averaging the 100 random realizations. Triplets with a delay greater than one year are not considered in this plot.}
    \label{fig:bestnexttriplet}
\end{figure}

\begin{figure}
    \includegraphics[width=0.85\columnwidth,right]{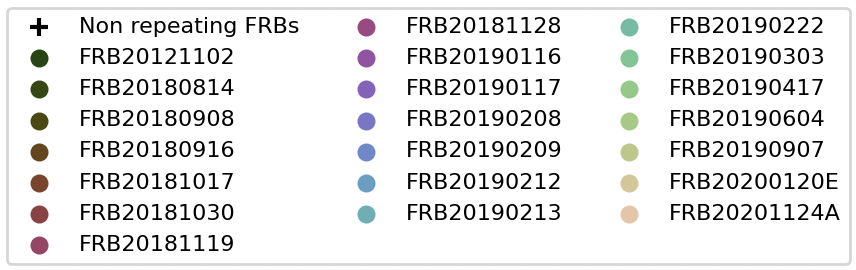}
    \includegraphics[width=\columnwidth,right]{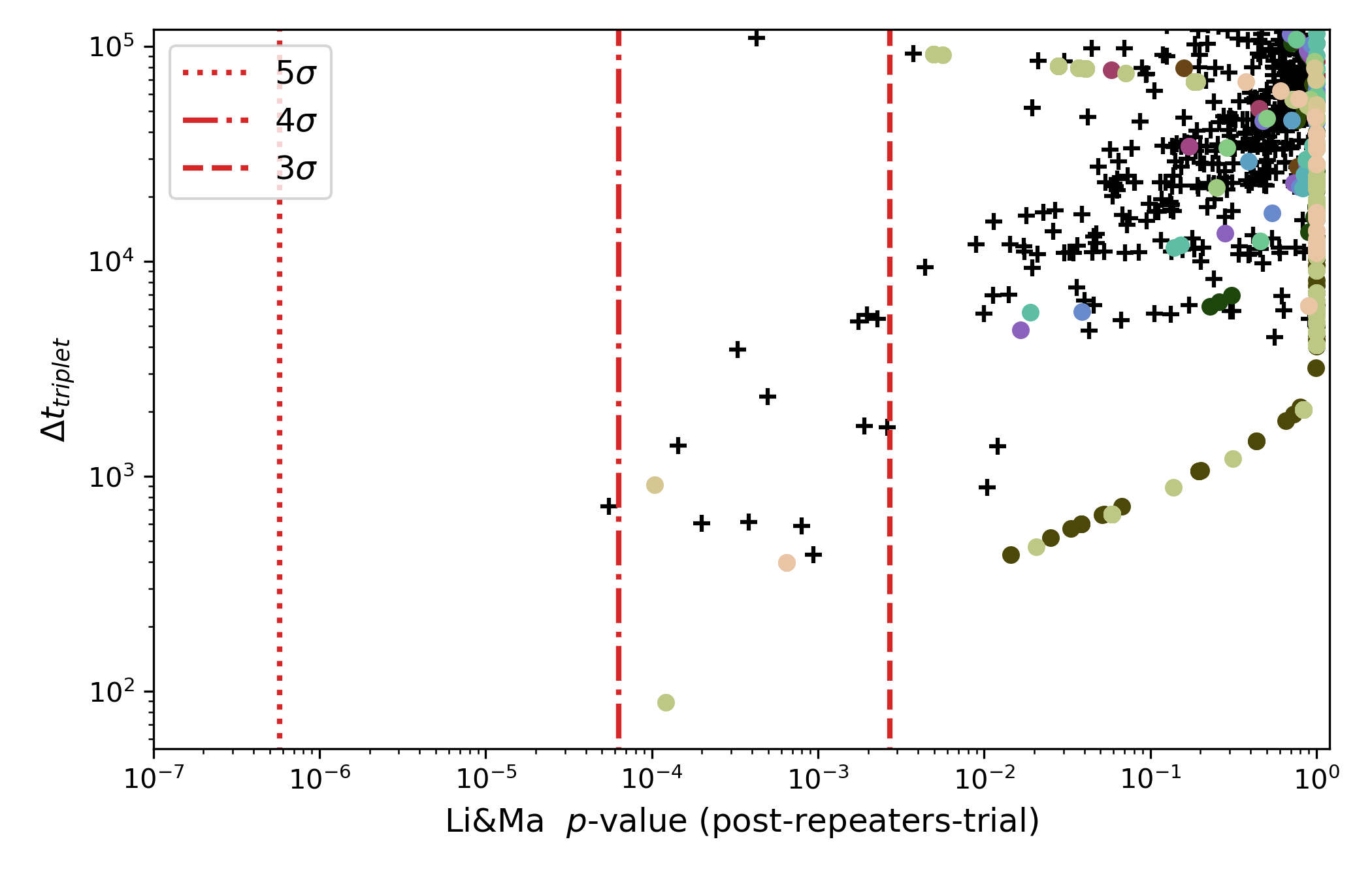}
    \includegraphics[width=\columnwidth,right]{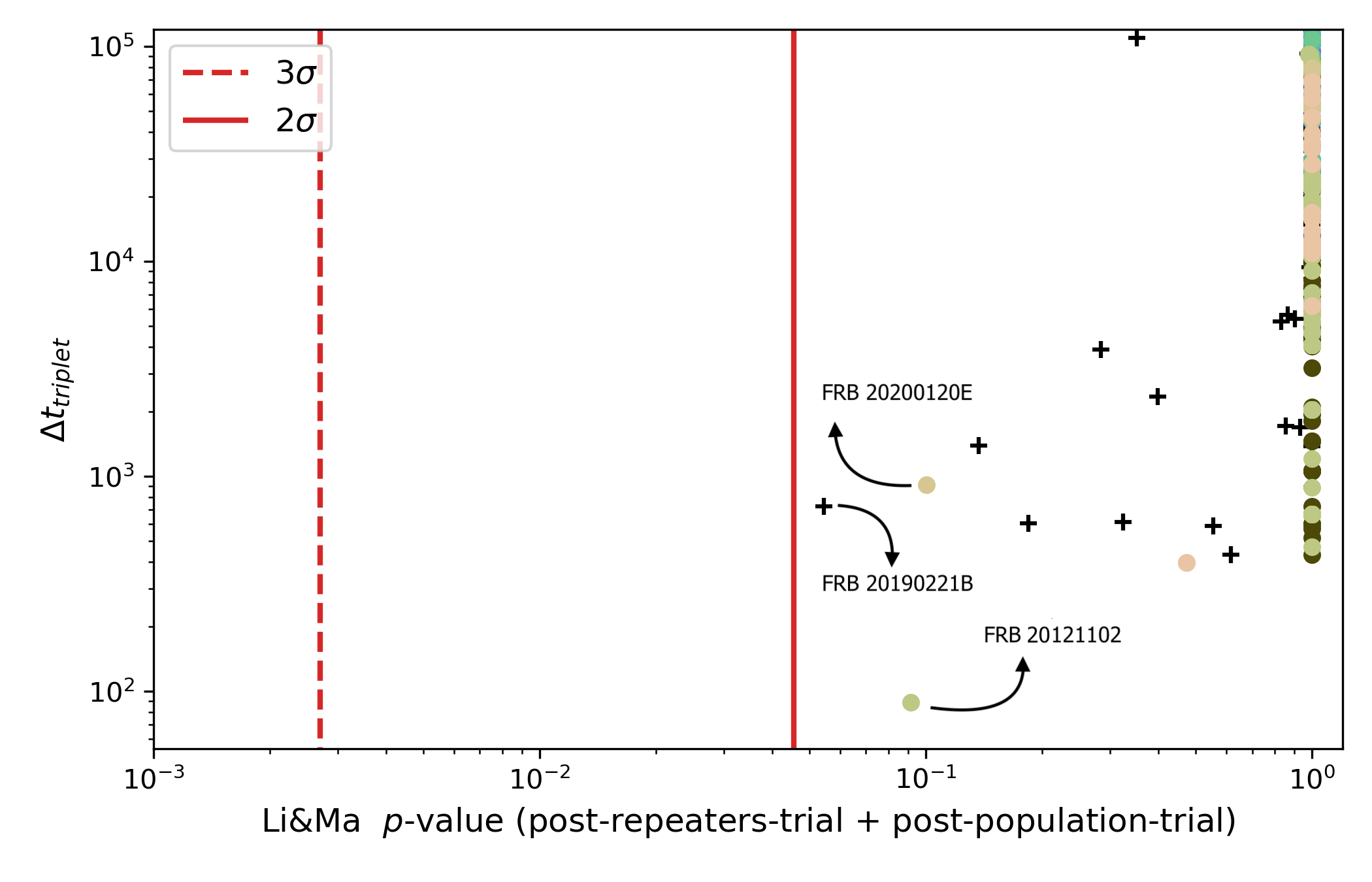}

    \caption{Demographics of the next subsequent triplet after the FRB-event trigger time. We refer to the text for the description of the plotted quantities. Each point corresponds to one FRB trigger. We mark with black crosses the non-repeating FRB sources, while the repeaters are marked in different colors according to the legend on the top panel. The red solid, dashed, dot-dashed, and dotted lines mark the confidence levels corresponding to 2, 3, 4 and 5$\sigma$, respectively.
    }
    \label{fig:nexttriplet}
\end{figure}

Fig.~\ref{fig:bestnexttriplet} and Fig.~\ref{fig:nexttriplet}  show the results of the study for the \textit{next-best triplet} and \textit{next triplet}, respectively. 
The former has a straightforward interpretation: the distribution of delays of the next-best triplet with respect to the FRB-event trigger has no statistically significant deviation from the null-distribution. This tells us that the \textit{next-best} triplets found after the trigger time are likely not related to the FRB. 

Regarding the \textit{next triplet} analysis, which is run for both single-event FRBs and FRB repeaters, we note that the minimum lag between the FRB-event trigger time and the arrival time of the first photon is of the order of 50 s for the sample of non repeaters, while it is 9 s for repeaters. 

We observe that no FRB events show a \textit{post-trial} $p$-value above 5$\sigma$. Such $p$-value is expected to be further worsened if we consider the additional \textit{population-trial factor}, which accounts for the total number of FRB events considered in this study. For example, as shown in Fig.~\ref{fig:nexttriplet} (bottom panel), the most significant FRB is the non-repeating event FRB~20190221B, with a Li\&Ma significance of 4.03$\sigma$. For this event the $p$-value after accounting for the total number of FRB events in our sample (1020) is only 0.055, namely slightly less than 2 sigma. All the other FRBs scale accordingly to lower significance than FRB~20190221B after accounting for the same \textit{population-trial factor}.

\subsubsection{\textit{Fermi}-LAT sensitivity on ms time scales}
\label{sec:lat_ms_sensitivity}
As mentioned above, no photons have been detected from the FRB directions within a time window of few seconds. While in Sect. \ref{sec:analysis_single_source}--\ref{sec:stacking_analysis} we derived limits starting from 10 s, in order to estimate the sensitivity of the LAT to events on a shorter time scale, down to the ms scale, we performed a single photon sensitivity study. For small time windows ${\rm \delta}_T\leq$1 s the background is negligible, and the total number of events that are expected to be detected by the LAT from a source with a photon spectrum described by Eq.~\ref{eq:power_law} can be estimated as:
\begin{equation}\label{eq:nexp}
{\rm N_{exp}(\theta,\gamma) = \delta_T \times \int_{E_{min}}^{E_{max}} N_0 \left(\frac{E}{E_b}\right)^{-\gamma} \times A_{\rm eff}({E,\theta}) \, dE }
\end{equation}
where ${\rm A_{eff}({E,\theta})}$ is the effective area of the \Fermi-\LAT instrument \citep{2018arXiv181011394B} and $\theta$ is the off-axis angle. Note that since we integrate over the entire energy range and over an ROI which is much larger than the PSF of the instrument, we can omit the integration over the PSF and over the energy dispersion.
Furthermore, considering that the 95\% probability to observe at least one event is reached when the expected number of events is three (using a Poisson probability density function), we can compute the 95\% photon flux UL by deriving the normalisation N$_{\rm 0}$ by equating Eq.~\ref{eq:nexp} to three, and integrating Eq.~\ref{eq:power_law} between 100\,MeV and 1\,TeV. In Tab.~\ref{tab:sensi} we summarise the UL values for different IRFs, incident angles and photon indexes: for these time scales, where the background is negligible, the event classes with the largest effective area (\texttt{P8R3\_TRANSIENT010E\_V3} and \texttt{P8R3\_TRANSIENT020E\_V3}) provide the most constraining ULs. 

\begin{table}[ht]
    \centering
    \begin{threeparttable}
    \begin{tabular}{l|c|ccc}
     \hline\hline
    IRF                      & $\Gamma$ & \multicolumn{3}{c}{F$_{\gamma}$}\tnote{a} \\
                                 &          & $\theta=0\adeg$ & $\theta=30\adeg$ & $\theta=60\adeg$ \\
        \hline
                                        & 1.5 & 0.32    & 0.34       & 0.8\\
       \texttt{P8R3\_SOURCE\_V3}        & 2 & 0.30       & 0.32       & 0.73 \\
                                        & 2.5 & 0.29    & 0.31       & 0.71\\
       \hline
                                        & 1.5 & 0.29    & 0.32      & 0.77 \\
       \texttt{P8R3\_TRANSIENT010E\_V3} & 2 & 0.27      & 0.30      &  0.71 \\
                                        & 2.5 & 0.27    & 0.30      & 0.69 \\
        \hline
                                        & 1.5 & 0.28    & 0.30      & 0.74 \\
       \texttt{P8R3\_TRANSIENT020E\_V3} & 2 & 0.27      & 0.29      & 0.69 \\
                                        & 2.5 & 0.26    & 0.29      & 0.68 \\
    \end{tabular}
    \caption{Millisecond-time scale \textit{Fermi}-LAT sensitivity.}
    \label{tab:sensi}
    \begin{tablenotes}
    \item{$^a$} Photon flux value for different IRFs, incident angles and photon indexes $\Gamma$. The values of the UL are in [ph cm$^{-2}$ s$^{-1}$] and are obtained for a $\delta_T$ of 1\,ms, and therefore they can be re-scaled by (1 \,ms/$\delta_T$) to obtain ULs for different exposure times.
    \end{tablenotes}
    \end{threeparttable}
\end{table}

\begin{figure}
\centering
\includegraphics[width=1.0\columnwidth]{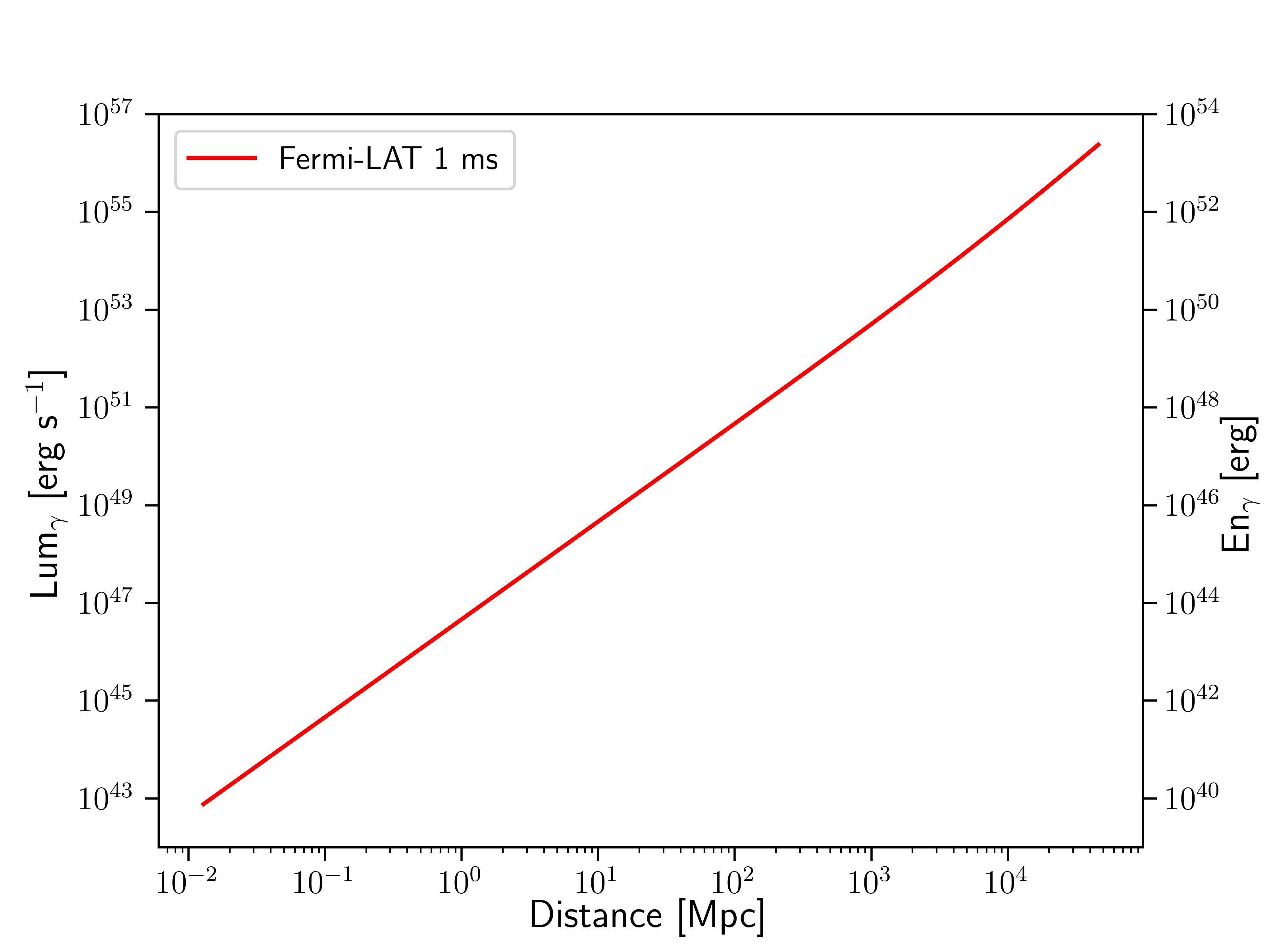}
\caption{\small \label{fig:ul_lum_en_ms}
The UL of the ms gamma-ray luminosity and energy as a function of luminosity distance.}
\end{figure}

\noindent In Appendix.~\ref{appendix:sensi_study} we plot the behavior of the flux UL as a function of the off-axis angle for the three IRFs and for the three different photon indexes considered here. The corresponding limits on energy flux are of the order of 240 erg cm$^{-2}$ s$^{-1}$.
Fig. \ref{fig:ul_lum_en_ms} shows the derived UL on ms gamma-ray luminosity and energy as a function of luminosity distance. 

\noindent In particular the UL on the luminosity for a ms time scale is of the order of:
\begin{equation}
    L_{\gamma, \delta_T = 1 \textrm{ms}} \lesssim 10^{50} \left(\frac{\textrm{D}_L}{150 \,\textrm{Mpc}}\right)^{2} \textrm{erg}\, \textrm{s}^{-1} \, ,
\end{equation}

which corresponds to an emitted energy of $E_{\gamma, \delta_T = 1 \textrm{ms}} \lesssim 10^{47} (\textrm{D}_L/150 \textrm{Mpc})^2$ erg.

\section{Discussion}
FRBs are one of the most intriguing and compelling current enigmas in astrophysics \citep{2019ARA&A..57..417C,2019A&ARv..27....4P}. 
They are thought to originate from some high-energy process not yet understood.
Despite there being hundreds of proposed models for FRB progenitors \citep{2019PhR...821....1P}, with those invoking magnetars being the most popular \citep[see e.g.,][]{2013arXiv1307.4924P,2019MNRAS.485.4091M}, their origin and emission mechanism are still a mystery \citep{2020Natur.587...45Z,2021SCPMA..6449501X,2021Univ....7...56L,2022arXiv221203972Z}. 

The recent discovery of coincident FRB-like and X-ray emission from the magnetar SGR\,1935+2154 \citep{2020Natur.587...54C,2020ApJ...898L..29M,2021NatAs...5..414K,2021NatAs...5..401T} implies that at least some FRBs may arise from magnetar activities, supporting the idea of a magnetar scenario for the origin of cosmological FRBs. 
Several models invoke strongly magnetized neutron stars and predict high-energy emission associated to the FRB event \citep[see e.g.][]{2013arXiv1307.4924P,2014MNRAS.442L...9L,2014ApJ...780L..21Z,2019MNRAS.485.4091M,2020ApJ...899L..27M,2021ApJ...919...89Y}. 
If we assume that FRBs are produced by magnetar flares we would expect afterglow emission peaking at gamma-ray energies E$_{\textrm{peak}} >$ MeV–GeV \citep{2019MNRAS.485.4091M}, with a predicted gamma-ray luminosity $L_{{\rm \gamma}} {\rm \sim} 10^{45}-10^{46}$ erg s$^{-1}$, on a time-scale comparable to the FRB itself. 
In addition, any FRB should be accompanied by X-ray/gamma-ray emission with an energy at least a factor $\eta \geq 10^{4}$ larger than the emitted radio energy \citep{2020ApJ...899L..27M}, or a gamma-ray-to-radio fluence ratio of 10$^{5}$ -- 10$^{6}$ \citep{2014MNRAS.442L...9L}.

\begin{figure}
\centering
\includegraphics[width=1.1\columnwidth]{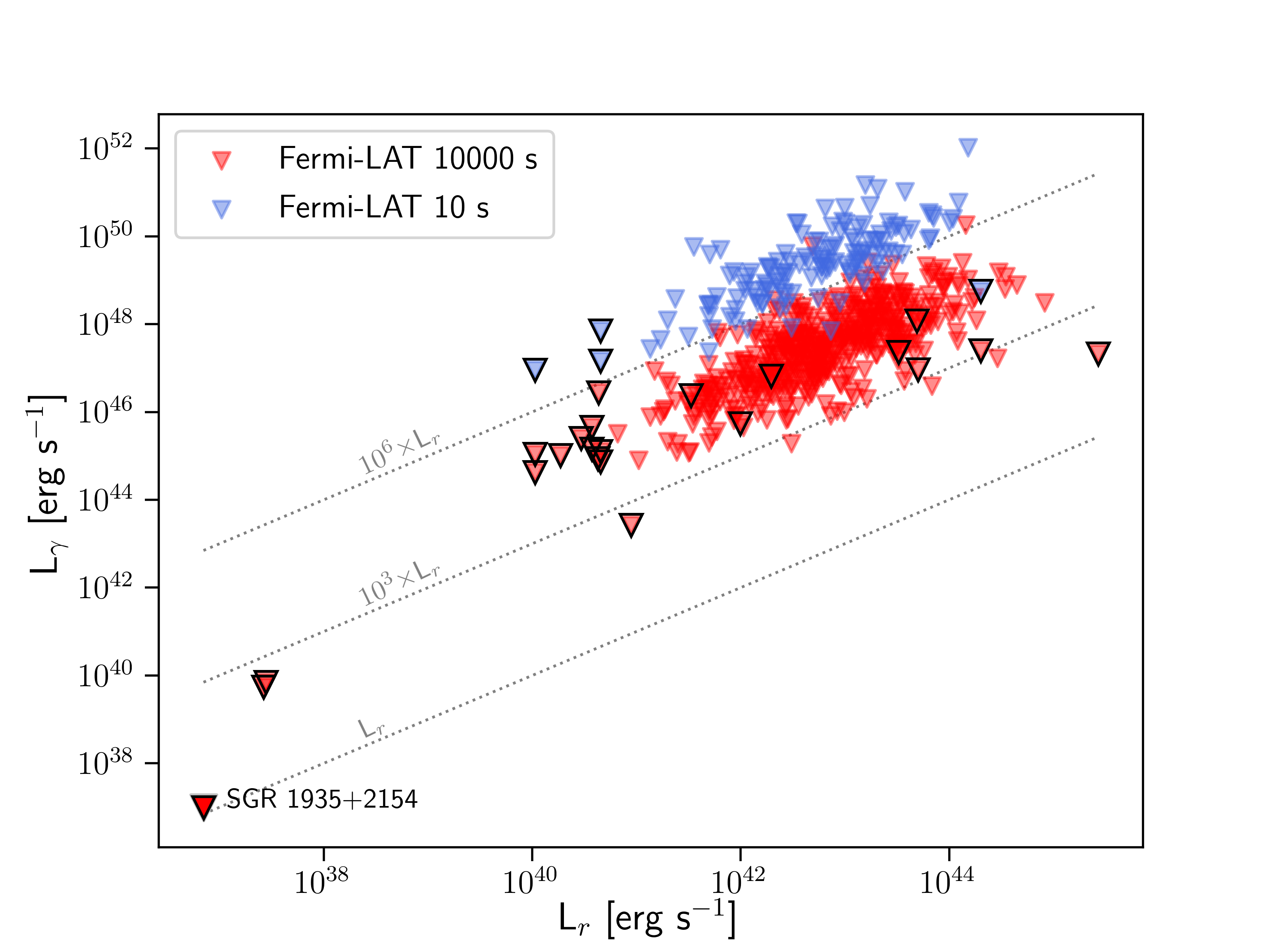}
\includegraphics[width=1.1\columnwidth]{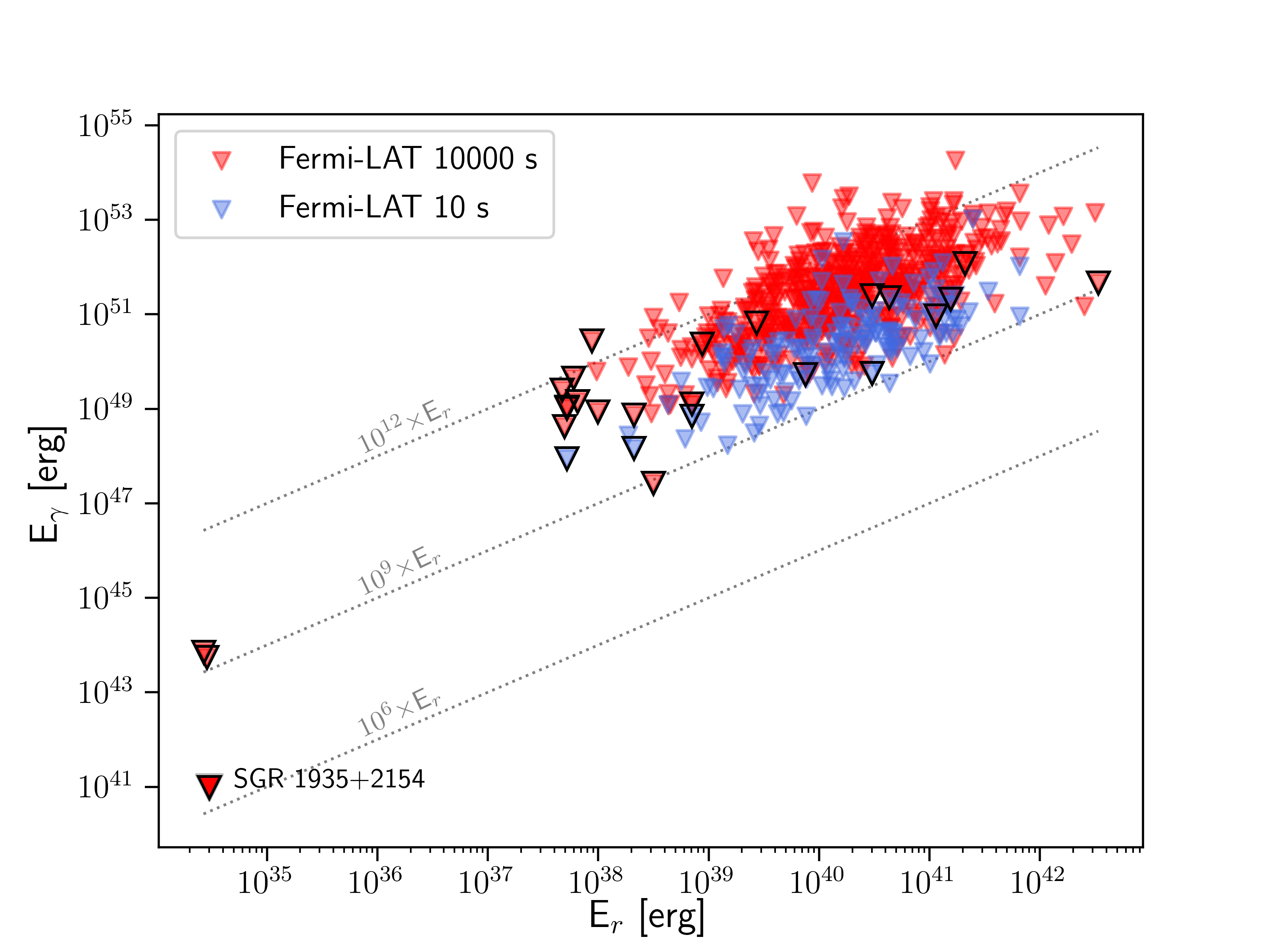}
\caption{\small \label{fig:lum_en_dist2}
Top: the ULs on the gamma-ray luminosity as a function of radio luminosity. Bottom: the ULs on the emitted gamma-ray energy as a function of radio energy. The FRB events reported are those with available information on their radio flux density and with estimated gamma-ray flux (sufficient gamma-ray exposure). Gamma-ray upper limits from the analysis over 10 s (10000 s) are plotted in blue (red). FRB sources with known host-galaxy are outlined in black.}
\end{figure}

We constrained the gamma-ray emission for the FRB events in our sample.
The top (bottom) panel of Fig. \ref{fig:lum_en_dist2} shows the \textit{Fermi}-LAT ULs on the gamma-ray luminosity (energy) compared to that emitted in radio, over time windows of 10 s and 10000 s centred on the FRB episode. 

\noindent The resulting ULs on the gamma-ray luminosity are in the range $7\times 10^{41}$--$10^{52}$ erg s$^{-1}$, for the analysis with $\delta_T=$10 s, and $9\times 10^{36}$--$10^{50}$ erg s$^{-1}$ for $\delta_T=$10000 s. 
While the FRB event related to SGR\,1935+2154 was not observed for short time scales (it entered the LAT FoV only after 2000 s), the UL we obtained at a longer time scale ($\delta_T=$10000 s) is, as of today, the most stringent at gamma-ray energies ($L_\gamma <  10^{37}$ erg s$^{-1}$). This value is close to the measured radio luminosity ($7 \times 10^{36}$ erg s$^{-1}$), ruling out a gamma-ray luminosity for this source larger than the radio one for $\delta_T=10000$ s.
As for the analysis with $\delta_T=$10 s, the most stringent ULs ($\sim 10^{42}$ erg s$^{-1}$) were obtained for two events related to the repeaters FRB\,20200120E (at 3.6 Mpc). 
Unfortunately, no information on the radio burst parameters other than localisation, event time and significance has yet been published for these two events \footnote{https://www.chime-frb.ca/repeaters/FRB20200120E}, precluding a direct comparison with the radio band.
For more distant bursts (e.g. the periodic FRB\,20180916B, at 150 Mpc) our ULs for $\delta_T=$10 s and $\delta_T=$1000 s are of the order of 10$^{46}$ erg s$^{-1}$ and 10$^{44}$ erg s$^{-1}$, respectively.

The ULs on the gamma-ray energy have been obtained by multiplying the UL on the luminosity by the corresponding time window ($\delta_T$).
Similarly to the results obtained for the luminosity, our most stringent limits on the gamma-ray energy are of the order of $\sim 10^{43}$ erg (FRB\,20200120E) and $10^{40}$ erg (SGR\,1935+2154), for the analyses with $\delta_T =$10 s and $\delta_T =$10000 s, respectively.
The latter limit implies a gamma-ray-to-radio energy ratio smaller than $3\times10^{6}$.
For more distant bursts our ULs on the gamma-ray energy for both the analyses with $\delta_T =$10 s and $\delta_T =$10000 s are of the order of 10$^{47}$ erg, excluding gamma-ray energies larger than $10^{9}$ times those emitted in radio.

Similar results have been obtained using the stacking analysis on the entire sample of FRB events (see Sect. \ref{sec:stacking_energy_luminosity}). The cumulative ULs on the gamma-ray energy are of the order of $10^{46}$ erg, two orders of magnitude smaller than the average ULs derived from individual events.

Considering shorter time intervals ($\delta_T<1$ s), we derived the LAT sensitivity on milli-second scales (see Sect. \ref{sec:lat_ms_sensitivity}). For the FRB sources in the LAT FoV at the time of the event, the ULs on the emitted gamma-ray energy are of the order of $E_{\gamma, \delta_T = 1 \textrm{ms}} \lesssim 10^{47} (\textrm{D}_L/150 \textrm{Mpc})^2$ erg. Since no photons have been detected from the FRB locations, by comparing our limits with the measured radio energy we can rule out energies greater than $10^{9}$ times the radio values at millisecond time scales.

The non-detection of gamma-ray emission from FRB events for any of our analysis methods implies that the gamma-ray-to-radio energy ratio has to be smaller than $3\times10^{6}$ -- 10$^{9}$ on time windows ranging from ms to 10000 s. These values are still in agreement with the predictions of \citet{2019MNRAS.485.4091M,2020ApJ...902L..22M}, while being close to the predicted gamma-ray emission by \citet{2014MNRAS.442L...9L}.
Our constraints could also be an indication that models for which the gamma-ray emission is strongly attenuated may be preferred.
This can be the case of different beaming angles for radio and high-energy emission \citep[e.g.,][]{2021MNRAS.501.3184S}, or hypernebulae origin for FRBs, for which gamma-rays are strongly attenuated within the nebula primarily by the Breit-Wheeler pair production process $\gamma \gamma \rightarrow e^{+} e^{-}$ \citep{2022arXiv221211236S}.

\section{Conclusion}
The goal of our work was to study the high-energy emission of FRB counterparts, performing the broadest and deepest search for gamma-rays from FRB sources to date. To this end we considered 13 years of \textit{Fermi}-LAT data for a sample of 1020 FRB events. We combined different compilations of FRB events, resulting in the largest sample used so far for a gamma-ray study. 
First, we analysed the gamma-ray data of each source individually for different time scales (from 10 to 10000 s) to look for a significant detection, while for the 22 repeaters included in our sample we also searched for steady emission using 13 years of LAT data. Then, we performed a stacking analysis of all undetected sources in order to search for a cumulative signal. Additionally, we also searched for triplets of photons possibly associated with the transient radio events.

\noindent The main results can be summarised as follows:
\begin{itemize}
\item{No significant emission has been observed from any of the analyses described above. }
\item{For the gamma-ray emission from individual FRB events our most stringent ULs on the energy flux are 1.3$\times10^{-8}$, 6.3$\times10^{-10}$, 1.2$\times10^{-10}$, 3.1$\times10^{-11}$ erg cm$^{-2}$ s$^{-1}$ for the analysis with $\delta_T=$10, 100, 1000, 10000 s, respectively. These limits are one order of magnitude more stringent than all values previously reported.}
\item{We did not detect significant steady emission (over 13 years time-scale) from any of the repeaters, obtaining ULs on the energy flux of the order of $10^{-13}$ erg cm$^{-2}$ s$^{-1}$. This value represents the most stringent UL on the search for steady gamma-ray emission from repeating FRB sources to date.}
\item{We did not detect any significant emission from FRB\,20180916B during the 5.4-day active-phase window. The UL obtained ($\nu F \nu_{\gamma} < 2.1 \times 10^{-12}$ erg cm$^{-2}$ s$^{-1}$) is the most stringent to date.}
\item{For the first time, we performed a stacking analysis of the undetected FRB events using LAT data. We did not find any statistically significant signal, neither for all the individual events with $\delta_T=$10, 100, 1000 and 10000 s (ranging from $2.3 \times 10^{-7}$ to $6.4 \times 10^{-10}$ ph. cm$^{-2}$ s$^{-1}$),
nor for the steady emission of the 22 repeaters with 13 years of data ($8.1 \times 10^{-11}$ ph cm$^{-2}$ s$^{-1}$). The cumulative ULs on the gamma-ray energy are of the order of $10^{46}$ erg, two orders of magnitude lower than the average value resulting from individual events.}
\item{For the photon triplets search, no FRB events show a post-trial p-value above 5$\sigma$, with the most significant event (FRB\,20190221B) reaching only a post-trial p-value of 0.055 (slightly less than 2$\sigma$).}
\item{Converting the obtained results into ULs on the gamma-ray energy associated to FRB events enables a direct comparison with the prediction by \citet{2014MNRAS.442L...9L} and \citet{2020ApJ...899L..27M}, in which any FRB event should be accompanied by afterglow gamma-ray emission with energies at least 10$^{6}$ (10$^{4}$ for the latter) larger than the energy emitted in radio. As a result we can rule out gamma-ray-to-radio energy ratio larger than $3\times10^{6}$ -- 10$^{9}$, on time scales of $\delta_T$ 10--10000 s.}
\item{Since no photons have been observed from the FRB directions with $\delta_T<1$ s, we calculated the LAT ms-sensitivity to be approximately 0.3 ph cm$^{-2}$ s$^{-1}$. This value can be converted to a limit on the gamma-ray emission $E_{\gamma, \delta_T = 1 \textrm{ms}} \lesssim 10^{47} (\textrm{D}_L/150 \textrm{Mpc})^2$ erg, also ruling out gamma-ray-to-radio energy ratios greater than $10^{9}$ on ms time-scale.}
\item{Comparing the location of the FRB sources with the 4FGL-DR2 sources, using a radius of $\sim 0.05^{\circ}$ similar to the LAT positional error, finds three FRB sources which might be associated to \textit{Fermi}-LAT sources. While the large DM value of FRB\,20180309 tends to exclude its association with the millisecond pulsar 4FGL\,J2124.7-3358, further MWL observations are needed to either confirm or reject the possible association of FRB\,20180924A and FRB\,20190220A with the two BCUs 4FGL\,J0221.8+3730 and 4FGL\,J1549.4+7409, respectively.}
\end{itemize}

Our results provide the most constraining limits to date on the gamma-ray emission (0.1–1000 GeV) emitted from FRB events on short time scales, from millisecond to 10000 s, as well as for persistent gamma-ray emission from the FRB repeaters. These limits rule out extreme scenarios, $ E_\gamma / E_{radio} > 3\times10^{6}$ -- 10$^{9}$ on ms to 10000 s time scale, but are still compatible with many emission models, including magnetar flares. 
The motivation for the lack of detection includes limited short-time scale sensitivity of gamma-ray instruments in the GeV domain, as well as possible suppression of the gamma-ray emission due to, e.g., different beaming angles for the radio and high-energy emission \citep[e.g.,][]{2021MNRAS.501.3184S} or a FRB scenario in which gamma rays are strongly attenuated by the Breit-Wheeler $\gamma \gamma \rightarrow e^{+} e^{-}$ in the local environment \citep{2022arXiv221211236S}.
Future searches for high-energy emission from FRBs will enable further insights for constraining their origin and emission mechanism.
In particular, very high energy observations of FRB sources by current IACT telescopes or by more sensitive instruments like the upcoming Cherenkov Telescope Array \citep{2019scta.book.....C,2022icrc.confE...5Z} are of fundamental interest: thanks to their larger effective area and better short-time scale sensitivity, these experiments may provide promising results on the search of gamma-ray counterparts to FRBs on short time scales.
In addition, the advent of wide-field radio surveys will enable a more complete picture of FRBs. A dramatic increase in the number of FRBs events will be enabled by the Square Kilometre Array \cite{2009IEEEP..97.1482D} as well as the new and expanded interferometer CHIME/CHORD \citep{2019clrp.2020...28V}, which will expand the samples of events with accurate localisation and will improve the characterization of their radio properties as well as the investigation of their local environments \citep{2020MNRAS.497.4107H}.

\subsection*{ACKNOWLEDGMENTS}
The effort of the LAT-team Calibration \& Analysis Working Group to develop Pass 8, and also the LAT MM-Transient Working Group for the support given to this project, are gratefully acknowledged.

GP acknowledges the partial support by ICSC – Centro Nazionale di Ricerca in High Performance Computing, Big Data and Quantum Computing, funded by European Union – NextGenerationEU.
MN acknowledges the support by NASA under award number 80GSFC21M0002.

The Fermi LAT Collaboration acknowledges generous ongoing support from a number of agencies and institutes that have supported both the development and the operation of the LAT as well as scientific data analysis. These include the National Aeronautics and Space Administration and the Department of Energy in the United States, the Commissariat à l'Energie Atomique and the Centre National de la Recherche Scientifique / Institut National de Physique Nucléaire et de Physique des Particules in France, the Agenzia Spaziale Italiana and the Istituto Nazionale di Fisica Nucleare in Italy, the Ministry of Education, Culture, Sports, Science and Technology (MEXT), High Energy Accelerator Research Organization (KEK) and Japan Aerospace Exploration Agency (JAXA) in Japan, and the K. A. Wallenberg Foundation, the Swedish Research Council and the Swedish National Space Board in Sweden. 

Additional support for science analysis during the operations phase is gratefully acknowledged from the Istituto Nazionale di Astrofisica in Italy and the Centre National d'Etudes Spatiales in France. This work is performed in part under DOE Contract DE-AC02-76SF00515.

\subsection*{Data Availability}
The data \footnote{\textit{Fermi}-LAT data: \url{https://fermi.gsfc.nasa.gov/ssc/data/access/}} and tools (\textit{Fermitools}\footnote{\url{https://fermi.gsfc.nasa.gov/ssc/data/analysis/software/}} and \textit{fermipy} \footnote{\url{https://fermipy.readthedocs.io/en/latest/}}) underlying this article are publicly available.


\bibliography{bibliography}  

\begin{thebibliography}{89}
\expandafter\ifx\csname natexlab\endcsname\relax\def\natexlab#1{#1}\fi

\bibitem[{{Aartsen} {et~al.}(2018){Aartsen}, {Ackermann}, {Adams}, {Aguilar},
  {Ahlers}, {Ahrens}, {Samarai}, {Altmann}, {Andeen}, {Anderson}, {Ansseau},
  {Anton}, {Arg{\"u}elles}, {Auffenberg}, {Axani}, {Bagherpour}, {Bai},
  {Barron}, {Barwick}, {Baum}, {Bay}, {Beatty}, {Becker Tjus}, {Becker},
  {BenZvi}, {Berley}, {Bernardini}, {Besson}, {Binder}, {Bindig}, {Blaufuss},
  {Blot}, {Bohm}, {B{\"o}rner}, {Bos}, {B{\"o}ser}, {Botner}, {Bourbeau},
  {Bourbeau}, {Bradascio}, {Braun}, {Brenzke}, {Bretz}, {Bron},
  {Brostean-Kaiser}, {Burgman}, {Busse}, {Carver}, {Cheung}, {Chirkin},
  {Christov}, {Clark}, {Classen}, {Collin}, {Conrad}, {Coppin}, {Correa},
  {Cowen}, {Cross}, {Dave}, {Day}, {de Andr{\'e}}, {De Clercq}, {DeLaunay},
  {Dembinski}, {De Ridder}, {Desiati}, {de Vries}, {de Wasseige}, {de With},
  {DeYoung}, {D{\'\i}az-V{\'e}lez}, {di Lorenzo}, {Dujmovic}, {Dumm},
  {Dunkman}, {Dvorak}, {Eberhardt}, {Ehrhardt}, {Eichmann}, {Eller}, {Evenson},
  {Fahey}, {Fazely}, {Felde}, {Filimonov}, {Finley}, {Flis}, {Franckowiak},
  {Friedman}, {Fritz}, {Gaisser}, {Gallagher}, {Gerhardt}, {Ghorbani}, {Giang},
  {Glauch}, {Gl{\"u}senkamp}, {Goldschmidt}, {Gonzalez}, {Grant}, {Griffith},
  {Haack}, {Hallgren}, {Halzen}, {Hanson}, {Hebecker}, {Heereman}, {Helbing},
  {Hellauer}, {Hickford}, {Hignight}, {Hill}, {Hoffman}, {Hoffmann}, {Hoinka},
  {Hokanson-Fasig}, {Hoshina}, {Huang}, {Huber}, {Hultqvist}, {H{\"u}nnefeld},
  {Hussain}, {In}, {Iovine}, {Ishihara}, {Jacobi}, {Japaridze}, {Jeong},
  {Jero}, {Jones}, {Kalaczynski}, {Kang}, {Kappes}, {Kappesser}, {Karg},
  {Karle}, {Katz}, {Kauer}, {Keivani}, {Kelley}, {Kheirandish}, {Kim}, {Kim},
  {Kintscher}, {Kiryluk}, {Kittler}, {Klein}, {Koirala}, {Kolanoski},
  {K{\"o}pke}, {Kopper}, {Kopper}, {Koschinsky}, {Koskinen}, {Kowalski},
  {Krings}, {Kroll}, {Kr{\"u}ckl}, {Kunwar}, {Kurahashi}, {Kuwabara},
  {Kyriacou}, {Labare}, {Lanfranchi}, {Larson}, {Lauber}, {Leonard},
  {Lesiak-Bzdak}, {Leuermann}, {Liu}, {Lozano Mariscal}, {Lu}, {L{\"u}nemann},
  {Luszczak}, {Madsen}, {Maggi}, {Mahn}, {Mancina}, {Maruyama}, {Mase},
  {Maunu}, {Meagher}, {Medici}, {Meier}, {Menne}, {Merino}, {Meures},
  {Miarecki}, {Micallef}, {Moment{\'e}}, {Montaruli}, {Moore}, {Moulai},
  {Nahnhauer}, {Nakarmi}, {Naumann}, {Neer}, {Niederhausen}, {Nowicki},
  {Nygren}, {Obertacke Pollmann}, {Olivas}, {O'Murchadha}, {O'Sullivan},
  {Palczewski}, {Pandya}, {Pankova}, {Peiffer}, {Pepper}, {P{\'e}rez de los
  Heros}, {Pieloth}, {Pinat}, {Plum}, {Price}, {Przybylski}, {Raab},
  {R{\"a}del}, {Rameez}, {Rauch}, {Rawlins}, {Rea}, {Reimann}, {Relethford},
  {Relich}, {Resconi}, {Rhode}, {Richman}, {Robertson}, {Rongen}, {Rott},
  {Ruhe}, {Ryckbosch}, {Rysewyk}, {Safa}, {S{\"a}lzer}, {Sanchez Herrera},
  {Sandrock}, {Sandroos}, {Santander}, {Sarkar}, {Sarkar}, {Satalecka},
  {Schlunder}, {Schmidt}, {Schneider}, {Schoenen}, {Sch{\"o}neberg},
  {Schumacher}, {Sclafani}, {Seckel}, {Seunarine}, {Soedingrekso}, {Soldin},
  {Song}, {Spiczak}, {Spiering}, {Stachurska}, {Stamatikos}, {Stanev},
  {Stasik}, {Stein}, {Stettner}, {Steuer}, {Stezelberger}, {Stokstad},
  {St{\"o}{\ss}l}, {Strotjohann}, {Stuttard}, {Sullivan}, {Sutherland},
  {Taboada}, {Tatar}, {Tenholt}, {Ter-Antonyan}, {Terliuk}, {Tilav}, {Toale},
  {Tobin}, {T{\"o}nnis}, {Toscano}, {Tosi}, {Tselengidou}, {Tung}, {Turcati},
  {Turley}, {Ty}, {Unger}, {Usner}, {Vandenbroucke}, {Van Driessche}, {van
  Eijk}, {van Eijndhoven}, {Vanheule}, {van Santen}, {Vogel}, {Vraeghe},
  {Walck}, {Wallace}, {Wallraff}, {Wandler}, {Wandkowsky}, {Waza}, {Weaver},
  {Weiss}, {Wendt}, {Werthebach}, {Westerhoff}, {Whelan}, {Wiebe}, {Wiebusch},
  {Wille}, {Williams}, {Wills}, {Wolf}, {Wood}, {Wood}, {Woolsey}, {Woschnagg},
  {Xu}, {Xu}, {Xu}, {Yanez}, {Yodh}, {Yoshida}, {Yuan}, \& {IceCube
  Collaboration}}]{2018ApJ...857..117A}
{Aartsen}, M.~G., {Ackermann}, M., {Adams}, J., {et~al.} 2018, \apj, 857, 117

\bibitem[{{Abbasi} {et~al.}(2023){Abbasi}, {Ackermann}, {Adams}, {Agarwalla},
  {Aguilar}, {Ahlers}, {Alameddine}, {Amin}, {Andeen}, {Anton},
  {Arg{\"u}elles}, {Ashida}, {Athanasiadou}, {Axani}, {Bai}, {Balagopal},
  {Baricevic}, {Barwick}, {Basu}, {Bay}, {Beatty}, {Becker}, {Becker Tjus},
  {Beise}, {Bellenghi}, {BenZvi}, {Berley}, {Bernardini}, {Besson}, {Binder},
  {Bindig}, {Blaufuss}, {Blot}, {Bontempo}, {Book}, {Boscolo Meneguolo},
  {B{\"o}ser}, {Botner}, {B{\"o}ttcher}, {Bourbeau}, {Braun}, {Brinson},
  {Brostean-Kaiser}, {Burley}, {Busse}, {Butterfield}, {Campana}, {Carloni},
  {Carnie-Bronca}, {Chattopadhyay}, {Chen}, {Chen}, {Chirkin}, {Choi}, {Clark},
  {Classen}, {Coleman}, {Collin}, {Connolly}, {Conrad}, {Coppin}, {Correa},
  {Countryman}, {Cowen}, {Dave}, {De Clercq}, {DeLaunay}, {Delgado L{\'o}pez},
  {Dembinski}, {Deoskar}, {Desai}, {Desiati}, {de Vries}, {de Wasseige},
  {DeYoung}, {Diaz}, {D{\'\i}az-V{\'e}lez}, {Dittmer}, {Domi}, {Dujmovic},
  {DuVernois}, {Ehrhardt}, {Eller}, {Engel}, {Erpenbeck}, {Evans}, {Evenson},
  {Fan}, {Fang}, {Fazely}, {Fedynitch}, {Feigl}, {Fiedlschuster}, {Finley},
  {Fischer}, {Fox}, {Franckowiak}, {Friedman}, {Fritz}, {F{\"u}rst}, {Gaisser},
  {Gallagher}, {Ganster}, {Garcia}, {Garrappa}, {Gerhardt}, {Ghadimi},
  {Glaser}, {Glauch}, {Gl{\"u}senkamp}, {Goehlke}, {Gonzalez}, {Goswami},
  {Grant}, {Gray}, {Griffin}, {Griswold}, {G{\"u}nther}, {Gutjahr}, {Haack},
  {Hallgren}, {Halliday}, {Halve}, {Halzen}, {Hamdaoui}, {Ha Minh}, {Hanson},
  {Hardin}, {Harnisch}, {Hatch}, {Haungs}, {Helbing}, {Hellrung}, {Henningsen},
  {Heuermann}, {Hickford}, {Hidvegi}, {Hill}, {Hill}, {Hoffman}, {Hoshina},
  {Hou}, {Huber}, {Hultqvist}, {H{\"u}nnefeld}, {Hussain}, {Hymon}, {In},
  {Iovine}, {Ishihara}, {Jacquart}, {Jansson}, {Japaridze}, {Jayakumar},
  {Jeong}, {Jin}, {Jones}, {Kang}, {Kang}, {Kang}, {Kappes}, {Kappesser},
  {Kardum}, {Karg}, {Karl}, {Karle}, {Katz}, {Kauer}, {Kelley}, {Khatee
  Zathul}, {Kheirandish}, {Kin}, {Kiryluk}, {Klein}, {Kochocki}, {Koirala},
  {Kolanoski}, {Kontrimas}, {K{\"o}pke}, {Kopper}, {Koskinen}, {Koundal},
  {Kovacevich}, {Kowalski}, {Kozynets}, {Kruiswijk}, {Krupczak}, {Kumar},
  {Kun}, {Kurahashi}, {Lad}, {Lagunas Gualda}, {Lamoureux}, {Larson}, {Lauber},
  {Lazar}, {Lee}, {DeHolton}, {Leszczy{\'n}ska}, {Lincetto}, {Liu},
  {Liubarska}, {Lohfink}, {Love}, {Mariscal}, {Lu}, {Lucarelli}, {Ludwig},
  {Luszczak}, {Lyu}, {Ma}, {Madsen}, {Mahn}, {Makino}, {Mancina}, {Sainte},
  {Mari{\c{s}}}, {Marka}, {Marka}, {Marsee}, {Martinez-Soler}, {Maruyama},
  {Mayhew}, {McElroy}, {McNally}, {Mead}, {Meagher}, {Mechbal}, {Medina},
  {Meier}, {Meighen-Berger}, {Merckx}, {Merten}, {Micallef}, {Mockler},
  {Montaruli}, {Moore}, {Morii}, {Morse}, {Moulai}, {Mukherjee}, {Naab},
  {Nagai}, {Nakos}, {Naumann}, {Necker}, {Neumann}, {Niederhausen}, {Nisa},
  {Noell}, {Nowicki}, {Obertacke Pollmann}, {O'Dell}, {Oehler}, {Oeyen},
  {Olivas}, {Orsoe}, {Osborn}, {O'Sullivan}, {Pandya}, {Park}, {Parker},
  {Paudel}, {Paul}, {P{\'e}rez de los Heros}, {Peterson}, {Philippen},
  {Pieper}, {Pizzuto}, {Plum}, {Popovych}, {Prado Rodriguez}, {Pries},
  {Procter-Murphy}, {Przybylski}, {Raab}, {Rack-Helleis}, {Rawlins}, {Rechav},
  {Rehman}, {Reichherzer}, {Renzi}, {Resconi}, {Reusch}, {Rhode}, {Richman},
  {Riedel}, {Roberts}, {Robertson}, {Rodan}, {Roellinghoff}, {Rongen}, {Rott},
  {Ruhe}, {Ruohan}, {Ryckbosch}, {Athanasiadou}, {Safa}, {Saffer},
  {Salazar-Gallegos}, {Sampathkumar}, {Sanchez Herrera}, {Sandrock},
  {Santander}, {Sarkar}, {Sarkar}, {Savelberg}, {Savina}, {Schaufel},
  {Schieler}, {Schindler}, {Schl{\"u}ter}, {Schmidt}, {Schneider},
  {Schr{\"o}der}, {Schumacher}, {Schwefer}, {Sclafani}, {Seckel}, {Seunarine},
  {Sharma}, {Shefali}, {Shimizu}, {Silva}, {Skrzypek}, {Smithers}, {Snihur},
  {Soedingrekso}, {S{\o}gaard}, {Soldin}, {Sommani}, {Spannfellner}, {Spiczak},
  {Spiering}, {Stamatikos}, {Stanev}, {Stein}, {Stezelberger}, {St{\"u}rwald},
  {Stuttard}, {Sullivan}, {Taboada}, {Ter-Antonyan}, {Thompson}, {Thwaites},
  {Tilav}, {Tollefson}, {T{\"o}nnis}, {Toscano}, {Tosi}, {Trettin}, {Tung},
  {Turcotte}, {Twagirayezu}, {Ty}, {Unland Elorrieta}, {Upadhyay}, {Upshaw},
  {Valtonen-Mattila}, {Vandenbroucke}, {van Eijndhoven}, {Vannerom}, {van
  Santen}, {Vara}, {Veitch-Michaelis}, {Venugopal}, {Verpoest}, {Veske},
  {Walck}, {Watson}, {Weaver}, {Weigel}, {Weindl}, {Weldert}, {Wendt},
  {Werthebach}, {Weyrauch}, {Whitehorn}, {Wiebusch}, {Willey}, {Williams},
  {Wolf}, {Wrede}, {Wulff}, {Xu}, {Yanez}, {Yildizci}, {Yoshida}, {Yu}, {Yu},
  {Yuan}, {Zhang}, {Zhelnin}, \& {IceCube Collaboration}}]{2023ApJ...946...80A}
{Abbasi}, R., {Ackermann}, M., {Adams}, J., {et~al.} 2023, \apj, 946, 80

\bibitem[{{Abdo} {et~al.}(2009){Abdo}, {Ackermann}, {Ajello}, {Atwood},
  {Baldini}, {Ballet}, {Barbiellini}, {Bastieri}, {Baughman}, {Bechtol},
  {Bellazzini}, {Berenji}, {Bloom}, {Bonamente}, {Borgland}, {Bouvier},
  {Bregeon}, {Brez}, {Brigida}, {Bruel}, {Buehler}, {Burnett}, {Buson},
  {Caliandro}, {Cameron}, {Caraveo}, {Casandjian}, {Cecchi}, {{\c C}elik},
  {Charles}, {Chekhtman}, {Chiang}, {Ciprini}, {Claus}, {Cohen-Tanugi},
  {Conrad}, {de Palma}, {Digel}, {Do Couto E Silva}, {Drell}, {Dubois},
  {Dumora}, {Farnier}, {Favuzzi}, {Fegan}, {Focke}, {Fortin}, {Frailis},
  {Fukazawa}, {Funk}, {Fusco}, {Gargano}, {Gehrels}, {Germani}, {Giebels},
  {Giglietto}, {Giordano}, {Glanzman}, {Godfrey}, {Grenier}, {Grondin},
  {Grove}, {Guillemot}, {Guiriec}, {Hays}, {Horan}, {Hughes},
  {J{\'o}hannesson}, {Johnson}, {Johnson}, {Johnson}, {Kamae}, {Katagiri},
  {Kataoka}, {Kawai}, {Kerr}, {Kn{\"o}dlseder}, {Kuss}, {Lande}, {Latronico},
  {Lemoine-Goumard}, {Longo}, {Loparco}, {Lott}, {Lovellette}, {Lubrano},
  {Makeev}, {Mazziotta}, {McEnery}, {Meurer}, {Michelson}, {Mitthumsiri},
  {Mizuno}, {Monte}, {Monzani}, {Morselli}, {Moskalenko}, {Murgia}, {Nolan},
  {Norris}, {Nuss}, {Ohsugi}, {Okumura}, {Omodei}, {Orlando}, {Ormes},
  {Paneque}, {Panetta}, {Parent}, {Pelassa}, {Pepe}, {Pesce-Rollins}, {Piron},
  {Porter}, {Rain{\`o}}, {Rando}, {Razzano}, {Reimer}, {Reimer}, {Reposeur},
  {Rochester}, {Rodriguez}, {Roth}, {Sadrozinski}, {Sander}, {Saz Parkinson},
  {Sgr{\`o}}, {Share}, {Siskind}, {Smith}, {Smith}, {Spandre}, {Spinelli},
  {Strickman}, {Suson}, {Takahashi}, {Tanaka}, {Thayer}, {Thayer}, {Thompson},
  {Tibaldo}, {Torres}, {Tosti}, {Tramacere}, {Uchiyama}, {Usher}, {Vasileiou},
  {Vilchez}, {Vitale}, {Waite}, {Wang}, {Winer}, {Wood}, {Ylinen}, \&
  {Ziegler}}]{2009PhRvD..80l2004A}
{Abdo}, A.~A., {Ackermann}, M., {Ajello}, M., {et~al.} 2009, \prd, 80, 122004

\bibitem[{{Abdollahi} {et~al.}(2020){Abdollahi}, {Acero}, {Ackermann},
  {Ajello}, {Atwood}, {Axelsson}, {Baldini}, {Ballet}, {Barbiellini},
  {Bastieri}, {Becerra Gonzalez}, {Bellazzini}, {Berretta}, {Bissaldi}, {Bland
  ford}, {Bloom}, {Bonino}, {Bottacini}, {Brandt}, {Bregeon}, {Bruel},
  {Buehler}, {Burnett}, {Buson}, {Cameron}, {Caputo}, {Caraveo}, {Casandjian},
  {Castro}, {Cavazzuti}, {Charles}, {Chaty}, {Chen}, {Cheung}, {Chiaro},
  {Ciprini}, {Cohen-Tanugi}, {Cominsky}, {Coronado-Bl{\'a}zquez}, {Costantin},
  {Cuoco}, {Cutini}, {D'Ammando}, {DeKlotz}, {Torre Luque}, {de Palma},
  {Desai}, {Digel}, {Lalla}, {Mauro}, {Venere}, {Dom{\'\i}nguez}, {Dumora},
  {Dirirsa}, {Fegan}, {Ferrara}, {Franckowiak}, {Fukazawa}, {Funk}, {Fusco},
  {Gargano}, {Gasparrini}, {Giglietto}, {Giommi}, {Giordano}, {Giroletti},
  {Glanzman}, {Green}, {Grenier}, {Griffin}, {Grondin}, {Grove}, {Guiriec},
  {Harding}, {Hayashi}, {Hays}, {Hewitt}, {Horan}, {J{\'o}hannesson},
  {Johnson}, {Kamae}, {Kerr}, {Kocevski}, {Kovac'evic'}, {Kuss}, {Landriu},
  {Larsson}, {Latronico}, {Lemoine-Goumard}, {Li}, {Liodakis}, {Longo},
  {Loparco}, {Lott}, {Lovellette}, {Lubrano}, {Madejski}, {Maldera},
  {Malyshev}, {Manfreda}, {Marchesini}, {Marcotulli}, {Mart{\'\i}-Devesa},
  {Martin}, {Massaro}, {Mazziotta}, {McEnery}, {Mereu}, {Meyer}, {Michelson},
  {Mirabal}, {Mizuno}, {Monzani}, {Morselli}, {Moskalenko}, {Negro}, {Nuss},
  {Ojha}, {Omodei}, {Orienti}, {Orlando}, {Ormes}, {Palatiello}, {Paliya},
  {Paneque}, {Pei}, {Pe{\~n}a-Herazo}, {Perkins}, {Persic}, {Pesce-Rollins},
  {Petrosian}, {Petrov}, {Piron}, {Poon}, {Porter}, {Principe}, {Rain{\`o}},
  {Rando}, {Razzano}, {Razzaque}, {Reimer}, {Reimer}, {Remy}, {Reposeur},
  {Romani}, {Parkinson}, {Schinzel}, {Serini}, {Sgr{\`o}}, {Siskind}, {Smith},
  {Spandre}, {Spinelli}, {Strong}, {Suson}, {Tajima}, {Takahashi}, {Tak},
  {Thayer}, {Thompson}, {Tibaldo}, {Torres}, {Torresi}, {Valverde}, {Klaveren},
  {Zyl}, {Wood}, {Yassine}, \& {Zaharijas}}]{2020ApJS..247...33A}
{Abdollahi}, S., {Acero}, F., {Ackermann}, M., {et~al.} 2020, \apjs, 247, 33

\bibitem[{{Aggarwal} {et~al.}(2021){Aggarwal}, {Burke-Spolaor}, {Tejos},
  {Pignata}, {Xavier Prochaska}, {Ravi}, {Kaczmarek}, \&
  {Os{\l}owski}}]{2021ApJ...913...78A}
{Aggarwal}, K., {Burke-Spolaor}, S., {Tejos}, N., {et~al.} 2021, \apj, 913, 78

\bibitem[{{Ajello} {et~al.}(2021){Ajello}, {Atwood}, {Axelsson}, {Bagagli},
  {Bagni}, {Baldini}, {Bastieri}, {Bellardi}, {Bellazzini}, {Bissaldi},
  {Bloom}, {Bonino}, {Bregeon}, {Brez}, {Bruel}, {Buehler}, {Buson}, {Cameron},
  {Caraveo}, {Cavazzuti}, {Ceccanti}, {Chen}, {Cheung}, {Ciprini}, {Cognard},
  {Cohen-Tanugi}, {Cutini}, {D'Ammando}, {de la Torre Luque}, {de Palma},
  {Digel}, {Dirirsa}, {Di Lalla}, {Di Venere}, {Dom{\'\i}nguez}, {Fabiani},
  {Ferrara}, {Fiori}, {Foglia}, {Fukazawa}, {Fusco}, {Gargano}, {Gasparrini},
  {Giroletti}, {Glanzman}, {Green}, {Griffin}, {Grondin}, {Grove}, {Guillemot},
  {Guiriec}, {Gustafsson}, {Hays}, {Horan}, {J{\'o}hannesson}, {Johnson},
  {Kamae}, {Kerr}, {Kuss}, {Larsson}, {Latronico}, {Lemoine-Goumard}, {Li},
  {Liodakis}, {Longo}, {Loparco}, {Lovellette}, {Lubrano}, {Maldera},
  {Manfreda}, {Mart{\'\i}-Devesa}, {Mazziotta}, {Menon}, {Mereu}, {Meyer},
  {Michelson}, {Minuti}, {Mitthumsiri}, {Mizuno}, {Mongelli}, {Monzani},
  {Moskalenko}, {Negro}, {Nuss}, {Ojha}, {Orienti}, {Orlando}, {Paccagnella},
  {Paliya}, {Paneque}, {Pei}, {Perkins}, {Pesce-Rollins}, {Pinchera}, {Piron},
  {Poon}, {Porter}, {Primavera}, {Principe}, {Racusin}, {Rain{\`o}}, {Rando},
  {Rani}, {Rapposelli}, {Razzano}, {Razzaque}, {Reimer}, {Reimer}, {Russell},
  {Saggini}, {Saz Parkinson}, {Scolieri}, {Serini}, {Sgr{\`o}}, {Siskind},
  {Smith}, {Spandre}, {Spinelli}, {Suson}, {Tajima}, {Thayer}, {Thompson},
  {Tibaldo}, {Torres}, {Tosti}, {Valverde}, {Vigiani}, \&
  {Zaharijas}}]{2021ApJS..256...12A}
{Ajello}, M., {Atwood}, W.~B., {Axelsson}, M., {et~al.} 2021, \apjs, 256, 12

\bibitem[{{Atwood} {et~al.}(2009){Atwood}, {Abdo}, {Ackermann}, {Althouse},
  {Anderson}, {Axelsson}, {Baldini}, {Ballet}, {Band}, {Barbiellini}, \&
  et~al.}]{2009ApJ...697.1071A}
{Atwood}, W.~B., {Abdo}, A.~A., {Ackermann}, M., {et~al.} 2009, \apj, 697, 1071

\bibitem[{{Batten}(2019)}]{2019JOSS....4.1399B}
{Batten}, A. 2019, The Journal of Open Source Software, 4, 1399

\bibitem[{{Bera} {et~al.}(2016){Bera}, {Bhattacharyya}, {Bharadwaj}, {Bhat}, \&
  {Chengalur}}]{2016MNRAS.457.2530B}
{Bera}, A., {Bhattacharyya}, S., {Bharadwaj}, S., {Bhat}, N.~D.~R., \&
  {Chengalur}, J.~N. 2016, \mnras, 457, 2530

\bibitem[{{Bhandari} {et~al.}(2020){Bhandari}, {Sadler}, {Prochaska}, {Simha},
  {Ryder}, {Marnoch}, {Bannister}, {Macquart}, {Flynn}, {Shannon}, {Tejos},
  {Corro-Guerra}, {Day}, {Deller}, {Ekers}, {Lopez}, {Mahony}, {Nu{\~n}ez}, \&
  {Phillips}}]{2020ApJ...895L..37B}
{Bhandari}, S., {Sadler}, E.~M., {Prochaska}, J.~X., {et~al.} 2020, \apjl, 895,
  L37

\bibitem[{{Bhardwaj} {et~al.}(2021){Bhardwaj}, {Gaensler}, {Kaspi},
  {Landecker}, {Mckinven}, {Michilli}, {Pleunis}, {Tendulkar}, {Andersen},
  {Boyle}, {Cassanelli}, {Chawla}, {Cook}, {Dobbs}, {Fonseca}, {Kaczmarek},
  {Leung}, {Masui}, {Mnchmeyer}, {Ng}, {Rafiei-Ravandi}, {Scholz}, {Shin},
  {Smith}, {Stairs}, \& {Zwaniga}}]{2021ApJ...910L..18B}
{Bhardwaj}, M., {Gaensler}, B.~M., {Kaspi}, V.~M., {et~al.} 2021, \apjl, 910,
  L18

\bibitem[{{Bruel} {et~al.}(2018){Bruel}, {Burnett}, {Digel}, {Johannesson},
  {Omodei}, \& {Wood}}]{2018arXiv181011394B}
{Bruel}, P., {Burnett}, T.~H., {Digel}, S.~W., {et~al.} 2018, arXiv e-prints,
  arXiv:1810.11394

\bibitem[{{Caleb} {et~al.}(2018){Caleb}, {Spitler}, \&
  {Stappers}}]{2018NatAs...2..839C}
{Caleb}, M., {Spitler}, L.~G., \& {Stappers}, B.~W. 2018, Nature Astronomy, 2,
  839

\bibitem[{{Chatterjee} {et~al.}(2017){Chatterjee}, {Law}, {Wharton},
  {Burke-Spolaor}, {Hessels}, {Bower}, {Cordes}, {Tendulkar}, {Bassa},
  {Demorest}, {Butler}, {Seymour}, {Scholz}, {Abruzzo}, {Bogdanov}, {Kaspi},
  {Keimpema}, {Lazio}, {Marcote}, {McLaughlin}, {Paragi}, {Ransom}, {Rupen},
  {Spitler}, \& {van Langevelde}}]{2017Natur.541...58C}
{Chatterjee}, S., {Law}, C.~J., {Wharton}, R.~S., {et~al.} 2017, \nat, 541, 58

\bibitem[{{Cherenkov Telescope Array Consortium} {et~al.}(2019){Cherenkov
  Telescope Array Consortium}, {Acharya}, {Agudo}, {Al Samarai}, {Alfaro},
  {Alfaro}, {Alispach}, {Alves Batista}, {Amans}, {Amato}, {Ambrosi},
  {Antolini}, {Antonelli}, {Aramo}, {Araya}, {Armstrong}, {Arqueros},
  {Arrabito}, {Asano}, {Ashley}, {Backes}, {Balazs}, {Balbo}, {Ballester},
  {Ballet}, {Bamba}, {Barkov}, {Barres de Almeida}, {Barrio}, {Bastieri},
  {Becherini}, {Belfiore}, {Benbow}, {Berge}, {Bernardini}, {Bernardini},
  {Bernardos}, {Bernl{\"o}hr}, {Bertucci}, {Biasuzzi}, {Bigongiari}, {Biland},
  {Bissaldi}, {Biteau}, {Blanch}, {Blazek}, {Boisson}, {Bolmont}, {Bonanno},
  {Bonardi}, {Bonavolont{\`a}}, {Bonnoli}, {Bosnjak}, {B{\"o}ttcher},
  {Braiding}, {Bregeon}, {Brill}, {Brown}, {Brun}, {Brunetti}, {Buanes},
  {Buckley}, {Bugaev}, {B{\"u}hler}, {Bulgarelli}, {Bulik}, {Burton},
  {Burtovoi}, {Busetto}, {Canestrari}, {Capalbi}, {Capitanio}, {Caproni},
  {Caraveo}, {C{\'a}rdenas}, {Carlile}, {Carosi}, {Carqu{\'\i}n}, {Carr},
  {Casanova}, {Cascone}, {Catalani}, {Catalano}, {Cauz}, {Cerruti}, {Chadwick},
  {Chaty}, {Chaves}, {Chen}, {Chen}, {Chernyakova}, {Chikawa}, {Christov},
  {Chudoba}, {Cie{\'s}lar}, {Coco}, {Colafrancesco}, {Colin}, {Conforti},
  {Connaughton}, {Conrad}, {Contreras}, {Cortina}, {Costa}, {Costantini},
  {Cotter}, {Covino}, {Crocker}, {Cuadra}, {Cuevas}, {Cumani}, {D'A{\`\i}},
  {D'Ammando}, {D'Avanzo}, {D'Urso}, {Daniel}, {Davids}, {Dawson}, {Dazzi}, {De
  Angelis}, {de C{\'a}ssia dos Anjos}, {De Cesare}, {De Franco}, {de Gouveia
  Dal Pino}, {de la Calle}, {de los Reyes Lopez}, {De Lotto}, {De Luca}, {De
  Lucia}, {de Naurois}, {de O{\~n}a Wilhelmi}, {De Palma}, {De Persio}, {de
  Souza}, {Deil}, {Del Santo}, {Delgado}, {della Volpe}, {Di Girolamo}, {Di
  Pierro}, {Di Venere}, {D{\'\i}az}, {Dib}, {Diebold}, {Djannati-Ata{\"\i}},
  {Dom{\'\i}nguez}, {Dominis Prester}, {Dorner}, {Doro}, {Drass}, {Dravins},
  {Dubus}, {Dwarkadas}, {Ebr}, {Eckner}, {Egberts}, {Einecke}, {Ekoume},
  {Els{\"a}sser}, {Ernenwein}, {Espinoza}, {Evoli}, {Fairbairn},
  {Falceta-Goncalves}, {Falcone}, {Farnier}, {Fasola}, {Fedorova}, {Fegan},
  {Fernandez-Alonso}, {Fern{\'a}ndez-Barral}, {Ferrand}, {Fesquet},
  {Filipovic}, {Fioretti}, {Fontaine}, {Fornasa}, {Fortson}, {Freixas
  Coromina}, {Fruck}, {Fujita}, {Fukazawa}, {Funk}, {F{\"u}{\ss}ling},
  {Gabici}, {Gadola}, {Gallant}, {Garcia}, {Garcia L{\'o}pez}, {Garczarczyk},
  {Gaskins}, {Gasparetto}, {Gaug}, {Gerard}, {Giavitto}, {Giglietto}, {Giommi},
  {Giordano}, {Giro}, {Giroletti}, {Giuliani}, {Glicenstein}, {Gnatyk},
  {Godinovic}, {Goldoni}, {G{\'o}mez-Vargas}, {Gonz{\'a}lez}, {Gonz{\'a}lez},
  {G{\"o}tz}, {Graham}, {Grandi}, {Granot}, {Green}, {Greenshaw}, {Griffiths},
  {Gunji}, {Hadasch}, {Hara}, {Hardcastle}, {Hassan}, {Hayashi}, {Hayashida},
  {Heller}, {Helo}, {Hermann}, {Hinton}, {Hnatyk}, {Hofmann}, {Holder},
  {Horan}, {H{\"o}randel}, {Horns}, {Horvath}, {Hovatta}, {Hrabovsky},
  {Hrupec}, {Humensky}, {H{\"u}tten}, {Iarlori}, {Inada}, {Inome}, {Inoue},
  {Inoue}, {Inoue}, {Iocco}, {Ioka}, {Iori}, {Ishio}, {Iwamura}, {Jamrozy},
  {Janecek}, {Jankowsky}, {Jean}, {Jung-Richardt}, {Jurysek}, {Kaaret},
  {Karkar}, {Katagiri}, {Katz}, {Kawanaka}, {Kazanas}, {Kh{\'e}lifi}, {Kieda},
  {Kimeswenger}, {Kimura}, {Kisaka}, {Knapp}, {Kn{\"o}dlseder}, {Koch},
  {Kohri}, {Komin}, {Kosack}, {Kraus}, {Krause}, {Krau{\ss}}, {Kubo}, {Kukec
  Mezek}, {Kuroda}, {Kushida}, {La Palombara}, {Lamanna}, {Lang}, {Lapington},
  {Le Blanc}, {Leach}, {Lees}, {Lefaucheur}, {Leigui de Oliveira}, {Lenain},
  {Lico}, {Limon}, {Lindfors}, {Lohse}, {Lombardi}, {Longo}, {L{\'o}pez},
  {L{\'o}pez-Coto}, {Lu}, {Lucarelli}, {Luque-Escamilla}, {Lyard}, {Maccarone},
  {Maier}, {Majumdar}, {Malaguti}, {Mandat}, {Maneva}, {Manganaro}, {Mangano},
  {Marcowith}, {Mar{\'\i}n}, {Markoff}, {Mart{\'\i}}, {Martin},
  {Mart{\'\i}nez}, {Mart{\'\i}nez}, {Masetti}, {Masuda}, {Maurin}, {Maxted},
  {Mazin}, {Medina}, {Melandri}, {Mereghetti}, {Meyer}, {Minaya}, {Mirabal},
  {Mirzoyan}, {Mitchell}, {Mizuno}, {Moderski}, {Mohammed}, {Mohrmann},
  {Montaruli}, {Moralejo}, {Morcuende-Parrilla}, {Mori}, {Morlino}, {Morris},
  {Morselli}, {Moulin}, {Mukherjee}, {Mundell}, {Murach}, {Muraishi}, {Murase},
  {Nagai}, {Nagataki}, {Nagayoshi}, {Naito}, {Nakamori}, {Nakamura}, {Niemiec},
  {Nieto}, {Niko{\l}ajuk}, {Nishijima}, {Noda}, {Nosek}, {Novosyadlyj},
  {Nozaki}, {O'Brien}, {Oakes}, {Ohira}, {Ohishi}, {Ohm}, {Okazaki}, {Okumura},
  {Ong}, {Orienti}, {Orito}, {Osborne}, {Ostrowski}, {Otte}, {Oya}, {Padovani},
  {Paizis}, {Palatiello}, {Palatka}, {Paoletti}, {Paredes}, {Pareschi},
  {Parsons}, {Pe'er}, {Pech}, {Pedaletti}, {Perri}, {Persic}, {Petrashyk},
  {Petrucci}, {Petruk}, {Peyaud}, {Pfeifer}, {Piano}, {Pisarski}, {Pita},
  {Pohl}, {Polo}, {Pozo}, {Prandini}, {Prast}, {Principe}, {Prokhorov},
  {Prokoph}, {Prouza}, {P{\"u}hlhofer}, {Punch}, {P{\"u}rckhauer}, {Queiroz},
  {Quirrenbach}, {Rain{\`o}}, {Razzaque}, {Reimer}, {Reimer}, {Reisenegger},
  {Renaud}, {Rezaeian}, {Rhode}, {Ribeiro}, {Rib{\'o}}, {Richtler}, {Rico},
  {Rieger}, {Riquelme}, {Rivoire}, {Rizi}, {Rodriguez}, {Rodriguez Fernandez},
  {Rodr{\'\i}guez V{\'a}zquez}, {Rojas}, {Romano}, {Romeo}, {Rosado}, {Rovero},
  {Rowell}, {Rudak}, {Rugliancich}, {Rulten}, {Sadeh}, {Safi-Harb}, {Saito},
  {Sakaki}, {Sakurai}, {Salina}, {S{\'a}nchez-Conde}, {Sandaker}, {Sandoval},
  {Sangiorgi}, {Sanguillon}, {Sano}, {Santander}, {Sarkar}, {Satalecka},
  {Saturni}, {Schioppa}, {Schlenstedt}, {Schneider}, {Schoorlemmer},
  {Schovanek}, {Schulz}, {Schussler}, {Schwanke}, {Sciacca}, {Scuderi},
  {Seitenzahl}, {Semikoz}, {Sergijenko}, {Servillat}, {Shalchi}, {Shellard},
  {Sidoli}, {Siejkowski}, {Sillanp{\"a}{\"a}}, {Sironi}, {Sitarek}, {Sliusar},
  {Slowikowska}, {Sol}, {Stamerra}, {Stani{\v{c}}}, {Starling}, {Stawarz},
  {Stefanik}, {Stephan}, {Stolarczyk}, {Stratta}, {Straumann}, {Suomijarvi},
  {Supanitsky}, {Tagliaferri}, {Tajima}, {Tavani}, {Tavecchio}, {Tavernet},
  {Tayabaly}, {Tejedor}, {Temnikov}, {Terada}, {Terrier}, {Terzic}, {Teshima},
  {Testa}, {Thoudam}, {Tian}, {Tibaldo}, {Tluczykont}, {Todero Peixoto},
  {Tokanai}, {Tomastik}, {Tonev}, {Tornikoski}, {Torres}, {Torresi}, {Tosti},
  {Tothill}, {Tovmassian}, {Travnicek}, {Trichard}, {Trifoglio}, {Troyano
  Pujadas}, {Tsujimoto}, {Umana}, {Vagelli}, {Vagnetti}, {Valentino},
  {Vallania}, {Valore}, {van Eldik}, {Vandenbroucke}, {Varner}, {Vasileiadis},
  {Vassiliev}, {V{\'a}zquez Acosta}, {Vecchi}, {Vega}, {Vercellone}, {Veres},
  {Vergani}, {Verzi}, {Vettolani}, {Viana}, {Vigorito}, {Villanueva}, {Voelk},
  {Vollhardt}, {Vorobiov}, {Vrastil}, {Vuillaume}, {Wagner}, {Wagner},
  {Walter}, {Ward}, {Warren}, {Watson}, {Werner}, {White}, {White},
  {Wierzcholska}, {Wilcox}, {Will}, {Williams}, {Wischnewski}, {Wood},
  {Yamamoto}, {Yamazaki}, {Yanagita}, {Yang}, {Yoshida}, {Yoshiike},
  {Yoshikoshi}, {Zacharias}, {Zaharijas}, {Zampieri}, {Zandanel}, {Zanin},
  {Zavrtanik}, {Zavrtanik}, {Zdziarski}, {Zech}, {Zechlin}, {Zhdanov},
  {Ziegler}, \& {Zorn}}]{2019scta.book.....C}
{Cherenkov Telescope Array Consortium}, {Acharya}, B.~S., {Agudo}, I., {et~al.}
  2019, {Science with the Cherenkov Telescope Array} (World Scientific
  Publishing Co. Pte. Ltd.)

\bibitem[{{CHIME/FRB Collaboration} {et~al.}(2021){CHIME/FRB Collaboration},
  {Amiri}, {Andersen}, {Bandura}, {Berger}, {Bhardwaj}, {Boyce}, {Boyle},
  {Brar}, {Breitman}, {Cassanelli}, {Chawla}, {Chen}, {Cliche}, {Cook},
  {Cubranic}, {Curtin}, {Deng}, {Dobbs}, {Dong}, {Eadie}, {Fandino}, {Fonseca},
  {Gaensler}, {Giri}, {Good}, {Halpern}, {Hill}, {Hinshaw}, {Josephy},
  {Kaczmarek}, {Kader}, {Kania}, {Kaspi}, {Landecker}, {Lang}, {Leung}, {Li},
  {Lin}, {Masui}, {McKinven}, {Mena-Parra}, {Merryfield}, {Meyers}, {Michilli},
  {Milutinovic}, {Mirhosseini}, {M{\"u}nchmeyer}, {Naidu}, {Newburgh}, {Ng},
  {Patel}, {Pen}, {Petroff}, {Pinsonneault-Marotte}, {Pleunis},
  {Rafiei-Ravandi}, {Rahman}, {Ransom}, {Renard}, {Sanghavi}, {Scholz}, {Shaw},
  {Shin}, {Siegel}, {Sikora}, {Singh}, {Smith}, {Stairs}, {Tan}, {Tendulkar},
  {Vanderlinde}, {Wang}, {Wulf}, \& {Zwaniga}}]{2021ApJS..257...59C}
{CHIME/FRB Collaboration}, {Amiri}, M., {Andersen}, B.~C., {et~al.} 2021,
  \apjs, 257, 59

\bibitem[{{CHIME/FRB Collaboration} {et~al.}(2020{\natexlab{a}}){CHIME/FRB
  Collaboration}, {Amiri}, {Andersen}, {Bandura}, {Bhardwaj}, {Boyle}, {Brar},
  {Chawla}, {Chen}, {Cliche}, {Cubranic}, {Deng}, {Denman}, {Dobbs}, {Dong},
  {Fandino}, {Fonseca}, {Gaensler}, {Giri}, {Good}, {Halpern}, {Hessels},
  {Hill}, {H{\"o}fer}, {Josephy}, {Kania}, {Karuppusamy}, {Kaspi}, {Keimpema},
  {Kirsten}, {Landecker}, {Lang}, {Leung}, {Li}, {Lin}, {Marcote}, {Masui},
  {McKinven}, {Mena-Parra}, {Merryfield}, {Michilli}, {Milutinovic},
  {Mirhosseini}, {Naidu}, {Newburgh}, {Ng}, {Nimmo}, {Paragi}, {Patel}, {Pen},
  {Pinsonneault-Marotte}, {Pleunis}, {Rafiei-Ravandi}, {Rahman}, {Ransom},
  {Renard}, {Sanghavi}, {Scholz}, {Shaw}, {Shin}, {Siegel}, {Singh}, {Smegal},
  {Smith}, {Stairs}, {Tendulkar}, {Tretyakov}, {Vanderlinde}, {Wang}, {Wang},
  {Wulf}, {Yadav}, \& {Zwaniga}}]{2020Natur.582..351C}
{CHIME/FRB Collaboration}, {Amiri}, M., {Andersen}, B.~C., {et~al.}
  2020{\natexlab{a}}, \nat, 582, 351

\bibitem[{{CHIME/FRB Collaboration} {et~al.}(2018){CHIME/FRB Collaboration},
  {Amiri}, {Bandura}, {Berger}, {Bhardwaj}, {Boyce}, {Boyle}, {Brar},
  {Burhanpurkar}, {Chawla}, {Chowdhury}, {Cliche}, {Cranmer}, {Cubranic},
  {Deng}, {Denman}, {Dobbs}, {Fandino}, {Fonseca}, {Gaensler}, {Giri},
  {Gilbert}, {Good}, {Guliani}, {Halpern}, {Hinshaw}, {H{\"o}fer}, {Josephy},
  {Kaspi}, {Landecker}, {Lang}, {Liao}, {Masui}, {Mena-Parra}, {Naidu},
  {Newburgh}, {Ng}, {Patel}, {Pen}, {Pinsonneault-Marotte}, {Pleunis}, {Rafiei
  Ravandi}, {Ransom}, {Renard}, {Scholz}, {Sigurdson}, {Siegel}, {Smith},
  {Stairs}, {Tendulkar}, {Vanderlinde}, \& {Wiebe}}]{2018ApJ...863...48C}
{CHIME/FRB Collaboration}, {Amiri}, M., {Bandura}, K., {et~al.} 2018, \apj,
  863, 48

\bibitem[{{CHIME/FRB Collaboration} {et~al.}(2020{\natexlab{b}}){CHIME/FRB
  Collaboration}, {Andersen}, {Bandura}, {Bhardwaj}, {Bij}, {Boyce}, {Boyle},
  {Brar}, {Cassanelli}, {Chawla}, {Chen}, {Cliche}, {Cook}, {Cubranic},
  {Curtin}, {Denman}, {Dobbs}, {Dong}, {Fandino}, {Fonseca}, {Gaensler},
  {Giri}, {Good}, {Halpern}, {Hill}, {Hinshaw}, {H{\"o}fer}, {Josephy},
  {Kania}, {Kaspi}, {Landecker}, {Leung}, {Li}, {Lin}, {Masui}, {McKinven},
  {Mena-Parra}, {Merryfield}, {Meyers}, {Michilli}, {Milutinovic},
  {Mirhosseini}, {M{\"u}nchmeyer}, {Naidu}, {Newburgh}, {Ng}, {Patel}, {Pen},
  {Pinsonneault-Marotte}, {Pleunis}, {Quine}, {Rafiei-Ravandi}, {Rahman},
  {Ransom}, {Renard}, {Sanghavi}, {Scholz}, {Shaw}, {Shin}, {Siegel}, {Singh},
  {Smegal}, {Smith}, {Stairs}, {Tan}, {Tendulkar}, {Tretyakov}, {Vanderlinde},
  {Wang}, {Wulf}, \& {Zwaniga}}]{2020Natur.587...54C}
{CHIME/FRB Collaboration}, {Andersen}, B.~C., {Bandura}, K.~M., {et~al.}
  2020{\natexlab{b}}, \nat, 587, 54

\bibitem[{{CHIME/FRB Collaboration} {et~al.}(2022){CHIME/FRB Collaboration},
  {Bandura}, {Bhardwaj}, {Boyle}, {Brar}, {Breitman}, {Cassanelli},
  {Chatterjee}, {Chawla}, {Cliche}, {Cubranic}, {Curtin}, {Deng}, {Dobbs},
  {Dong}, {Fonseca}, {Gaensler}, {Giri}, {Good}, {Hill}, {Josephy},
  {Kaczmarek}, {Kader}, {Kania}, {Kaspi}, {Leung}, {Li}, {Lin}, {Masui},
  {McKinven}, {Mena-Parra}, {Merryfield}, {Meyers}, {Michilli}, {Naidu},
  {Newburgh}, {Ng}, {Ordog}, {Patel}, {Pearlman}, {Pen}, {Petroff}, {Pleunis},
  {Rafiei-Ravandi}, {Rahman}, {Ransom}, {Renard}, {Sanghavi}, {Scholz}, {Shaw},
  {Shin}, {Siegel}, {Singh}, {Smith}, {Stairs}, {Tan}, {Tendulkar},
  {Vanderlinde}, {Wiebe}, {Wulf}, \& {Zwaniga}}]{2022Natur.607..256C}
{CHIME/FRB Collaboration}, Andersen, B.~C., {Bandura}, K., {Bhardwaj}, M.,
  {et~al.} 2022, \nat, 607, 256

\bibitem[{{Cordes} \& {Chatterjee}(2019)}]{2019ARA&A..57..417C}
{Cordes}, J.~M. \& {Chatterjee}, S. 2019, \araa, 57, 417

\bibitem[{{Cunningham} {et~al.}(2019){Cunningham}, {Cenko}, {Burns},
  {Goldstein}, {Lien}, {Kocevski}, {Briggs}, {Connaughton}, {Miller},
  {Racusin}, \& {Stanbro}}]{2019ApJ...879...40C}
{Cunningham}, V., {Cenko}, S.~B., {Burns}, E., {et~al.} 2019, \apj, 879, 40

\bibitem[{{Dewdney} {et~al.}(2009){Dewdney}, {Hall}, {Schilizzi}, \&
  {Lazio}}]{2009IEEEP..97.1482D}
{Dewdney}, P.~E., {Hall}, P.~J., {Schilizzi}, R.~T., \& {Lazio}, T.~J.~L.~W.
  2009, IEEE Proceedings, 97, 1482

\bibitem[{{Dong} \& {CHIME/FRB Collaboration}(2022)}]{2022ATel15681....1D}
{Dong}, F.~A. \& {CHIME/FRB Collaboration}. 2022, The Astronomer's Telegram,
  15681, 1

\bibitem[{{Falcke} \& {Rezzolla}(2014)}]{2014A&A...562A.137F}
{Falcke}, H. \& {Rezzolla}, L. 2014, \aap, 562, A137

\bibitem[{{\textit{Fermi}-LAT Collaboration} {et~al.}(2021){\textit{Fermi}-LAT
  Collaboration}, {Ajello}, {Atwood}, {Axelsson}, {Baldini}, {Barbiellini},
  {Baring}, {Bastieri}, {Bellazzini}, {Berretta}, {Bissaldi}, {Blandford},
  {Bonino}, {Bregeon}, {Bruel}, {Buehler}, {Burns}, {Buson}, {Cameron},
  {Caraveo}, {Cavazzuti}, {Chen}, {Cheung}, {Chiaro}, {Ciprini}, {Costantin},
  {Crnogorcevic}, {Cutini}, {D'Ammando}, {de la Torre Luque}, {de Palma},
  {Digel}, {Di Lalla}, {Di Venere}, {Dirirsa}, {Fukazawa}, {Funk}, {Fusco},
  {Gargano}, {Giglietto}, {Gill}, {Giordano}, {Giroletti}, {Granot}, {Green},
  {Grenier}, {Griffin}, {Guiriec}, {Hays}, {Horan}, {J{\'o}hannesson}, {Kerr},
  {Kova{\v{c}}evi{\'c}}, {Kuss}, {Larsson}, {Latronico}, {Li}, {Longo},
  {Loparco}, {Lovellette}, {Lubrano}, {Maldera}, {Manfreda},
  {Mart{\'\i}-Devesa}, {Mazziotta}, {McEnery}, {Mereu}, {Michelson}, {Mizuno},
  {Monzani}, {Morselli}, {Moskalenko}, {Negro}, {Omodei}, {Orienti}, {Orlando},
  {Paliya}, {Paneque}, {Pei}, {Pesce-Rollins}, {Piron}, {Poon}, {Porter},
  {Principe}, {Racusin}, {Rain{\`o}}, {Rando}, {Rani}, {Razzaque}, {Reimer},
  {Reimer}, {Saz Parkinson}, {Scargle}, {Scotton}, {Serini}, {Sgr{\`o}},
  {Siskind}, {Spandre}, {Spinelli}, {Tajima}, {Takahashi}, {Tak}, {Torres},
  {Tosti}, {Troja}, {Wadiasingh}, {Wood}, {Yassine}, {Yusafzai}, \&
  {Zaharijas}}]{2021NatAs.tmp...11F}
{\textit{Fermi}-LAT Collaboration}, {Ajello}, M., {Atwood}, W.~B., {et~al.}
  2021, Nature Astronomy

\bibitem[{{Finke} {et~al.}(2010){Finke}, {Razzaque}, \&
  {Dermer}}]{2010ApJ...712..238F}
{Finke}, J.~D., {Razzaque}, S., \& {Dermer}, C.~D. 2010, \apj, 712, 238

\bibitem[{{Frederiks} {et~al.}(2022){Frederiks}, {Ridnaia}, {Svinkin},
  {Lysenko}, {Ulanov}, \& {Tsvetkova}}]{2022ATel15686....1F}
{Frederiks}, D., {Ridnaia}, A., {Svinkin}, D., {et~al.} 2022, The Astronomer's
  Telegram, 15686, 1

\bibitem[{{Guidorzi} {et~al.}(2020){Guidorzi}, {Marongiu}, {Martone},
  {Nicastro}, {Xiong}, {Liao}, {Li}, {Zhang}, {Amati}, {Frontera}, {Orlandini},
  {Rosati}, {Virgilli}, {Zhang}, {Bu}, {Cai}, {Cao}, {Chang}, {Chen}, {Chen},
  {Chen}, {Chen}, {Chen}, {Cui}, {Cui}, {Deng}, {Dong}, {Du}, {Fu}, {Gao},
  {Gao}, {Gao}, {Ge}, {Gu}, {Guan}, {Guo}, {Han}, {Huang}, {Huo}, {Jia},
  {Jiang}, {Jiang}, {Jin}, {Jin}, {Kong}, {Li}, {Li}, {Li}, {Li}, {Li}, {Li},
  {Li}, {Li}, {Li}, {Li}, {Liang}, {Liu}, {Liu}, {Liu}, {Liu}, {Liu}, {Liu},
  {Lu}, {Lu}, {Lu}, {Luo}, {Luo}, {Ma}, {Ma}, {Meng}, {Nang}, {Nie}, {Ou},
  {Qu}, {Sai}, {Shang}, {Song}, {Song}, {Sun}, {Tan}, {Tao}, {Tuo}, {Wang},
  {Wang}, {Wang}, {Wang}, {Wang}, {Wen}, {Wu}, {Wu}, {Wu}, {Xiao}, {Xiao},
  {Xu}, {Yang}, {Yang}, {Yang}, {Yi}, {Yin}, {You}, {Zhang}, {Zhang}, {Zhang},
  {Zhang}, {Zhang}, {Zhang}, {Zhang}, {Zhang}, {Zhang}, {Zhang}, {Zhang},
  {Zhang}, {Zhang}, {Zhang}, {Zhang}, {Zhang}, {Zhang}, {Zhang}, {Zheng},
  {Zhou}, {Zhou}, {Zhu}, {Zhu}, \& {Zhuang}}]{2020A&A...637A..69G}
{Guidorzi}, C., {Marongiu}, M., {Martone}, R., {et~al.} 2020, \aap, 637, A69

\bibitem[{{Hashimoto} {et~al.}(2020){Hashimoto}, {Goto}, {On}, {Lu}, {Santos},
  {Ho}, {Wang}, {Kim}, \& {Hsiao}}]{2020MNRAS.497.4107H}
{Hashimoto}, T., {Goto}, T., {On}, A. Y.~L., {et~al.} 2020, \mnras, 497, 4107

\bibitem[{{Heintz} {et~al.}(2020){Heintz}, {Prochaska}, {Simha}, {Platts},
  {Fong}, {Tejos}, {Ryder}, {Aggerwal}, {Bhandari}, {Day}, {Deller},
  {Kilpatrick}, {Law}, {Macquart}, {Mannings}, {Marnoch}, {Sadler}, \&
  {Shannon}}]{2020ApJ...903..152H}
{Heintz}, K.~E., {Prochaska}, J.~X., {Simha}, S., {et~al.} 2020, \apj, 903, 152

\bibitem[{{Kirsten} {et~al.}(2022){Kirsten}, {Marcote}, {Nimmo}, {Hessels},
  {Bhardwaj}, {Tendulkar}, {Keimpema}, {Yang}, {Snelders}, {Scholz},
  {Pearlman}, {Law}, {Peters}, {Giroletti}, {Paragi}, {Bassa}, {Hewitt},
  {Bach}, {Bezrukovs}, {Burgay}, {Buttaccio}, {Conway}, {Corongiu}, {Feiler},
  {Forss{\'e}n}, {Gawro{\'n}ski}, {Karuppusamy}, {Kharinov}, {Lindqvist},
  {Maccaferri}, {Melnikov}, {Ould-Boukattine}, {Possenti}, {Surcis}, {Wang},
  {Yuan}, {Aggarwal}, {Anna-Thomas}, {Bower}, {Blaauw}, {Burke-Spolaor},
  {Cassanelli}, {Clarke}, {Fonseca}, {Gaensler}, {Gopinath}, {Kaspi}, {Kassim},
  {Lazio}, {Leung}, {Li}, {Lin}, {Masui}, {Mckinven}, {Michilli}, {Mikhailov},
  {Ng}, {Orbidans}, {Pen}, {Petroff}, {Rahman}, {Ransom}, {Shin}, {Smith},
  {Stairs}, \& {Vlemmings}}]{2022Natur.602..585K}
{Kirsten}, F., {Marcote}, B., {Nimmo}, K., {et~al.} 2022, \nat, 602, 585

\bibitem[{{Kirsten} {et~al.}(2021){Kirsten}, {Snelders}, {Jenkins}, {Nimmo},
  {van den Eijnden}, {Hessels}, {Gawro{\'n}ski}, \&
  {Yang}}]{2021NatAs...5..414K}
{Kirsten}, F., {Snelders}, M.~P., {Jenkins}, M., {et~al.} 2021, Nature
  Astronomy, 5, 414

\bibitem[{{Li} {et~al.}(2022){Li}, {Cai}, {Xiong}, {Ge}, {Liu}, {Li}, \&
  {Zhang}}]{2022ATel15698....1L}
{Li}, C.~K., {Cai}, C., {Xiong}, S.~L., {et~al.} 2022, The Astronomer's
  Telegram, 15698, 1

\bibitem[{{Li} {et~al.}(2021){Li}, {Lin}, {Xiong}, {Ge}, {Li}, {Li}, {Lu},
  {Zhang}, {Tuo}, {Nang}, {Zhang}, {Xiao}, {Chen}, {Song}, {Xu}, {Liu}, {Jia},
  {Cao}, {Qu}, {Zhang}, {Gu}, {Liao}, {Zhao}, {Tan}, {Nie}, {Zhao}, {Zheng},
  {Zheng}, {Luo}, {Cai}, {Li}, {Xue}, {Bu}, {Chang}, {Chen}, {Chen}, {Chen},
  {Chen}, {Chen}, {Cui}, {Cui}, {Deng}, {Dong}, {Du}, {Fu}, {Gao}, {Gao},
  {Gao}, {Gu}, {Guan}, {Guo}, {Han}, {Huang}, {Huo}, {Jiang}, {Jiang}, {Jin},
  {Jin}, {Kong}, {Li}, {Li}, {Li}, {Li}, {Li}, {Li}, {Li}, {Liang}, {Liu},
  {Liu}, {Liu}, {Liu}, {Liu}, {Lu}, {Lu}, {Luo}, {Ma}, {Meng}, {Ou}, {Sai},
  {Shang}, {Song}, {Sun}, {Tao}, {Wang}, {Wang}, {Wang}, {Wang}, {Wang}, {Wen},
  {Wu}, {Wu}, {Wu}, {Xiao}, {Xu}, {Yang}, {Yang}, {Yang}, {Yang}, {Yi}, {Yin},
  {You}, {Zhang}, {Zhang}, {Zhang}, {Zhang}, {Zhang}, {Zhang}, {Zhang},
  {Zhang}, {Zhang}, {Zhang}, {Zhang}, {Zhang}, {Zhang}, {Zhang}, {Zhang},
  {Zhang}, {Zhou}, {Zhou}, {Zhu}, {Zhu}, \& {Zhuang}}]{2021NatAs...5..378L}
{Li}, C.~K., {Lin}, L., {Xiong}, S.~L., {et~al.} 2021, Nature Astronomy, 5, 378

\bibitem[{{Li} \& {Ma}(1983)}]{LiMa}
{Li}, T.~P. \& {Ma}, Y.~Q. 1983, \apj, 272, 317

\bibitem[{{Lorimer} {et~al.}(2007){Lorimer}, {Bailes}, {McLaughlin},
  {Narkevic}, \& {Crawford}}]{2007Sci...318..777L}
{Lorimer}, D.~R., {Bailes}, M., {McLaughlin}, M.~A., {Narkevic}, D.~J., \&
  {Crawford}, F. 2007, Science, 318, 777

\bibitem[{{Lyubarsky}(2014)}]{2014MNRAS.442L...9L}
{Lyubarsky}, Y. 2014, \mnras, 442, L9

\bibitem[{{Lyubarsky}(2021)}]{2021Univ....7...56L}
{Lyubarsky}, Y. 2021, Universe, 7, 56

\bibitem[{{Macquart} {et~al.}(2020){Macquart}, {Prochaska}, {McQuinn},
  {Bannister}, {Bhandari}, {Day}, {Deller}, {Ekers}, {James}, {Marnoch},
  {Os{\l}owski}, {Phillips}, {Ryder}, {Scott}, {Shannon}, \&
  {Tejos}}]{2020Natur.581..391M}
{Macquart}, J.~P., {Prochaska}, J.~X., {McQuinn}, M., {et~al.} 2020, \nat, 581,
  391

\bibitem[{{Manchester} {et~al.}(2005){Manchester}, {Hobbs}, {Teoh}, \&
  {Hobbs}}]{2005AJ....129.1993M}
{Manchester}, R.~N., {Hobbs}, G.~B., {Teoh}, A., \& {Hobbs}, M. 2005, \aj, 129,
  1993

\bibitem[{{Mannings} {et~al.}(2021){Mannings}, {Fong}, {Simha}, {Prochaska},
  {Rafelski}, {Kilpatrick}, {Tejos}, {Heintz}, {Bannister}, {Bhandari}, {Day},
  {Deller}, {Ryder}, {Shannon}, \& {Tendulkar}}]{2021ApJ...917...75M}
{Mannings}, A.~G., {Fong}, W.-f., {Simha}, S., {et~al.} 2021, \apj, 917, 75

\bibitem[{{Marcote} {et~al.}(2020){Marcote}, {Nimmo}, {Hessels}, {Tendulkar},
  {Bassa}, {Paragi}, {Keimpema}, {Bhardwaj}, {Karuppusamy}, {Kaspi}, {Law},
  {Michilli}, {Aggarwal}, {Andersen}, {Archibald}, {Bandura}, {Bower}, {Boyle},
  {Brar}, {Burke-Spolaor}, {Butler}, {Cassanelli}, {Chawla}, {Demorest},
  {Dobbs}, {Fonseca}, {Giri}, {Good}, {Gourdji}, {Josephy}, {Kirichenko},
  {Kirsten}, {Landecker}, {Lang}, {Lazio}, {Li}, {Lin}, {Linford}, {Masui},
  {Mena-Parra}, {Naidu}, {Ng}, {Patel}, {Pen}, {Pleunis}, {Rafiei-Ravandi},
  {Rahman}, {Renard}, {Scholz}, {Siegel}, {Smith}, {Stairs}, {Vanderlinde}, \&
  {Zwaniga}}]{2020Natur.577..190M}
{Marcote}, B., {Nimmo}, K., {Hessels}, J.~W.~T., {et~al.} 2020, \nat, 577, 190

\bibitem[{{Marcote} {et~al.}(2017){Marcote}, {Paragi}, {Hessels}, {Keimpema},
  {van Langevelde}, {Huang}, {Bassa}, {Bogdanov}, {Bower}, {Burke-Spolaor},
  {Butler}, {Campbell}, {Chatterjee}, {Cordes}, {Demorest}, {Garrett}, {Ghosh},
  {Kaspi}, {Law}, {Lazio}, {McLaughlin}, {Ransom}, {Salter}, {Scholz},
  {Seymour}, {Siemion}, {Spitler}, {Tendulkar}, \&
  {Wharton}}]{2017ApJ...834L...8M}
{Marcote}, B., {Paragi}, Z., {Hessels}, J.~W.~T., {et~al.} 2017, \apjl, 834, L8

\bibitem[{{Margalit} {et~al.}(2020){Margalit}, {Beniamini}, {Sridhar}, \&
  {Metzger}}]{2020ApJ...899L..27M}
{Margalit}, B., {Beniamini}, P., {Sridhar}, N., \& {Metzger}, B.~D. 2020,
  \apjl, 899, L27

\bibitem[{{Mattox} {et~al.}(1996){Mattox}, {Bertsch}, {Chiang}, {Dingus},
  {Digel}, {Esposito}, {Fierro}, {Hartman}, {Hunter}, {Kanbach}, {Kniffen},
  {Lin}, {Macomb}, {Mayer-Hasselwander}, {Michelson}, {von Montigny},
  {Mukherjee}, {Nolan}, {Ramanamurthy}, {Schneid}, {Sreekumar}, {Thompson}, \&
  {Willis}}]{1996ApJ...461..396M}
{Mattox}, J.~R., {Bertsch}, D.~L., {Chiang}, J., {et~al.} 1996, \apj, 461, 396

\bibitem[{{Mereghetti} {et~al.}(2020){Mereghetti}, {Savchenko}, {Ferrigno},
  {G{\"o}tz}, {Rigoselli}, {Tiengo}, {Bazzano}, {Bozzo}, {Coleiro},
  {Courvoisier}, {Doyle}, {Goldwurm}, {Hanlon}, {Jourdain}, {von Kienlin},
  {Lutovinov}, {Martin-Carrillo}, {Molkov}, {Natalucci}, {Onori}, {Panessa},
  {Rodi}, {Rodriguez}, {S{\'a}nchez-Fern{\'a}ndez}, {Sunyaev}, \&
  {Ubertini}}]{2020ApJ...898L..29M}
{Mereghetti}, S., {Savchenko}, V., {Ferrigno}, C., {et~al.} 2020, \apjl, 898,
  L29

\bibitem[{{Metzger} {et~al.}(2020){Metzger}, {Fang}, \&
  {Margalit}}]{2020ApJ...902L..22M}
{Metzger}, B.~D., {Fang}, K., \& {Margalit}, B. 2020, \apjl, 902, L22

\bibitem[{{Metzger} {et~al.}(2019){Metzger}, {Margalit}, \&
  {Sironi}}]{2019MNRAS.485.4091M}
{Metzger}, B.~D., {Margalit}, B., \& {Sironi}, L. 2019, \mnras, 485, 4091

\bibitem[{{Michilli} {et~al.}(2022){Michilli}, {Bhardwaj}, {Brar}, {Patel},
  {Gaensler}, {Kaspi}, {Kirichenko}, {Masui}, {Sand}, {Scholz}, {Shin},
  {Stairs}, {Cassanelli}, {Cook}, {Dobbs}, {Dong}, {Fonseca}, {Ibik},
  {Kaczmarek}, {Leung}, {Pearlman}, {Petroff}, {Pleunis}, {Rafiei-Ravandi},
  {Sanghavi}, \& {Tendulkar}}]{2022arXiv221211941M}
{Michilli}, D., {Bhardwaj}, M., {Brar}, C., {et~al.} 2022, arXiv e-prints,
  arXiv:2212.11941

\bibitem[{{Murase} {et~al.}(2017){Murase}, {M{\'e}sz{\'a}ros}, \&
  {Fox}}]{2017ApJ...836L...6M}
{Murase}, K., {M{\'e}sz{\'a}ros}, P., \& {Fox}, D.~B. 2017, \apjl, 836, L6

\bibitem[{{Nicastro} {et~al.}(2021){Nicastro}, {Guidorzi}, {Palazzi},
  {Zampieri}, {Turatto}, \& {Gardini}}]{2021Univ....7...76N}
{Nicastro}, L., {Guidorzi}, C., {Palazzi}, E., {et~al.} 2021, Universe, 7, 76

\bibitem[{{Niu} {et~al.}(2022){Niu}, {Aggarwal}, {Li}, {Zhang}, {Chatterjee},
  {Tsai}, {Yu}, {Law}, {Burke-Spolaor}, {Cordes}, {Zhang}, {Ocker}, {Yao},
  {Wang}, {Feng}, {Niino}, {Bochenek}, {Cruces}, {Connor}, {Jiang}, {Dai},
  {Luo}, {Li}, {Miao}, {Niu}, {Anna-Thomas}, {Sydnor}, {Stern}, {Wang}, {Yuan},
  {Yue}, {Zhou}, {Yan}, {Zhu}, \& {Zhang}}]{2022Natur.606..873N}
{Niu}, C.~H., {Aggarwal}, K., {Li}, D., {et~al.} 2022, \nat, 606, 873

\bibitem[{{Petroff} {et~al.}(2016){Petroff}, {Barr}, {Jameson}, {Keane},
  {Bailes}, {Kramer}, {Morello}, {Tabbara}, \& {van
  Straten}}]{2016PASA...33...45P}
{Petroff}, E., {Barr}, E.~D., {Jameson}, A., {et~al.} 2016, \pasa, 33, e045

\bibitem[{{Petroff} {et~al.}(2019){Petroff}, {Hessels}, \&
  {Lorimer}}]{2019A&ARv..27....4P}
{Petroff}, E., {Hessels}, J.~W.~T., \& {Lorimer}, D.~R. 2019, \aapr, 27, 4

\bibitem[{{Planck Collaboration} {et~al.}(2020){Planck Collaboration},
  {Aghanim}, {Akrami}, {Ashdown}, {Aumont}, {Baccigalupi}, {Ballardini},
  {Banday}, {Barreiro}, {Bartolo}, {Basak}, {Battye}, {Benabed}, {Bernard},
  {Bersanelli}, {Bielewicz}, {Bock}, {Bond}, {Borrill}, {Bouchet}, {Boulanger},
  {Bucher}, {Burigana}, {Butler}, {Calabrese}, {Cardoso}, {Carron},
  {Challinor}, {Chiang}, {Chluba}, {Colombo}, {Combet}, {Contreras}, {Crill},
  {Cuttaia}, {de Bernardis}, {de Zotti}, {Delabrouille}, {Delouis}, {Di
  Valentino}, {Diego}, {Dor{\'e}}, {Douspis}, {Ducout}, {Dupac}, {Dusini},
  {Efstathiou}, {Elsner}, {En{\ss}lin}, {Eriksen}, {Fantaye}, {Farhang},
  {Fergusson}, {Fernandez-Cobos}, {Finelli}, {Forastieri}, {Frailis},
  {Fraisse}, {Franceschi}, {Frolov}, {Galeotta}, {Galli}, {Ganga},
  {G{\'e}nova-Santos}, {Gerbino}, {Ghosh}, {Gonz{\'a}lez-Nuevo}, {G{\'o}rski},
  {Gratton}, {Gruppuso}, {Gudmundsson}, {Hamann}, {Handley}, {Hansen},
  {Herranz}, {Hildebrandt}, {Hivon}, {Huang}, {Jaffe}, {Jones}, {Karakci},
  {Keih{\"a}nen}, {Keskitalo}, {Kiiveri}, {Kim}, {Kisner}, {Knox},
  {Krachmalnicoff}, {Kunz}, {Kurki-Suonio}, {Lagache}, {Lamarre}, {Lasenby},
  {Lattanzi}, {Lawrence}, {Le Jeune}, {Lemos}, {Lesgourgues}, {Levrier},
  {Lewis}, {Liguori}, {Lilje}, {Lilley}, {Lindholm}, {L{\'o}pez-Caniego},
  {Lubin}, {Ma}, {Mac{\'\i}as-P{\'e}rez}, {Maggio}, {Maino}, {Mandolesi},
  {Mangilli}, {Marcos-Caballero}, {Maris}, {Martin}, {Martinelli},
  {Mart{\'\i}nez-Gonz{\'a}lez}, {Matarrese}, {Mauri}, {McEwen}, {Meinhold},
  {Melchiorri}, {Mennella}, {Migliaccio}, {Millea}, {Mitra},
  {Miville-Desch{\^e}nes}, {Molinari}, {Montier}, {Morgante}, {Moss}, {Natoli},
  {N{\o}rgaard-Nielsen}, {Pagano}, {Paoletti}, {Partridge}, {Patanchon},
  {Peiris}, {Perrotta}, {Pettorino}, {Piacentini}, {Polastri}, {Polenta},
  {Puget}, {Rachen}, {Reinecke}, {Remazeilles}, {Renzi}, {Rocha}, {Rosset},
  {Roudier}, {Rubi{\~n}o-Mart{\'\i}n}, {Ruiz-Granados}, {Salvati}, {Sandri},
  {Savelainen}, {Scott}, {Shellard}, {Sirignano}, {Sirri}, {Spencer},
  {Sunyaev}, {Suur-Uski}, {Tauber}, {Tavagnacco}, {Tenti}, {Toffolatti},
  {Tomasi}, {Trombetti}, {Valenziano}, {Valiviita}, {Van Tent}, {Vibert},
  {Vielva}, {Villa}, {Vittorio}, {Wandelt}, {Wehus}, {White}, {White},
  {Zacchei}, \& {Zonca}}]{2020A&A...641A...6P}
{Planck Collaboration}, {Aghanim}, N., {Akrami}, Y., {et~al.} 2020, \aap, 641,
  A6

\bibitem[{{Platts} {et~al.}(2019){Platts}, {Weltman}, {Walters}, {Tendulkar},
  {Gordin}, \& {Kandhai}}]{2019PhR...821....1P}
{Platts}, E., {Weltman}, A., {Walters}, A., {et~al.} 2019, \physrep, 821, 1

\bibitem[{{Pleunis} {et~al.}(2021){Pleunis}, {Good}, {Kaspi}, {Mckinven},
  {Ransom}, {Scholz}, {Bandura}, {Bhardwaj}, {Boyle}, {Brar}, {Cassanelli},
  {Chawla}, {(Adam) Dong}, {Fonseca}, {Gaensler}, {Josephy}, {Kaczmarek},
  {Leung}, {Lin}, {Masui}, {Mena-Parra}, {Michilli}, {Ng}, {Patel},
  {Rafiei-Ravandi}, {Rahman}, {Sanghavi}, {Shin}, {Smith}, {Stairs}, \&
  {Tendulkar}}]{2021ApJ...923....1P}
{Pleunis}, Z., {Good}, D.~C., {Kaspi}, V.~M., {et~al.} 2021, \apj, 923, 1

\bibitem[{{Popov} \& {Postnov}(2013)}]{2013arXiv1307.4924P}
{Popov}, S.~B. \& {Postnov}, K.~A. 2013, arXiv e-prints, arXiv:1307.4924

\bibitem[{{Principe} {et~al.}(2021){Principe}, {Di Venere}, {Orienti},
  {Migliori}, {D'Ammando}, {Mazziotta}, \& {Giroletti}}]{2021MNRAS.507.4564P}
{Principe}, G., {Di Venere}, L., {Orienti}, M., {et~al.} 2021, \mnras, 507,
  4564

\bibitem[{{Principe} {et~al.}(2018){Principe}, {Malyshev}, {Ballet}, \&
  {Funk}}]{2018A&A...618A..22P}
{Principe}, G., {Malyshev}, D., {Ballet}, J., \& {Funk}, S. 2018, \aap, 618,
  A22

\bibitem[{{Principe} {et~al.}(2019){Principe}, {Malyshev}, {Ballet}, \&
  {Funk}}]{2019RLSFN.tmp....7P}
{Principe}, G., {Malyshev}, D., {Ballet}, J., \& {Funk}, S. 2019, Rendiconti
  Lincei. Scienze Fisiche e Naturali, 7

\bibitem[{{Principe} {et~al.}(2022){Principe}, {Omodei}, {Longo}, {Di Venere},
  {Di Lalla}, \& {Fermi-LAT Collaboration}}]{2022icrc.confE.624P}
{Principe}, G., {Omodei}, N., {Longo}, F., {et~al.} 2022, in 37th International
  Cosmic Ray Conference, 624

\bibitem[{{Prochaska} {et~al.}(2019){Prochaska}, {Macquart}, {McQuinn},
  {Simha}, {Shannon}, {Day}, {Marnoch}, {Ryder}, {Deller}, {Bannister},
  {Bhandari}, {Bordoloi}, {Bunton}, {Cho}, {Flynn}, {Mahony}, {Phillips},
  {Qiu}, \& {Tejos}}]{2019Sci...366..231P}
{Prochaska}, J.~X., {Macquart}, J.-P., {McQuinn}, M., {et~al.} 2019, Science,
  366, 231

\bibitem[{{Rajwade} {et~al.}(2020){Rajwade}, {Mickaliger}, {Stappers},
  {Morello}, {Agarwal}, {Bassa}, {Breton}, {Caleb}, {Karastergiou}, {Keane}, \&
  {Lorimer}}]{2020MNRAS.495.3551R}
{Rajwade}, K.~M., {Mickaliger}, M.~B., {Stappers}, B.~W., {et~al.} 2020,
  \mnras, 495, 3551

\bibitem[{{Ravi} {et~al.}(2019){Ravi}, {Catha}, {D'Addario}, {Djorgovski},
  {Hallinan}, {Hobbs}, {Kocz}, {Kulkarni}, {Shi}, {Vedantham}, {Weinreb}, \&
  {Woody}}]{2019Natur.572..352R}
{Ravi}, V., {Catha}, M., {D'Addario}, L., {et~al.} 2019, \nat, 572, 352

\bibitem[{{Ridnaia} {et~al.}(2021){Ridnaia}, {Svinkin}, {Frederiks}, {Bykov},
  {Popov}, {Aptekar}, {Golenetskii}, {Lysenko}, {Tsvetkova}, {Ulanov}, \&
  {Cline}}]{2021NatAs...5..372R}
{Ridnaia}, A., {Svinkin}, D., {Frederiks}, D., {et~al.} 2021, Nature Astronomy,
  5, 372

\bibitem[{{Spitler} {et~al.}(2016){Spitler}, {Scholz}, {Hessels}, {Bogdanov},
  {Brazier}, {Camilo}, {Chatterjee}, {Cordes}, {Crawford}, {Deneva}, {Ferdman},
  {Freire}, {Kaspi}, {Lazarus}, {Lynch}, {Madsen}, {McLaughlin}, {Patel},
  {Ransom}, {Seymour}, {Stairs}, {Stappers}, {van Leeuwen}, \&
  {Zhu}}]{2016Natur.531..202S}
{Spitler}, L.~G., {Scholz}, P., {Hessels}, J.~W.~T., {et~al.} 2016, \nat, 531,
  202

\bibitem[{{Sridhar} {et~al.}(2022){Sridhar}, {Metzger}, \&
  {Fang}}]{2022arXiv221211236S}
{Sridhar}, N., {Metzger}, B.~D., \& {Fang}, K. 2022, arXiv e-prints,
  arXiv:2212.11236

\bibitem[{{Sridhar} {et~al.}(2021){Sridhar}, {Zrake}, {Metzger}, {Sironi}, \&
  {Giannios}}]{2021MNRAS.501.3184S}
{Sridhar}, N., {Zrake}, J., {Metzger}, B.~D., {Sironi}, L., \& {Giannios}, D.
  2021, \mnras, 501, 3184

\bibitem[{{Tavani} {et~al.}(2021){Tavani}, {Casentini}, {Ursi}, {Verrecchia},
  {Addis}, {Antonelli}, {Argan}, {Barbiellini}, {Baroncelli}, {Bernardi},
  {Bianchi}, {Bulgarelli}, {Caraveo}, {Cardillo}, {Cattaneo}, {Chen}, {Costa},
  {Del Monte}, {Di Cocco}, {Di Persio}, {Donnarumma}, {Evangelista}, {Feroci},
  {Ferrari}, {Fioretti}, {Fuschino}, {Galli}, {Gianotti}, {Giuliani},
  {Labanti}, {Lazzarotto}, {Lipari}, {Longo}, {Lucarelli}, {Magro},
  {Marisaldi}, {Mereghetti}, {Morelli}, {Morselli}, {Naldi}, {Pacciani},
  {Parmiggiani}, {Paoletti}, {Pellizzoni}, {Perri}, {Perotti}, {Piano},
  {Picozza}, {Pilia}, {Pittori}, {Puccetti}, {Pupillo}, {Rapisarda},
  {Rappoldi}, {Rubini}, {Setti}, {Soffitta}, {Trifoglio}, {Trois},
  {Vercellone}, {Vittorini}, {Giommi}, \& {D'Amico}}]{2021NatAs...5..401T}
{Tavani}, M., {Casentini}, C., {Ursi}, A., {et~al.} 2021, Nature Astronomy, 5,
  401

\bibitem[{{Tavani} {et~al.}(2020){Tavani}, {Verrecchia}, {Casentini}, {Perri},
  {Ursi}, {Pacciani}, {Pittori}, {Bulgarelli}, {Piano}, {Pilia}, {Bernardi},
  {Addis}, {Antonelli}, {Argan}, {Baroncelli}, {Caraveo}, {Cattaneo}, {Chen},
  {Costa}, {Di Persio}, {Donnarumma}, {Evangelista}, {Feroci}, {Ferrari},
  {Fioretti}, {Lazzarotto}, {Longo}, {Morselli}, {Paoletti}, {Parmiggiani},
  {Trois}, {Vercellone}, {Naldi}, {Pupillo}, {Bianchi}, \&
  {Puccetti}}]{2020ApJ...893L..42T}
{Tavani}, M., {Verrecchia}, F., {Casentini}, C., {et~al.} 2020, \apjl, 893, L42

\bibitem[{{Tendulkar} {et~al.}(2017){Tendulkar}, {Bassa}, {Cordes}, {Bower},
  {Law}, {Chatterjee}, {Adams}, {Bogdanov}, {Burke-Spolaor}, {Butler},
  {Demorest}, {Hessels}, {Kaspi}, {Lazio}, {Maddox}, {Marcote}, {McLaughlin},
  {Paragi}, {Ransom}, {Scholz}, {Seymour}, {Spitler}, {van Langevelde}, \&
  {Wharton}}]{2017ApJ...834L...7T}
{Tendulkar}, S.~P., {Bassa}, C.~G., {Cordes}, J.~M., {et~al.} 2017, \apjl, 834,
  L7

\bibitem[{{The LIGO Scientific Collaboration} {et~al.}(2022){The LIGO
  Scientific Collaboration}, {the Virgo Collaboration}, {the KAGRA
  Collaboration}, {the CHIME/FRB Collaboration}, {:}, \&
  {Abbott}.}]{2022arXiv220312038T}
{The LIGO Scientific Collaboration}, {the Virgo Collaboration}, {the KAGRA
  Collaboration}, {et~al.} 2022, arXiv e-prints, arXiv:2203.12038

\bibitem[{{Thornton} {et~al.}(2013){Thornton}, {Stappers}, {Bailes},
  {Barsdell}, {Bates}, {Bhat}, {Burgay}, {Burke-Spolaor}, {Champion}, {Coster},
  {D'Amico}, {Jameson}, {Johnston}, {Keith}, {Kramer}, {Levin}, {Milia}, {Ng},
  {Possenti}, \& {van Straten}}]{2013Sci...341...53T}
{Thornton}, D., {Stappers}, B., {Bailes}, M., {et~al.} 2013, Science, 341, 53

\bibitem[{{Vanderlinde} {et~al.}(2019){Vanderlinde}, {Liu}, {Gaensler}, {Bond},
  {Hinshaw}, {Ng}, {Chiang}, {Stairs}, {Brown}, {Sievers}, {Mena}, {Smith},
  {Bandura}, {Masui}, {Spekkens}, {Belostotski}, {Dobbs}, {Turok}, {Boyle},
  {Rupen}, {Landecker}, {Pen}, \& {Kaspi}}]{2019clrp.2020...28V}
{Vanderlinde}, K., {Liu}, A., {Gaensler}, B., {et~al.} 2019, in Canadian Long
  Range Plan for Astronomy and Astrophysics White Papers, Vol. 2020, 28

\bibitem[{{Verrecchia} {et~al.}(2021){Verrecchia}, {Casentini}, {Tavani},
  {Ursi}, {Mereghetti}, {Pilia}, {Cardillo}, {Addis}, {Barbiellini},
  {Baroncelli}, {Bulgarelli}, {Cattaneo}, {Chen}, {Costa}, {Del Monte}, {Di
  Piano}, {Ferrari}, {Fioretti}, {Longo}, {Lucarelli}, {Parmiggiani}, {Piano},
  {Pittori}, {Rappoldi}, \& {Vercellone}}]{2021ApJ...915..102V}
{Verrecchia}, F., {Casentini}, C., {Tavani}, M., {et~al.} 2021, \apj, 915, 102

\bibitem[{{Wang} {et~al.}(2022){Wang}, {Xiong}, {Zhang}, {Liu}, {Zheng}, {Xue},
  {Tan}, {Xie}, {Yi}, {Zhao}, {Wang}, {Cai}, {Xiao}, {Huang}, {Ma}, {Qiao},
  {Wang}, {Zhao}, {Zhang}, {Li}, {Wen}, {Peng}, {Song}, {Zheng}, {Du}, {Guo},
  {Li}, {Li}, {Liang}, {Lu}, {Wang}, {Wu}, {Song}, {Yu}, {Zhang}, {An}, {Feng},
  {Gao}, {Gong}, {Liu}, {Liu}, {Sun}, {Wang}, {Xu}, {Yang}, {Zhang}, {Zhang},
  {Li}, {Li}, {Liao}, {Chen}, {Lu}, {Zhang}, {Gecam Team}, \& {Hebs
  Team}}]{2022ATel15682....1W}
{Wang}, C.~W., {Xiong}, S.~L., {Zhang}, Y.~Q., {et~al.} 2022, The Astronomer's
  Telegram, 15682, 1

\bibitem[{{Wei} {et~al.}(2023){Wei}, {Zhang}, \&
  {Murase}}]{2023arXiv230110184W}
{Wei}, Y., {Zhang}, B.~T., \& {Murase}, K. 2023, arXiv e-prints,
  arXiv:2301.10184

\bibitem[{{Wood} {et~al.}(2017){Wood}, {Caputo}, {Charles}, {Di Mauro},
  {Magill}, \& {Jeremy Perkins for the Fermi-LAT
  Collaboration}}]{2017arXiv170709551W}
{Wood}, M., {Caputo}, R., {Charles}, E., {et~al.} 2017, ArXiv e-prints
  [\eprint[arXiv]{1707.09551}]

\bibitem[{{Xiao} {et~al.}(2021){Xiao}, {Wang}, \& {Dai}}]{2021SCPMA..6449501X}
{Xiao}, D., {Wang}, F., \& {Dai}, Z. 2021, Science China Physics, Mechanics,
  and Astronomy, 64, 249501

\bibitem[{{Xu} {et~al.}(2022){Xu}, {Niu}, {Chen}, {Lee}, {Zhu}, {Dong},
  {Zhang}, {Jiang}, {Wang}, {Xu}, {Zhang}, {Fu}, {Filippenko}, {Peng}, {Zhou},
  {Zhang}, {Wang}, {Feng}, {Li}, {Brink}, {Li}, {Lu}, {Yang}, {Caballero},
  {Cai}, {Chen}, {Dai}, {Djorgovski}, {Esamdin}, {Gan}, {Guhathakurta}, {Han},
  {Hao}, {Huang}, {Jiang}, {Li}, {Li}, {Li}, {Li}, {Li}, {Liu}, {Luo}, {Men},
  {Niu}, {Peng}, {Qian}, {Song}, {Stern}, {Stockton}, {Sun}, {Wang}, {Wang},
  {Wang}, {Wang}, {Wu}, {Xiao}, {Xiong}, {Xu}, {Xu}, {Yang}, {Yang}, {Yao},
  {Yi}, {Yue}, {Yu}, {Yu}, {Yuan}, {Zhang}, {Zhang}, {Zhang}, {Zhao}, {Zheng},
  {Zhu}, \& {Zou}}]{2022Natur.609..685X}
{Xu}, H., {Niu}, J.~R., {Chen}, P., {et~al.} 2022, \nat, 609, 685

\bibitem[{{Yang} \& {Zhang}(2021)}]{2021ApJ...919...89Y}
{Yang}, Y.-P. \& {Zhang}, B. 2021, \apj, 919, 89

\bibitem[{{Younes} {et~al.}(2022){Younes}, {Burns}, {Roberts}, {Wood}, {Veres},
  \& {Kouveliotou}}]{2022ATel15794....1Y}
{Younes}, G., {Burns}, E., {Roberts}, O.~J., {et~al.} 2022, The Astronomer's
  Telegram, 15794, 1

\bibitem[{{Zanin} {et~al.}(2022){Zanin}, {Abdalla}, {Abe}, {Abe}, {Abusleme},
  {Acero}, {Acharyya}, {Acin Portella}, {Ackley}, {Adam}, {Adams}, {Adhikari},
  {Aguado Ruesga}, {Agudo}, {Aguilera}, {Aguirre Santaella}, {Aharonian},
  {Alberdi}, {Alfaro}, {Alfaro}, {Alispach}, {Aloisio}, {Alves Batista},
  {Amans}, {Amati}, {Amato}, {Ambrogi}, {Ambrosi}, {Ambrosio}, {Ammendola},
  {Anderson}, {Anduze}, {Anguner}, {Antonelli}, {Antonuccio}, {Antoranz},
  {Anutarawiramkul}, {Aragunde Gutierrez}, {Aramo}, {Araudo}, {Araya}, {Arbet
  Engels}, {Arcaro}, {Arendt}, {Armand}, {Armstrong}, {Arqueros}, {Arrabito},
  {Arsioli}, {Artero}, {Asano}, {Ascasibar}, {Aschersleben}, {Ashley},
  {Attina}, {Aubert}, {Singh}, {Baack}, {Babic}, {Backes}, {Baena}, {Bajtlik},
  {Baktash}, {Balazs}, {Balbo}, {Ballester}, {Ballet}, {Balmaverde}, {Bamba},
  {Bandiera}, {Baquero Larriva}, {Barai}, {Barbier}, {Barbosa Martins},
  {Barcelo}, {Barkov}, {Barnard}, {Baroncelli}, {Barres de Almeida}, {Barrio},
  {Bastieri}, {Batista}, {Batkovic}, {Bauer}, {Bautista Gonz{\'a}lez},
  {Baxter}, {Becciani}, {Becerra Gonz{\'a}lez}, {Becherini}, {Beck}, {Becker
  Tjus}, {Bednarek}, {Belfiore}, {Bellizzi}, {Belmont}, {Benbow}, {Berge},
  {Bernardini}, {Bernardos}, {Bernl{\"o}hr}, {Berti}, {Berton}, {Bertucci},
  {Beshley}, {Bhatt}, {Bhattacharyya}, {Bhattacharyya}, {Bhattacharyya}, {Bi},
  {Bicknell}, {Biederbeck}, {Bigongiari}, {Biland}, {Bird}, {Bissaldi},
  {Biteau}, {Bitossi}, {Blanch}, {Blank}, {Blazek}, {Bobin}, {Boccato},
  {Bocchino}, {Boehm}, {Bohacova}, {Boisson}, {Boix}, {Bolle}, {Bolmont},
  {Bonanno}, {Bonavolont{\`a}}, {Bonneau Arbeletche}, {Bonnoli}, {Bordas},
  {Borkowski}, {Bose}, {Bose}, {Bosnjak}, {Bottacini}, {B{\"o}ttcher},
  {Botticella}, {Boutonnet}, {Bouyjou}, {Bozhilov}, {Bozzo}, {Brahimi},
  {Braiding}, {Brau Nogue}, {Breen}, {Bregeon}, {Breuhaus}, {Brill}, {Brisken},
  {Brocato}, {Brown}, {Br{\"u}gge}, {Brun}, {Brun}, {Brunetti}, {Brunetti},
  {Bruno}, {Bruno}, {Bruzzese}, {Bucciantini}, {Buckley}, {B{\"u}hler},
  {Bulgarelli}, {Bulik}, {B{\"u}nning}, {Bunse}, {Burton}, {Burtovoi},
  {Buscemi}, {Buschjager}, {Busetto}, {Buss}, {Byrum}, {Caccianiga}, {Cadoux},
  {Calanducci}, {Calderon}, {Calvo Tovar}, {Cameron}, {Campana}, {Canestrari},
  {Cangemi}, {Cantlay}, {Capalbi}, {Capasso}, {Cappi}, {Caproni}, {Capuzzo
  Dolcetta}, {Caraveo}, {C{\'a}rdenas}, {Cardiel}, {Cardillo}, {Carlile},
  {Caroff}, {Carosi}, {Carosi}, {Carquin}, {Carrere}, {Casandjian}, {Casanova},
  {Cassol}, {Catalani}, {Catalano}, {Cauz}, {Ceccanti}, {Celestino Silva},
  {Cerny}, {Cerruti}, {Chabanne}, {Chadwick}, {Chai}, {Chambery}, {Champion},
  {Chaty}, {Chen}, {Cheng}, {Chernyakova}, {Chiaro}, {Chiavassa}, {Chikawa},
  {Chitnis}, {Chudoba}, {Chytka}, {Cikota}, {Circiello}, {Clark}, {Colak},
  {Colombo}, {Colonges}, {Comastri}, {Compagnino}, {Conforti}, {Congiu},
  {Coniglione}, {Conrad}, {Conte}, {Contreras}, {Coppi}, {Cornat}, {Coronado
  Blazquez}, {Cortina}, {Costa}, {Costantini}, {Cotter}, {Courty}, {Covino},
  {Crestan}, {Cristofari}, {Crocker}, {Croston}, {Cubuk}, {Cuevas}, {Cui},
  {Cusumano}, {Cutini}, {D'Amico}, {D'Ammando}, {D'Avanzo}, {Da Vela},
  {Dadina}, {Dai}, {Dalchenko}, {Dall'Ora}, {Daniel}, {Dauguet}, {Davids},
  {Davies}, {Dawson}, {De Angelis}, {de Araujo Carvalho}, {de Bony de
  Lavergne}, {De Cesare}, {de Frondat}, {de la Calle}, {de Gouveia Dal Pino},
  {De Lotto}, {De Luca}, {De Martino}, {de Naurois}, {de Ona Wilhelmi}, {De
  Palma Persio}, {De Simone}, {de Souza Valle}, {Delagnes}, {Deleglise
  Reznicek}, {Delgado}, {Delgado Giler}, {Delgado Mengual Valle}, {della
  Volpe}, {Depaoli}, {Devin}, {Di Girolamo}, {Di Giulio Pierro}, {Di Venere},
  {D{\'\i}az}, {Dib}, {Diebold}, {Digel}, {Djannati Atai}, {Djuvsland},
  {Dmytriiev}, {Docher}, {Dom{\'\i}nguez}, {Dominis Prester}, {Donini},
  {Dorner}, {Doro}, {dos Anjos}, {Dournaux}, {Downes}, {Drake}, {Drass},
  {Dravins}, {Duangchan}, {Duara}, {Dubus}, {Ducci}, {Duffy}, {Dumora}, {Dundas
  Mora}, {Durkalec}, {Dwarkadas}, {Ebr}, {Eckner}, {Eder}, {Edy}, {Egberts},
  {Einecke}, {Eleftheriadis}, {Els{\"a}sser}, {Emery}, {Emmanoulopoulos},
  {Ernenwein}, {Errando}, {Escarate}, {Escudero}, {Espinoza}, {Ettori},
  {Eungwanichayapant}, {Evans}, {Evoli}, {Fairbairn}, {Falceta Goncalves},
  {Falcone}, {Fallah Ramazan{\i}}, {Falomo}, {Farakos}, {Fasola}, {Fattorini},
  {Favre}, {Fedora}, {Fedorova}, {Feijen}, {Feng}, {Ferrand}, {Ferrara},
  {Ferreira}, {Fesquet}, {Fiandrini}, {Fiasson}, {Filipovic}, {Fink}, {Finley},
  {Fioretti}, {Fiorillo}, {Fiorini}, {Flis}, {Flores}, {Foffano}, {Fohr},
  {Fonseca}, {Font}, {Fontaine}, {Fornieri}, {Fortin}, {Fortson}, {Fouque},
  {Fraga}, {Franceschini}, {Franco}, {Freixas Coromina}, {Fresnillo},
  {Fugazza}, {Fujita}, {Fukami}, {Fukazawa}, {Fulla}, {Funk}, {Furniss},
  {Gabici}, {Gaggero}, {Galanti}, {Galdemard}, {Gallant}, {Galloway},
  {Gallozzi}, {Gammaldi}, {Garcia}, {Garcia}, {Garcia Lopez}, {Gargano},
  {Gargano}, {Garozzo}, {Gascon}, {Gasparetto}, {Gasparrini}, {Gasparyan},
  {Gaug}, {Geffroy}, {Gent}, {Germani}, {Ghalumyan}, {Ghedina}, {Ghirlanda},
  {Gianotti}, {Giarrusso}, {Giarrusso}, {Giavitto}, {Giebels}, {Giglietto},
  {Gika}, {Gillardo}, {Gimenes}, {Giordano}, {Giro}, {Giroletti}, {Giuliani},
  {Gjaja}, {Glicenstein}, {Gliwny}, {Goksu}, {Goldoni}, {Gomez}, {Gonzalez},
  {Gonzalez}, {Gothe}, {Gotz Coelho}, {Grabarczyk}, {Graciani}, {Grandi},
  {Grasseau}, {Grasso}, {Green}, {Green}, {Greenshaw}, {Grespan}, {Grillo},
  {Grondin}, {Grube}, {Guarino}, {Guest}, {Gueta}, {G{\"u}nduz}, {Gunji},
  {Gyuk}, {Hackfeld}, {Hadasch}, {Hagge}, {Hahn}, {Hajlaoui}, {Halim}, {Hamal},
  {Hanlon}, {Harada}, {Hardcastle}, {Collado}, {Haubold}, {Haupt}, {Havelka},
  {Hayashi}, {Hayashi}, {Hayashida}, {He}, {Heckmann}, {Heller}, {Henault},
  {Henri}, {Hermann}, {Hern{\'a}ndez Cadena}, {Herrera Llorente}, {Hervet},
  {Hinton}, {Hiramatsu}, {Hirotani}, {Hnatyk}, {Hnatyk}, {Hoang}, {Hoffmann},
  {Hoischen}, {Holder}, {Holler}, {Hona}, {Horan}, {Horns}, {Horvath},
  {Houles}, {Hrabovsky}, {Hrupec}, {Huang}, {Huet}, {Hughes}, {Hull},
  {Humensky}, {H{\"u}tten}, {Iarlori}, {Illa}, {Imazawa}, {Inada}, {Incardona},
  {Ingallinera}, {Inoue}, {Inoue}, {Inoue}, {Iocco}, {Ioka}, {Ionica},
  {Iovenitti}, {Iriarte}, {Ishio}, {Ishizaki}, {Iwamura}, {Jacquemier},
  {Jacquemont}, {Jamrozy}, {Janecek}, {Jankowsky}, {JardinBlicq}, {Jarnot},
  {Mart{\'\i}nez}, {Jocou}, {Jordana}, {Josselin}, {JungRichardt}, {Junqueira},
  {Juramy Gilles}, {Kaaret}, {Kadowaki}, {Kagaya}, {Kankanyan}, {Kantzas},
  {Karas}, {Karastergiou}, {Karkar}, {Kasperek}, {Katagiri}, {Kataoka},
  {Katarzynski}, {Katsuda}, {Kawanaka}, {Kazanas}, {Kerszberg}, {Kh{\'e}lifi},
  {Kherlakian}, {Kian}, {Kieda}, {Kihm}, {Kim}, {Kisaka}, {Kissmann},
  {Kleijwegt}, {Kluge}, {Klu{\'z}niak}, {Knapp}, {Kobakhidze}, {Kobayashi},
  {Koch}, {Kocot}, {Kohri}, {Komin}, {Kong}, {Kosack}, {Krack}, {Krause},
  {Krennrich}, {Kubo}, {Kudryavtsev}, {Kunwar}, {Kushida}, {Kushwaha},
  {Parola}, {La Rosa}, {Lahmann}, {Lamastra}, {Landoni}, {Landriu}, {Lang},
  {Lapington}, {Laporte}, {Lason}, {Lasuik}, {Lazendic Galloway}, {Le Flour},
  {Le Sidaner}, {Leach}, {Lee}, {Lee}, {Oliveira}, {Lemiere}, {Lemoine
  Goumard}, {Lenain}, {Leone}, {Leray}, {Leto}, {Leuschner}, {Lindemann},
  {Lindfors}, {Linhoff}, {Liodakis}, {Lipniacka}, {Lobo}, {Lohse}, {Lombardi},
  {Lopez}, {Lopez}, {Lopez Coto}, {Louis}, {Louys}, {Lucarelli}, {Boudi},
  {Luque Escamilla}, {Maccarone}, {Mach}, {Maciejewski}, {Mackey}, {Maeght},
  {Maggio}, {Maier}, {Majumdar}, {Makariev}, {Mallamaci}, {Malta Nunes de
  Almeida}, {Malyshev}, {Malyshev}, {Mandat}, {Maneva}, {Manganaro}, {Manigot},
  {Mannheim}, {Maragos}, {Marano}, {Marconi}, {Marcowith}, {Marculewicz},
  {Marcun}, {Marin}, {Marinello}, {Marinos}, {Markoff}, {Marquez}, {Marsella},
  {Martin}, {Martin}, {Martinez}, {Martinez}, {Martinez}, {Martinez Huerta},
  {Marty}, {Marx}, {Masetti}, {Massimino}, {Matsumoto}, {Matthews}, {Maurin},
  {Moerbeck}, {Maxted}, {Mazziotta}, {Mazzola}, {Mbarubucyeye}, {Mc Comb},
  {McHardy}, {McKeague}, {McMuldroch}, {Medina}, {Medina Miranda}, {Melandri},
  {Melioli}, {Melkumyan}, {Menchiari}, {Mereghetti}, {Merino Arevalo},
  {Mestre}, {Meunier}, {Meures}, {Micanovic}, {Miceli}, {Michailidis},
  {Michalowski}, {Miener}, {Mievre}, {Miller}, {Mineo}, {Minev}, {Miranda},
  {Mitchell}, {Mizuno}, {Mode}, {Moderski}, {Mohrmann}, {Molinari},
  {Montaruli}, {Monteiro}, {Moore}, {Moralejo}, {Morcuende Parrilla},
  {Moretti}, {Mori}, {Moriarty}, {Morik}, {Morris}, {Morselli}, {Mosshammer},
  {Mukherjee}, {Muller}, {Mundell}, {Mundet}, {Murach}, {Muraczewski},
  {Muraishi}, {Musella}, {Musumarra}, {Nagai}, {Nagataki}, {Naito}, {Nakamori},
  {Nakashima}, {Nakayama}, {Nakhjiri}, {Naletto}, {Naumann}, {Nava}, {Nawaz},
  {Ndiyavala}, {Neise}, {Nellen}, {Nemmen}, {Neyroud}, {Ngernphat}, {Nguyen
  Trung}, {Nicastro}, {Nickel}, {Niemiec}, {Nieto}, {Nigro}, {Niko{\l}ajuk},
  {Ninci}, {Noda}, {Nogami}, {Nolan}, {Norris}, {Nosek}, {N{\"o}the},
  {Novotny}, {Nozaki}, {Nunio}, {O'Brien}, {Obara}, {Ohira}, {Ohishi}, {Ohm},
  {Oka}, {Okazaki}, {Okumura}, {Oliver}, {Olivera}, {Olmi}, {Orienti}, {Orito},
  {Orlandini}, {Orlando}, {Osborne}, {Ostrowski}, {Otte}, {Ovcharov}, {Owen},
  {Oya}, {Ozieblo}, {Padovani}, {Pagliaro}, {Paizis}, {Palatiello}, {Palatka},
  {Palazzi}, {Panazol}, {Paneque}, {Panny}, {Pantaleo}, {Panter}, {Paolillo},
  {Papitto}, {Paravac}, {Paredes}, {Pareschi}, {Parmiggiani}, {Parsons},
  {Pa{\'s}ko}, {Patel}, {Patricelli}, {Pavletic}, {Pavy}, {Peer}, {Pecimotika},
  {Pellegriti}, {Pe{\~n}il Del Campo}, {Pepato}, {Perard}, {Perennes},
  {Peresano}, {Perez Aguilera}, {Perez Romero}, {Perez Torres}, {Persic},
  {Petrucci}, {Petruk}, {Peyaud}, {Pfrang}, {Pian}, {Piatteli}, {Pietropaolo},
  {Pillera}, {Pimentel}, {Pintore}, {Garcia}, {Pirola}, {Piron}, {Pita},
  {Pohl}, {Poireau}, {Pollo}, {Polo}, {Pongkitivanichkul}, {Porthault},
  {Powell}, {Pozo}, {Prado}, {Prandini}, {Prast}, {Pressard}, {Principe},
  {Produit}, {Prokhorov}, {Prokoph}, {Przybilski}, {Pueschel}, {P{\"u}hlhofer},
  {Puljak}, {Pumo}, {Punch}, {Queiroz}, {Quinn}, {Quirrenbach}, {Rajda},
  {Rando}, {Razzaque}, {Recchia}, {Reichherzer}, {Reimer}, {Reisenegger},
  {Remy}, {Renaud}, {Reposeur}, {Reville}, {Reymond}, {Reynolds}, {Ribeiro},
  {Ribo}, {Richards}, {Rico}, {Rieger}, {Riitano}, {Riquelme}, {Riquelme},
  {Rivoire}, {Rizi}, {Roache}, {Roche}, {Rodriguez}, {Rodriguez Fernandez},
  {Rodriguez Ramirez}, {Rodriguez Vazquez}, {Rojas}, {Romano}, {Romeo Lobato},
  {Romoli}, {Roncadelli}, {Rosado}, {Rosales de Leon}, {Rowell}, {Rugliancich},
  {Ruiz del Mazo}, {Rulten}, {Russell}, {Russo Hatlen}, {Safi Harb}, {Saha},
  {Sahakian}, {Sailer}, {Saito}, {Sakaki}, {Sakurai}, {Salina}, {Salzmann},
  {Sanchez}, {Sandaker}, {Sandoval}, {Sangiorgi}, {Sanguillon}, {Sano},
  {Santander}, {Santangelo}, {Santos Lima}, {Sanuy}, {Sapozhnikov}, {Saric},
  {Sarkar}, {Sasaki}, {Sasaki}, {Sato}, {Saturni}, {Sawada}, {Schaefer},
  {Scherer}, {Scherpenberg}, {Schipani}, {Schleicher}, {Schmoll}, {Schneider},
  {Schoorlemmer}, {Schovanek}, {Schussler}, {Schwab}, {Schwanke}, {Schwarz},
  {Sciacca}, {Scuderi}, {Seglar Arroyo}, {Seitenzahl}, {Semikoz}, {Sergijenko},
  {Serna Franco}, {Seweryn}, {Sguera}, {Shalchi}, {Shang}, {Sharma}, {Sidoli},
  {Sieiro}, {Siejkowski}, {Sillanpaa}, {Singh}, {Singh}, {Sinha}, {Siqueira},
  {Sitarek}, {Sizun}, {Sliusar}, {Sobczynska}, {Sobrinho}, {Sol}, {Sottile},
  {Spackman}, {Spencer}, {Spengler}, {Spiga}, {Springer}, {Stamerra}, {Stanic},
  {Starling}, {Stawarz}, {Stefanik}, {Stegmann}, {Steiner}, {Steinmassl},
  {Stella}, {Sternberger}, {Sterzel}, {Stevens}, {Stevenson}, {Stolarczyk},
  {Stratta}, {Straumann}, {Striskovic}, {Strzys}, {Stuik}, {Suchenek},
  {Sunada}, {Suomijarvi}, {Suric}, {Suzuki}, {Swierk}, {Szepieniec},
  {Tachihara}, {Tagliaferri}, {Tajima}, {Tajima}, {Tak}, {Takahashi},
  {Takahashi}, {Takata}, {Takeishi}, {Tam}, {Tanaka}, {Tanaka}, {Tanaka},
  {Tavani}, {Tavecchio}, {Tavernier}, {Taylor}, {Tejedor}, {Temnikov},
  {Terauchi}, {Terrazas}, {Terrier}, {Terzic}, {Teshima}, {Thibaut},
  {Thocquenne}, {Tian}, {Tibaldo}, {Tiengo}, {Tluczykont}, {Todero Peixoto},
  {Toma}, {Tomankova}, {Tomastik}, {Tornikoski}, {Torres}, {Torresi}, {Tosti},
  {Tosti}, {Tothill}, {Toussenel}, {Tovmassian}, {Trichard}, {Trifoglio},
  {Trois}, {Truzzi}, {Tsiahina}, {Turk}, {Tutone}, {Uchiyama}, {Utayarat},
  {Vaclavek}, {Vacula}, {Vagelli}, {Vagnetti}, {Valdivia}, {Valentino},
  {Valio}, {Vallage}, {Vallania Quispe}, {van den Berg}, {van Driel}, {van
  Eldik}, {van Rensburg}, {van Soelen}, {Vandenbroucke}, {Vasileiadis},
  {Vassiliev}, {Vazquez Acosta}, {Vecchi}, {Vega}, {Veh}, {Veitch}, {Venter},
  {Ventura}, {Vercellone}, {Verguilov}, {Verna}, {Vernetto}, {Verzi},
  {Vettolani}, {Veyssiere}, {Viale}, {Viana}, {Viaux}, {Vignatti}, {Vigorito},
  {Villanueva}, {Vitale}, {Vittorini}, {Vodeb}, {Vogel}, {Voisin}, {Vorobiov},
  {Vrastil}, {Vuillaume}, {Wagner}, {Wagner}, {Wakazono}, {Wakely}, {Ward},
  {Warren}, {Watson}, {Wechakama}, {Wegner}, {Weinstein}, {Weniger}, {Werner},
  {Wetteskind}, {White}, {Wierzcholska}, {Wiesand}, {Wijers}, {Wilkinson},
  {Will}, {Williams}, {Williamson}, {Wolter}, {Wong}, {Wood}, {Yamamoto},
  {Yamamoto}, {Yamane}, {Yamazaki}, {Yanagita}, {Yang}, {Yoo}, {Yoshida},
  {Yoshikoshi}, {Yu}, {Yusafzai}, {Zacharias}, {Zaldivar}, {Zampieri}, {Zanin},
  {Zanmar Sanchez}, {Zaric}, {Zavrtanik}, {Zavrtanik}, {Zdziarski}, {Zech},
  {Zechlin}, {Zenin}, {Zerwekh}, {Zi{\k{e}}tara}, {Zink}, {Ziolkowski},
  {Zivec}, \& {Zmija}}]{2022icrc.confE...5Z}
{Zanin}, R., {Abdalla}, H., {Abe}, H., {et~al.} 2022, in 37th International
  Cosmic Ray Conference, 5

\bibitem[{{Zhang}(2014)}]{2014ApJ...780L..21Z}
{Zhang}, B. 2014, \apjl, 780, L21

\bibitem[{{Zhang}(2018)}]{2018ApJ...867L..21Z}
{Zhang}, B. 2018, \apjl, 867, L21

\bibitem[{{Zhang}(2020)}]{2020Natur.587...45Z}
{Zhang}, B. 2020, \nat, 587, 45

\bibitem[{{Zhang}(2022)}]{2022arXiv221203972Z}
{Zhang}, B. 2022, arXiv e-prints, arXiv:2212.03972

\end{thebibliography}

\begin{appendix}
\appendix
\section{The search for steady emission from FRB\,20190117}
\label{appendix:190117_analysis}
FRB\,20190117 (RA$= 331.82\pm0.13 ^{\circ}$ dec$=17.37 \pm 0.25 ^{\circ}$, DM=393.6) lies in the vicinity, separation $<0.85^{\circ}$, of a bright and strongly variable 4FGL source \citep[4FGL\,J2203.4+1725, sign.$>60 \sigma$, $\Gamma_{var}=940$,][]{2020ApJS..247...33A}. Our sample contains six events from this FRB starting from January 17, 2019. This nearby bright 4FGL source underwent many bright flaring episodes in the first eight years of observation, while only steady faint emission has been observed from this source since 2016. We therefore restricted our search for steady emission from the repeating FRB\,20190117 to times after January 2016, in order to avoid any contamination from the nearby bright flaring source.

\section{Sensitivity Study}
\label{appendix:sensi_study}
\begin{figure}[hb!]
    \centering
    \includegraphics[width=0.45\textwidth]{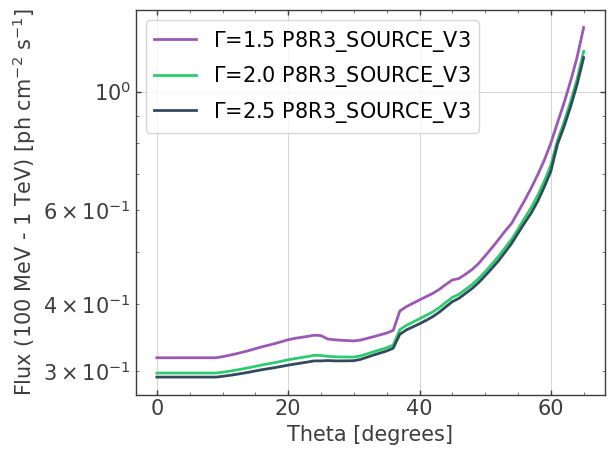}
    \includegraphics[width=0.45\textwidth]{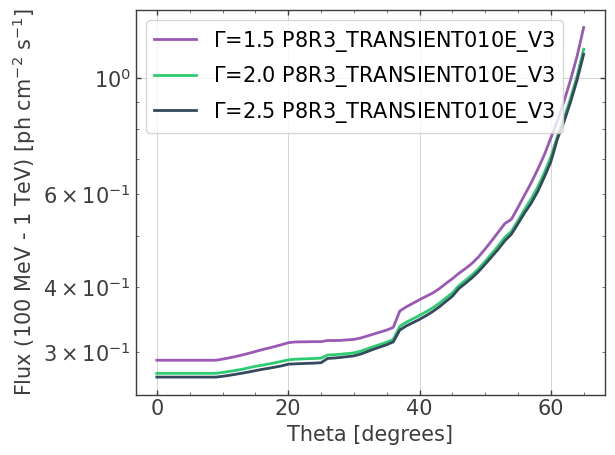}
    \includegraphics[width=0.45\textwidth]{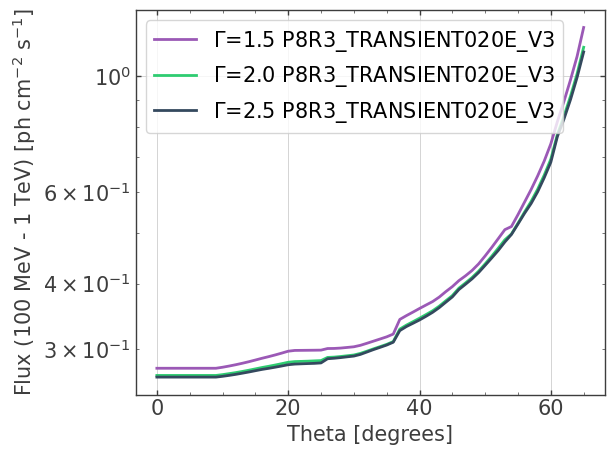}\\
    \caption{Values of the single photon 95\% flux upper limits for three different spectral indexes for \texttt{P8R3\_SOURCE\_V3} (left), \texttt{P8R3\_TRANSIENT010E\_V3} (center) and \texttt{P8R3\_TRANSIENT020E\_V3} (right).}
    \label{fig:sensi}
\end{figure}

The sensitivity of the \Fermi-\LAT can be parameterized through its effective area, which strongly depends on the incident angle $\theta$ with respect to the zenith axis of the instrument. In Sect.~\ref{sec:results_triplets} we describe the methodology to derive single-photon ULs and present some of the results in Tab.~\ref{tab:sensi} for different IRFs, spectral indexes and incident angles. The three panels of Fig.~\ref{fig:sensi} complement this study by showing, for three different IRFs, the UL dependence on the off-axis angle for three different photon indexes. The values are calculated for an exposure of 1~ms and they can be re-scaled by (1\,ms/$\Delta$\,T) to estimate the UL for different exposures (in the limit of negligible background, $\delta_T\leq$1s).


\section{Results format}
The format of the FRB-event sample results file is described in Table \ref{table:format_results}. All the parameter values are in \textsc{float} format a part the Name and Time parameters which are in \textsc{string} format. The values of the parameters of  the analysis on events outside the LAT field of view or with low exposure are set to \textsc{NaN}. The CSV format version of the results of our work is available at the webpage\footnote{\url{https://www-glast.stanford.edu/pub\_data/1807/}}. .

\newpage

\begin{table*}
\centering
\caption{\small \label{table:format_results} Table format of the FRB-event sample results.}
\small
\hspace*{-0.8cm}
\begin{tabular}{ccc}
\hline \hline \\
Column & Unit & Description\\
\\
\hline
\\
Name & & FRB name (name format: FRB\,YYYYMMDD)\\
RA & $^{\circ}$ & Right ascension angle\\
dec  & $^{\circ}$ & Declination angle\\
DM  & pc cm$^{-3}$ & Dispersion measure\\
z & & Redshift\\
Dist & Mpc & Luminosity distance\\
Flux\_radio & Jy & Radio flux density at peak time\\
Fluence\_radio & Jy ms & Radio fluence\\
Time &  & Time in UTC (YYYY-MM-DD hh:mm:ss.xx)\\
Width & s & FRB width\\
Lum\_radio & erg s$^{-1}$ & Radio luminosity\\
En\_radio & erg & Radio energy\\
\\
\hline
\\
TS\_10s & & Test Statistic ($\delta_T=10$ s).\\
Flux\_gamma\_10s & cm$^{-2}$ s$^{-1}$ & UL on the gamma-ray flux ($\delta_T=10$ s) \\
Eflux\_gamma\_10s & erg cm$^{-2}$ s$^{-1}$ & UL on the gamma-ray energy flux ($\delta_T=10$)\\
Lum\_gamma\_10s & erg s$^{-1}$ & UL on the gamma-ray luminosity ($\delta_T=10$ s)  \\
En\_gamma\_10s & erg & UL on the gamma-ray energy ($\delta_T=10$ s)\\

TS\_100s & & Test Statistic ($\delta_T=100$ s)\\
Flux\_gamma\_100s & cm$^{-2}$ s$^{-1}$ & UL on the gamma-ray flux ($\delta_T=100$ s) \\
Eflux\_gamma\_100s & erg cm$^{-2}$ s$^{-1}$ & UL on the gamma-ray energy flux ($\delta_T=100$ s)\\
Lum\_gamma\_100s & erg s$^{-1}$ & UL on the gamma-ray luminosity ($\delta_T=100$ s)  \\
En\_gamma\_100s & erg & UL on the gamma-ray energy ($\delta_T=100$ s)\\

TS\_1000s & & Test Statistic ($\delta_T=1000$ s)\\
Flux\_gamma\_1000s & cm$^{-2}$ s$^{-1}$ & UL on the gamma-ray flux ($\delta_T=1000$ s) \\
Eflux\_gamma\_1000s & erg cm$^{-2}$ s$^{-1}$ & UL on the gamma-ray energy flux ($\delta_T=1000$ s)\\
Lum\_gamma\_1000s & erg s$^{-1}$ & UL on the gamma-ray luminosity ($\delta_T=1000$ s)  \\
En\_gamma\_1000s & erg & UL on the gamma-ray energy ($\delta_T=1000$ s)\\

TS\_10000s & & Test Statistic ($\delta_T=10000$ s)\\
Flux\_gamma\_10000s & cm$^{-2}$ s$^{-1}$ & UL on the gamma-ray flux ($\delta_T=10000$ s) \\
Eflux\_gamma\_10000s & erg cm$^{-2}$ s$^{-1}$ & UL on the gamma-ray energy flux ($\delta_T=10000$ s)\\
Lum\_gamma\_10000s & erg s$^{-1}$ & UL on the gamma-ray luminosity ($\delta_T=10000$ s)  \\
En\_gamma\_10000s & erg & UL on the gamma-ray energy ($\delta_T=10000$ s)\\
 & & \\ 
Delay\_best\_triplet & s & Delay (since FRB time) of the shortest duration triplet\\
Triplet\_post\_trial\_p\_value & & P-value of the first triplet after the FRB event\\
Triplet\_duration & s & Interval of time between the first and third photon of the triplet\\
\\
\hline \\
\end{tabular}
\end{table*}


\end{appendix}

\end{document}